\documentclass{jfm}

\usepackage{graphicx}
\usepackage{newtxtext}
\usepackage{newtxmath}
\usepackage{natbib}
\usepackage{tikz}
\usepackage{hyperref}
\usepackage{longtable}
\usepackage{pdflscape}
\usepackage{afterpage}
\usepackage{cleveref}
\hypersetup{
    colorlinks = true,
    urlcolor   = blue,
    citecolor  = red,
}

\newcommand{\RomanNumeralCaps}[1]
\newcommand{\AxisRotator}[1][rotate=0]{%
    \tikz [x=0.25cm,y=0.7cm,line width=.1ex,-stealth,#1] \draw (0,0) arc (-150:150:1 and 1);%
}
\newcommand{\wT}{\overline{\langle wT \rangle} }
\newcommand{\T}{\overline{\langle T \rangle} }
\newcommand{\epsnu}{\overline{\langle \epsilon_\nu \rangle} }
\newcommand{\epsth}{\overline{\langle \epsilon_\theta \rangle} }
\newcommand{\lepsnu}{\langle \epsilon_\nu \rangle} 
\newcommand{\lepsth}{\langle \epsilon_\theta \rangle} 
\newcommand{\Fb}{\mathcal{F}_B }

\definecolor{colorbar15}{rgb}{0.636364,0.000000,0.000000}
\definecolor{mygrey}{rgb}{0.6,0.6,0.6}

\title{
Prandtl number dependence of rotating internally heated convection
}

\author{
Rodolfo Ostilla-M\'onico\aff{1}
  \corresp{\email{rodolfo.ostilla@uca.es}}
 \and Ali Arslan\aff{2}}

\affiliation{
\aff{1}Dpto.~Ing. Mec\'anica y Dise\~no Industrial, Escuela Superior de Ingenier\'ia, Universidad de C\'adiz, Av.~de la Universidad de C\'adiz 10, 11519 Puerto Real, Espa\~na
\aff{2}Institute of Geophysics, ETH Zurich, 8092 Zurich, Switzerland}

\begin{document}
\maketitle

\begin{abstract}
%\nolinenumbers
We investigate the influence of the Prandtl number ($Pr$) on penetrative internally heated convection (IHC) in both non-rotating and rotating regimes using three-dimensional direct numerical simulations. By varying $Pr$ between 0.1 and 100, we show that the global mean temperature $\T$ is not very sensitive to $Pr$, and is primarily controlled by the dynamics of the unstably stratified top boundary layer. In contrast, the Prandtl number dictates the behavior of the lower, stably stratified region and affects the vertical convective heat flux $\wT$. In the non-rotating case, low $Pr$ fluids exhibit a ``symmetry recovery'' where turbulent stirring agitates the stable layer, whereas high $Pr$ fluids transition toward a ``dead zone'' of suppressed fluctuations. Under rotation, we find that $\wT$ is enhanced across all Prandtl numbers, though global cooling efficiency, measured by the reduction in $\T$, is only improved for $Pr\ge1$ due to the emergence of Ekman pumping. These results demonstrate that while IHC shares some scaling similarities with Rayleigh-B\'enard convection at the top boundary, the internal stratification creates a unique sensitivity to $Pr$ that is critical for understanding heat transport in planetary and stellar interiors.
\end{abstract}

\begin{keywords}
%\nolinenumbers
Turbulent convection, Rotating flows
\end{keywords}
% \tableofcontents

%\nolinenumbers
\section{Introduction}
\label{sec:intro}

Turbulent convection is ubiquitous in nature. Driven by thermal or chemical gradients, this buoyancy-induced motion occurs throughout the atmospheres and interiors of rotating celestial bodies \citep{marshall1999,roberts2013}. In these systems, the competition between Coriolis-driven columnar structures and buoyant plumes dictates the mixing dynamics of atmospheres, mantles, and cores \citep{emanuel1994,aurnou2015}. While the relative strength of these forces is traditionally characterized by the Rayleigh ($R$) and Ekman ($E$) numbers, a fundamental challenge in accurately modeling these environments lies in the extreme diversity of the fluid properties themselves. In particular, in Nature there is a vast range of Prandtl numbers ($Pr$), i.e.~the ratio of viscous to thermal diffusion, ranging from being $O(1)$ in atmospheres and $10^{-1}$ to $10^{-2}$ in liquid iron cores to $\sim10^{23}$ in planetary mantles. Such variations fundamentally alter how heat is processed and how rotation constrains the flow.

The canonical model used to describe rotating convection is the Boussinesq approximation to the Navier-Stokes equations in a rotating frame, where the centrifugal force is neglected under either a constant density or small rotation assumptions \citep{ecke2023}. While the vast majority of literature focuses on convection driven by boundary temperature gradients in the classic Rayleigh-B\'enard (RB) problem \citep{chandrasekhar1953,veronis1959,grossmann2000scaling,vorobieff2002,stevens2013,Kunnen04052021,ecke2023,lohse2024ultimate}, we instead focus on internally heated convection (IHC) between two perfectly conducting horizontal boundaries, where the fluid is heated volumetrically \citep{roberts1967convection,tritton1967,thirlby1970, kulacki1972,goluskin2012convection,goluskin2016penetrative,bouillaut2021,hadjerci2024}. This configuration is fundamentally different from the classical Rayleigh–Bénard (RB) problem in several ways. Notably, IHC supports subcritical convection; the Rayleigh number determining the energy stability limit ($R_E\approx 26,926.6$ for no-slip boundaries) is significantly lower than the Rayleigh number determining the linear instability threshold ($R_L\approx37,325.2$), whereas RB convection onsets via a pitchfork bifurcation at $R_E=R_L\approx1,707.76$ \citep{Goluskin2016book, chandrasekhar1961hydrodynamic}. In addition, because heat is not transported from one boundary to the other, the standard Nusselt number used in RB studies \citep{lohse2023} is not applicable. Instead, two global quantities are used to assess the effectiveness of convection: the mean temperature $\T$, and the bottom heat flux heat fraction $\mathcal{F}_B$, where $\langle .\rangle$ denotes averaging in the horizontal directions and time, and $\overline{\phi}$ denote averages in the vertical direction. 

In IHC, turbulence serves a dual role: it homogenizes the bulk temperature field while simultaneously creating an asymmetry in the heat flux leaving the boundaries, a flux that is perfectly balanced in the purely conductive state \citep{Arslan2021}. This asymmetry results in an unstably stratified thermal boundary layer at the top and a stably stratified layer at the bottom. The distinct structures of these layers contribute to scaling laws for viscous dissipation that remain less understood than their thermal counterparts \citep{Wang2020,arslan2025a}, with evidence suggesting that the viscous dissipation scales differently between two- and three-dimensional systems \citep{goluskin2016penetrative,ostilla2025ihc}.

The disparity between the top and bottom boundary layers is further accentuated by rotation. In rotating RB convection, the relative thicknesses of thermal and viscous boundary layers serve as a key diagnostic for identifying the transition from buoyancy-dominated flow to geostrophic turbulence \citep{king2009b, maffei2021inverse}. A hallmark of this transition is the emergence and subsequent decoherence of Taylor columns as the rotation rate increases \citep{grooms2010, kunnen2016transition}. However, in rotating IHC, the stable bottom stratification interacts with the Ekman boundary layer, a phenomenon absent in boundary-driven systems which complicates the picture \citep{ostilla2025ihc}. The effect of $Pr$ on this phenomena is yet to be unveiled.

The morphology of these flows is also heavily dictated by the Prandtl number. In the limits of $Pr\to\infty$ and $Pr\to0$, the momentum equation approaches forced Stokes and Euler flows, respectively. While $Pr$ effects have been extensively documented for rotating RB convection \citep{King2013, horn2017prograde,aurnou2018rotating,vogt2018jump,abbate2023rotating,fan2024scaling,xu2025thermovelocimetric}, the low-$Pr$ regime is computationally demanding and remains relatively unexplored for IHC. One critical feature of rotating RB convection that vanishes at low $Pr$ is the rotational enhancement of heat transport. In RB systems, this enhancement is attributed to Ekman pumping, which becomes ineffective at low Pr as high thermal diffusivity allows heat to escape the boundary layers before it can be vertically transported \citep{Stevens_2010}. Whether this suppression of transport enhancement persists in internally heated systems, where the driving is volumetric rather than boundary-based, remains an open question. 

This manuscript builds upon the work of \cite{ostilla2025ihc} to provide the first direct numerical simulations (DNS) of low-$Pr$ through high-$Pr$ rotating IHC. We conduct a parametric study across a wide range of $R$ and $E$ to quantify the bulk responses, horizontally averaged profiles, and boundary layer dynamics of this unique system. The paper is organised as follows. $\S$\ref{sec:nummeth} describes the governing equations and simulation method. In $\S$\ref{eq:non_rotating_ihc}, we present results for the variation of $Pr$ on non-rotating IHC, while $\S$\ref{sec:rihc} contains results for the effect of $Pr$ on rotating IHC. Finally, $\S$\ref{sec:conc} briefly summarises the results obtained.

\section{Setup}
\label{sec:nummeth}
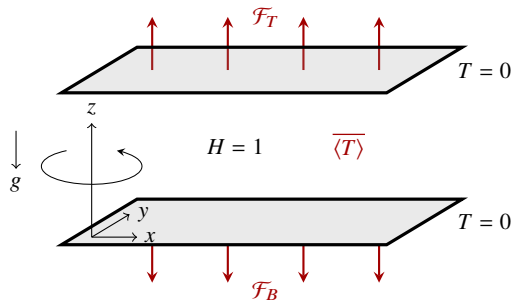
\begin{figure}
\centering
\begin{tikzpicture}[every node/.style={scale=0.95}]

    \draw [-stealth,colorbar15, thick] (-0.8,0) -- (-0.8,-0.5);
    \draw [-stealth,colorbar15, thick]  (0.2,0) -- (0.2,-0.5);
    \draw [-stealth,colorbar15, thick] (1.2,0) -- (1.2,-0.5);
    \draw [-stealth,colorbar15, thick]  (2.2,0) -- (2.2,-0.5);
    \draw [-stealth,colorbar15, thick]  (-0.8,2.3) -- (-0.8,3);
    \draw [-stealth,colorbar15, thick]  (0.2,2.3) -- (0.2,3);
    \draw [-stealth,colorbar15, thick]  (1.2,2.3) -- (1.2,3);
    \draw [-stealth,colorbar15, thick]  (2.2,2.3) -- (2.2,3);
    \draw[black,very thick, fill=mygrey, fill opacity = 0.2] (-2,2) -- (-1,2.6) -- (3.3,2.6) -- (2.3,2) -- cycle;
    \draw[black,very thick, fill=mygrey, fill opacity = 0.2] (-2,0) -- (-1,0.6) -- (3.3,0.6) -- (2.3,0) -- cycle;
    \draw[->] (-2.6,1.5) -- (-2.6,1) node [anchor=north] {$g$};
    \draw[->] (-1.6,0.1) -- (-1,0.1) node [anchor=west] {$x$};
    \draw[->] (-1.6,0.1) -- (-1.1,0.4) node [anchor=west] {$y$};
    \draw[->] (-1.6,0.1) -- (-1.6,1.6) node [anchor=south] {$z$};
    \node at (-1.6,1) {\AxisRotator[rotate=-90]};
    % \node at (-2.1,0.4) {$z=0$};
    % \node at (-2.1,2.4) {$z=1$};
    \node at (3.6,0.3) {$ T = 0 $};
    \node at (3.6,2.3) {$  T = 0 $};
    \node at (0.7,-0.6) {${\color{colorbar15}\mathcal{F}_B }  $};
    \node at (1.8,1.3) {${\color{colorbar15}\overline{\langle T\rangle}}  $};
    \node at (0.7,3) {${\color{colorbar15}\mathcal{F}_T }  $}; 
    \node at (0.3,1.3) {$ H = 1$};
    \end{tikzpicture}
\caption{A non-dimensional schematic diagram for rotating uniform internally heated convection. The upper and lower plates are at the same temperature, and the domain is periodic in the $x$ and $y$ directions and rotates about the $z$ axis. $\mathcal{F}_B$ and $\mathcal{F}_T$ are the mean heat fluxes out the bottom and top plates, $\overline{\langle T \rangle}$ the mean temperature, and $g$ is the acceleration due to gravity.}
\label{fig:schema}
\end{figure}

In this study, internally heated convection is examined within a fluid layer of depth $d$, bounded by two horizontal plates, as shown in Figure \ref{fig:schema}. We assume periodic boundary conditions in the horizontal dimensions with a domain aspect ratio of $\Gamma d$. The physical properties of the fluid, i.e.~the kinematic viscosity $\nu$, the thermal diffusivity $\kappa$, the density $\rho$, the specific heat $c_p$ and the thermal expansion coefficient $\alpha$, are assumed constant. The system undergoes uniform internal heating at a rate $H$, and rotates at a constant angular velocity $\Omega$ about the vertical axis, with gravity $g$ acting downward. We employ the Boussinesq approximation to account for density variations solely through the buoyancy term.

Following the scaling convention of \citet{roberts1967convection}, the system is non-dimensionalised using $d$ for length, $d^2/\kappa$ for time, and $d^2H/(\kappa \rho c_p)$ for temperature. Under this non-dimensionalisation, the evolution of the velocity field, $\boldsymbol{u}(\boldsymbol{x},t)=u(\boldsymbol{x},t)\boldsymbol{e}_1+v(\boldsymbol{x},t)\boldsymbol{e}_2+w(\boldsymbol{x},t)\boldsymbol{e}_3$ and temperature field $T(\boldsymbol{x},t)$  is governed by the following equations:

\begin{gather}
    \bnabla \cdot\boldsymbol{u}=0, \label{eq:navierstokes1}\\
    \partial_t \boldsymbol{u} + \boldsymbol{u}\cdot\bnabla\boldsymbol{u} + E^{-1} \boldsymbol{e}_3 \times\boldsymbol{u} = -\bnabla p + Pr \nabla^2\boldsymbol{u} + Pr R T\boldsymbol{e}_3, \label{eq:navierstokes2}\\
    \partial_tT + \boldsymbol{u}\cdot\bnabla T = \nabla^2 T +1 ,\label{eq:navierstokes3}
\end{gather}

\noindent the velocity and temperature fields are constrained by no-slip ($\boldsymbol{u}=0$) and isothermal ($T=0$) conditions at both the upper and lower boundaries. 

The dynamical response of the system is determined by three fundamental dimensionless groups: the Rayleigh number ($R$), the Ekman number ($E$), and the Prandtl number ($Pr$):

\begin{equation}
    R = \frac{g\alpha Hd^5}{\rho c_p \nu \kappa^2}, \qquad E=\frac{\nu}{2\Omega d^2}, \qquad Pr=\frac{\nu}{\kappa}.
\end{equation}

\noindent Additionally, we can obtain the convective Rossby number, $Ro=E\sqrt{R/Pr}$, to quantify the competition between buoyant and Coriolis forces.

The governing equations \ref{eq:navierstokes1}-\ref{eq:navierstokes3} are numerically integrated using AFiD \citep{van2015pencil}. This open-source code employs a second-order centered finite difference scheme and has been thoroughly benchmarked for internal heating scenarios \citep{goluskin2016penetrative, kazemi2022transition, ostilla2025ihc}.

In this study, we extend the results of \cite{ostilla2025ihc} by varying the Prandtl number to explore the ($Pr$,$R$,$E$) parameter space of rotationally-affected convection. We vary $Pr$ in the range $Pr\in[0.1,10]$, $R$ in the range $R\in[3.16\times 10^5,10^{10}]$ and $E\in[10^{-6},\infty]$. To be more specific, we extend the parameter space explored in \cite{ostilla2025ihc} by repeating the $(R,E)$ pairs for four different values of $Pr$: $Pr=0.1$, $0.3$, $3$ and $10$. We also run a series of non-rotating cases ($E=\infty$) cases for $Pr=30$ and $100$. 

The non-dimensional periodicity length $\Gamma$ is selected following the values of \cite{ostilla2025ihc} for the $Pr=1$ cases, except for $R=10^9$ where $\Gamma$ is reduced to 1. For $Pr\geq 1$, this ensures that $\T$ and $\wT$ remain independent of domain size. For $Pr\leq 1$, a dependence $\wT$ on the horizontal periodicity length $\Gamma$ is introduced. While this does not change the overall physics of the problem, we quantify its effect in Appendix \ref{sec:app}.

The resolution adequacy is measured by the exact relationships:

\begin{gather}
    \overline{\langle\epsilon_\nu\rangle}  \equiv \overline{\langle | \nabla \boldsymbol{u}|^2 \rangle} = R\overline{\langle wT\rangle}, \label{eq:exrel1}\\ 
    \overline{\langle\epsilon_\theta\rangle} \equiv \overline{\langle |\nabla T|^2\rangle} = \overline{\langle T \rangle} . \label{eq:exrel2}
\end{gather}

\noindent To guarantee numerical accuracy, we ensure that both sides of the equation do not deviate by more than 1\%. This criterion provides a more stringent resolution constraint than those based solely on the mean Kolmogorov or Batchelor scales, yet it is recognized as sufficient for characterizing thermal convection \citep{stevens2010radial}.

We base our resolutions on the values used \cite{ostilla2025ihc}, which satisfy the criteria above for all $Pr$. These yield grids from $288^2\times144$ points to the largest grid of $384^2\times512$. As a general rule, the number of points in the vertical direction increases as $E$ becomes smaller, while the number of points in the horizontal direction does not vary as much because the required periodicity length $\Gamma$ decreases with $R$, so grid resolution is increased despite keeping the same number of points. In the most adverse cases, the discretisation used in this manuscript has a maximum point distance of $1.5$ Kolmogorov length-scales. In the non-rotating case, the upper thermal boundary layer, whose extent can be estimated by $\langle\overline{T}\rangle/(\frac{1}{2}+\wT)$ is resolved by at least six points. As rotation increases, the number of points in the $z$ direction must be increased to resolve the increasingly thinner Ekman (velocity) boundary layers.  A full list of all the resolutions used for the new simulations can be found in Appendix \ref{sec:appb}.

For $Pr=0.3$ and $Pr=3$, simulations are initialised from initial conditions at $Pr=1$ with the same ($R$,$E$) parameter values, while for $Pr=0.1$ and $Pr=10$ they are started from the $Pr=0.3$ and $Pr=3$ cases respectively. After transients, statistics are collected for a time interval of length $t_{av}$ which ensure that the time-averaged values of $\T$ and $\wT$ are less than 1$\%$ different from those corresponding to half the run-time. In general $t_{av}$ depends on both $Pr$ and $R$, as the flow becomes faster with increasing $R$ (decreasing the necessary averaging time) but increases with $Pr$ as the flow becomes slower. 

\section{Non-rotating IHC}
\label{eq:non_rotating_ihc}

\subsection{Flow visualization}

\begin{figure}
\centering
 \includegraphics[width=.24\linewidth]{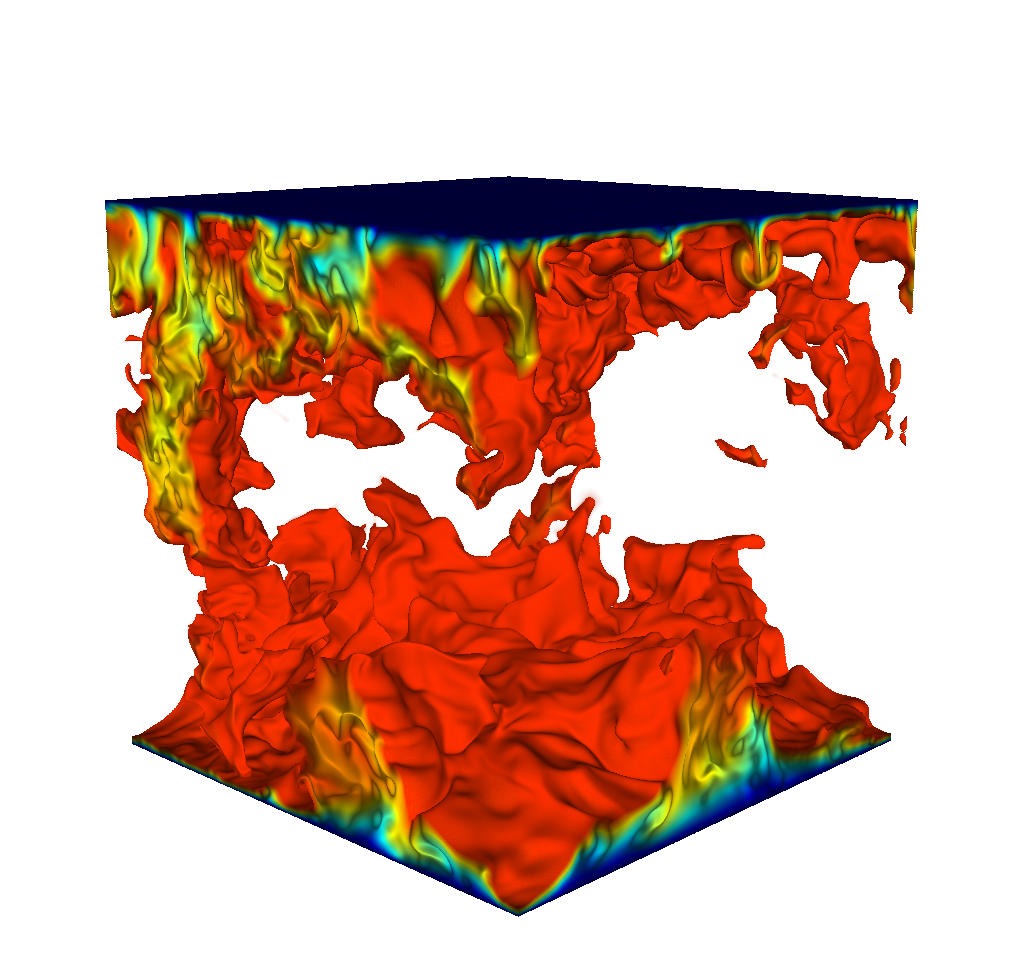}
 \includegraphics[width=.24\linewidth]{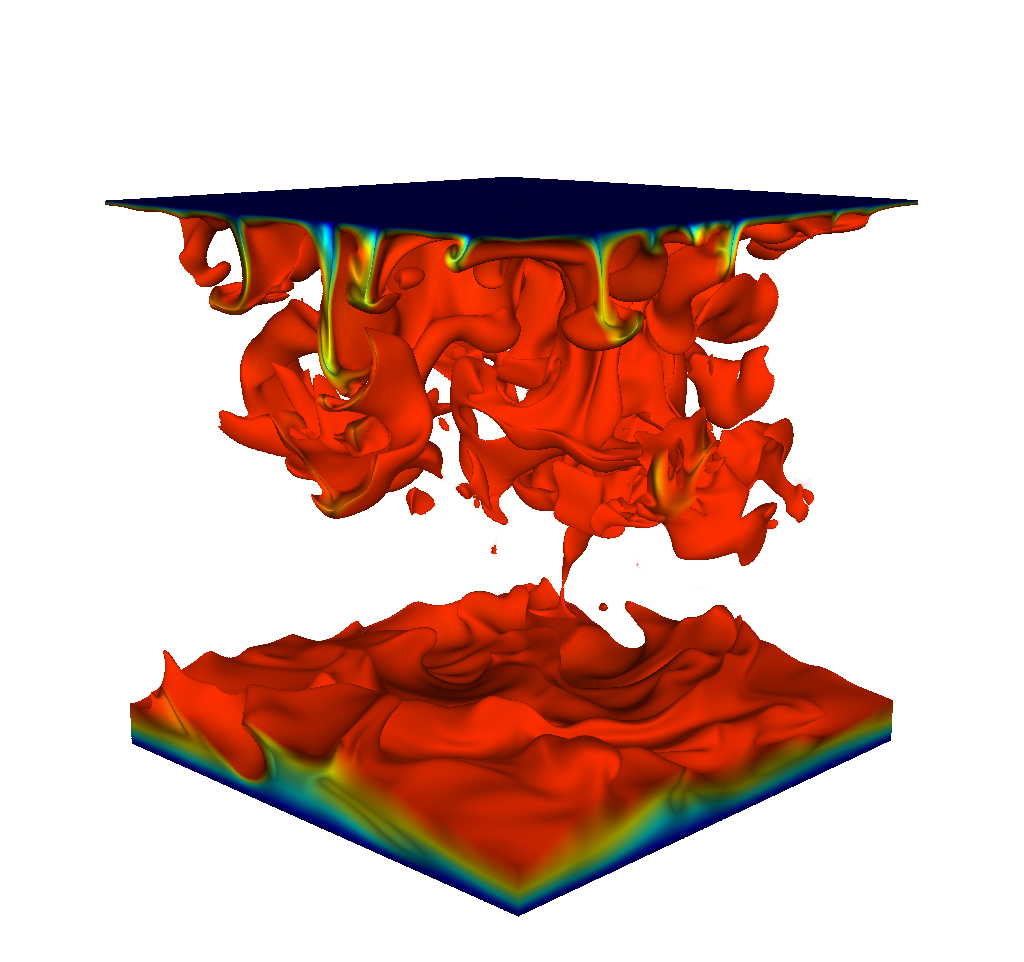}
 \includegraphics[width=.24\linewidth]{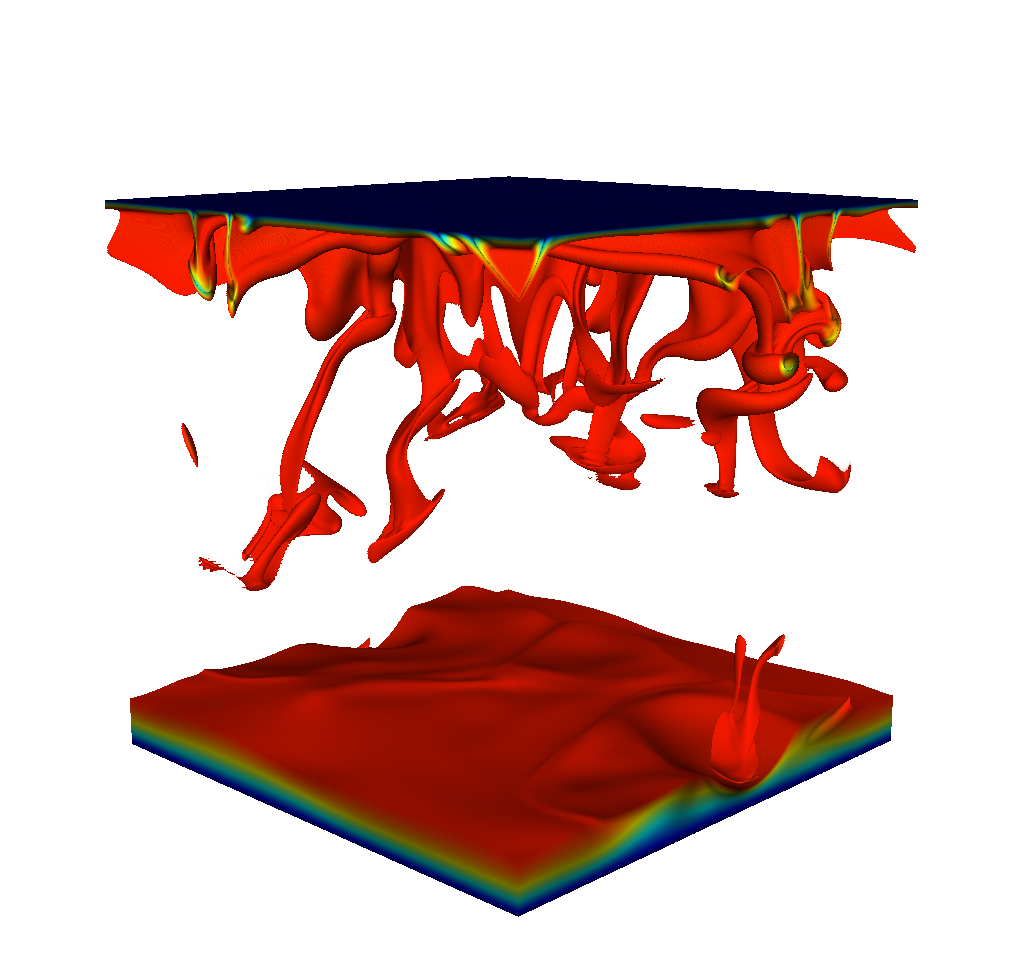}
 \includegraphics[width=.24\linewidth]{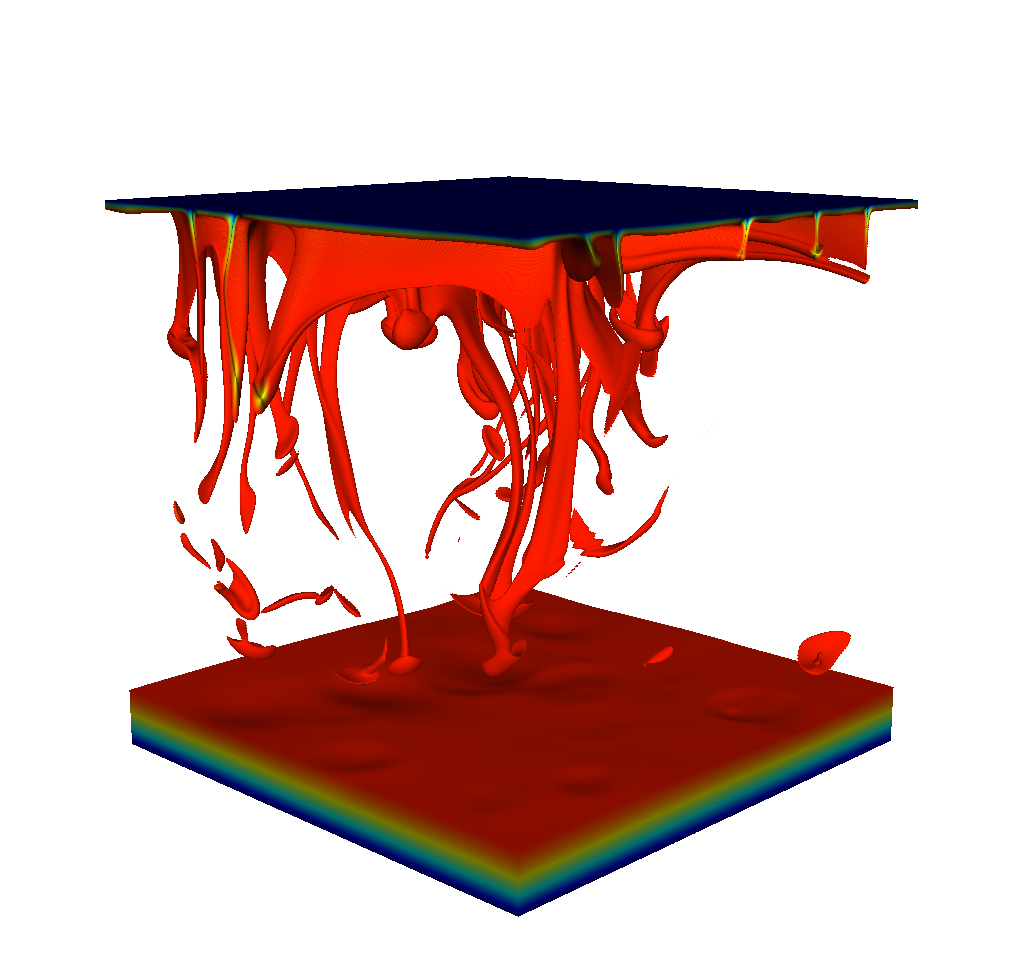}
\caption{Volumetric visualisation of the instantaneous temperature field without rotation for $R=10^{10}$ and from left to right: $Pr=0.1$, $Pr=1$, $Pr=10$ and $Pr=100$.}
\label{fig:flowvis-norot}
\end{figure}

We begin by examining cases without rotation to characterise the influence of the Prandtl number on the flow structure. Figure \ref{fig:flowvis-norot} illustrates the morphological evolution of the flow as $Pr$ is varied. In all instances, convective structures and plumes originate within the unstably stratified upper boundary layer before descending into the bulk. At lower Prandtl numbers ($Pr=0.1$ and $Pr=1$), these structures vigorously stir the lower boundary layer, generating widespread turbulence throughout the domain. Conversely, for $Pr=10$, the dynamics of the bottom boundary layer are noticeably diminished, and the flow field is dominated by coherent, elongated plume-like structures. For $Pr=100$, the plume structures become even thinner, while the bottom of the domain is practically quiescent.

\subsection{Global quantities}

To characterise the global system response, we consider three primary dimensionless parameters: the bottom heat flux fraction $\mathcal{F}_B$, the volume-averaged temperature $\T$ and the wind Reynolds number $Re_w=U_wd/\nu$. Here, the characteristic wind velocity $U_w$ is defined as the root-mean-square velocity:

\begin{equation}
 U_w = \overline{\langle u^2+v^2+w^2\rangle}^{1/2}.
\end{equation}

\begin{figure}
\centering
\includegraphics[width=.49\linewidth]{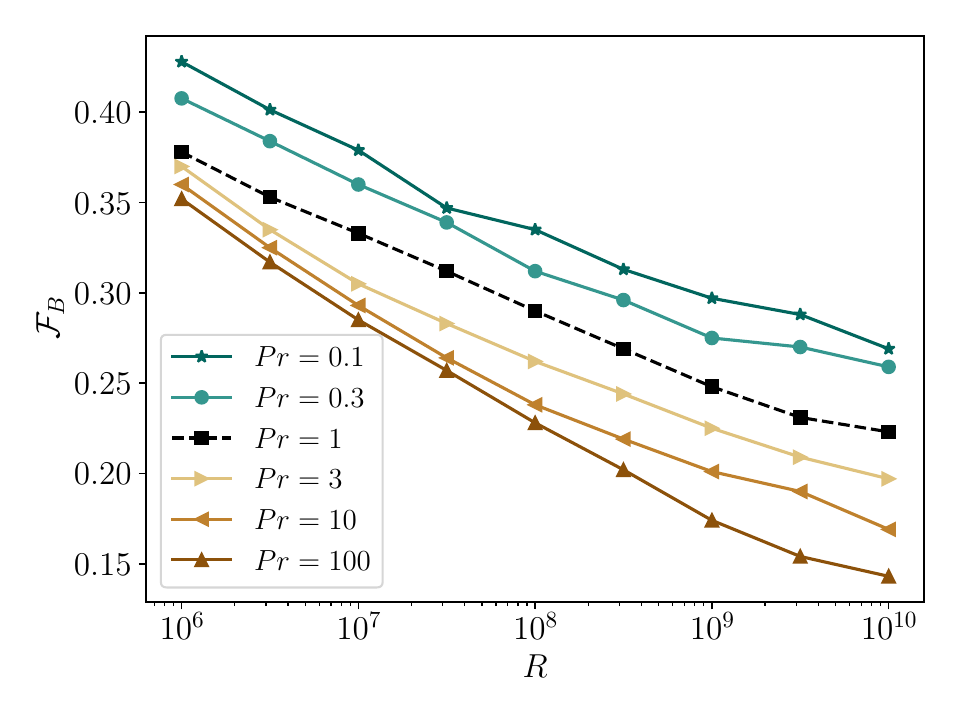}
\includegraphics[width=.49\linewidth]{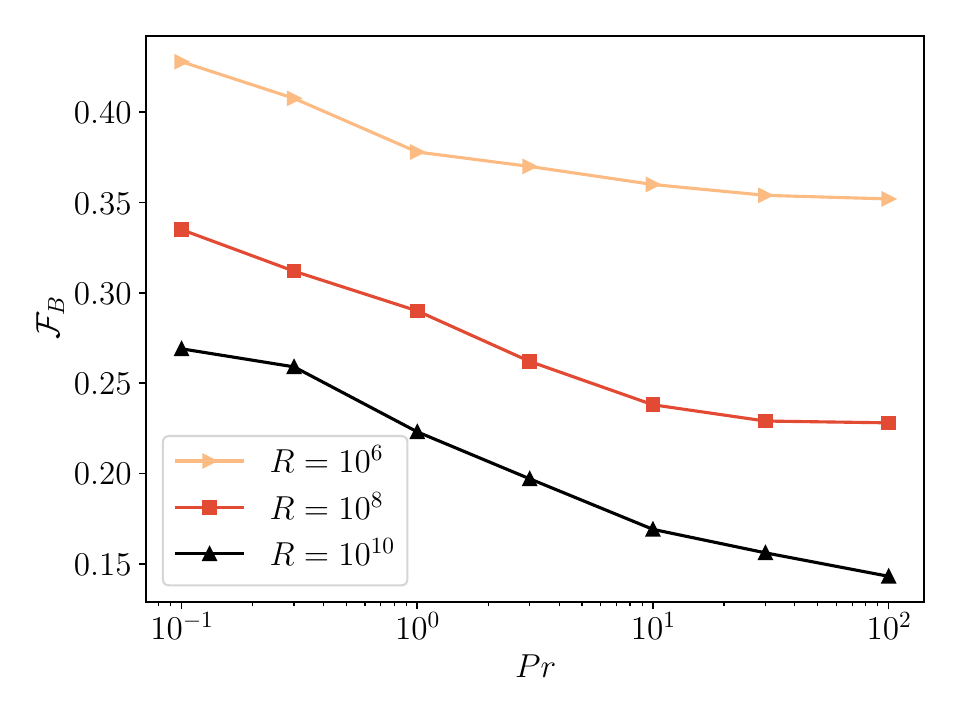}
\includegraphics[width=.49\linewidth]{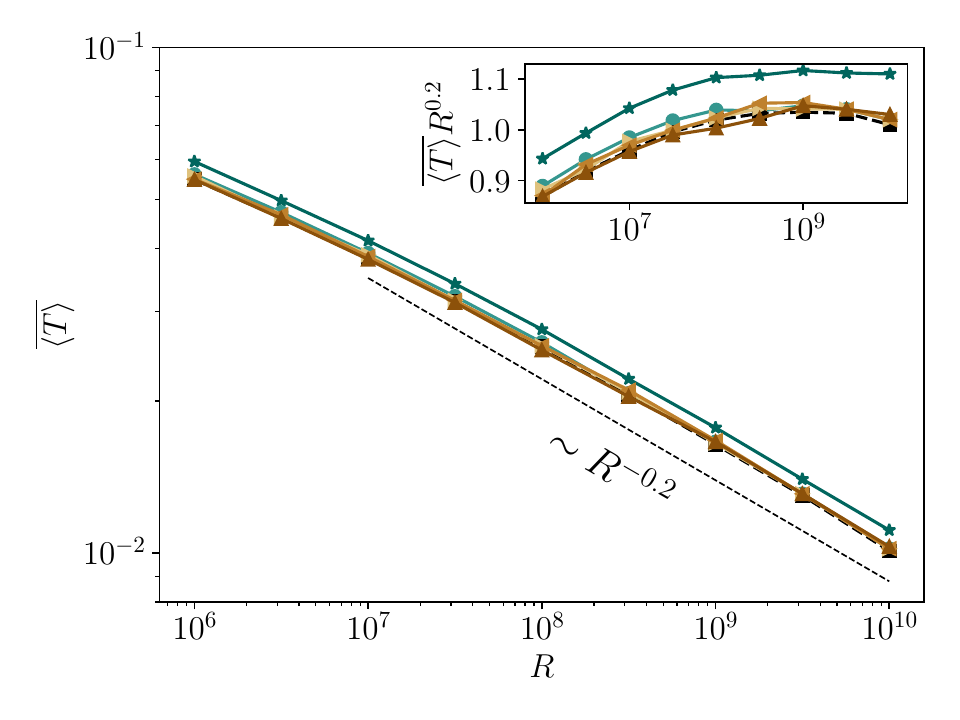}
\includegraphics[width=.49\linewidth]{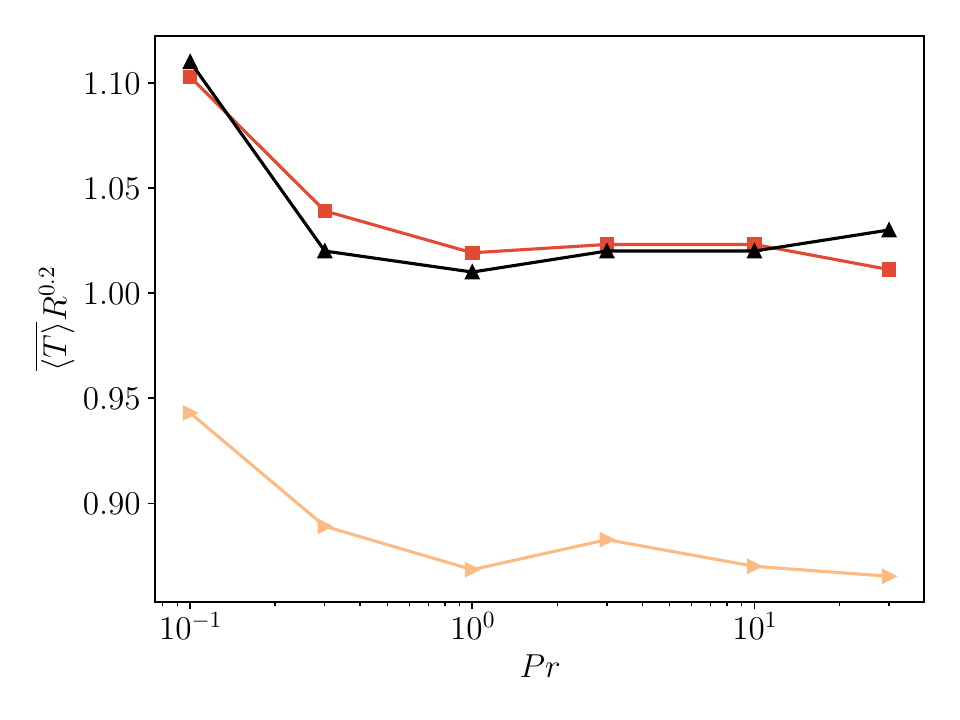}
\includegraphics[width=.49\linewidth]{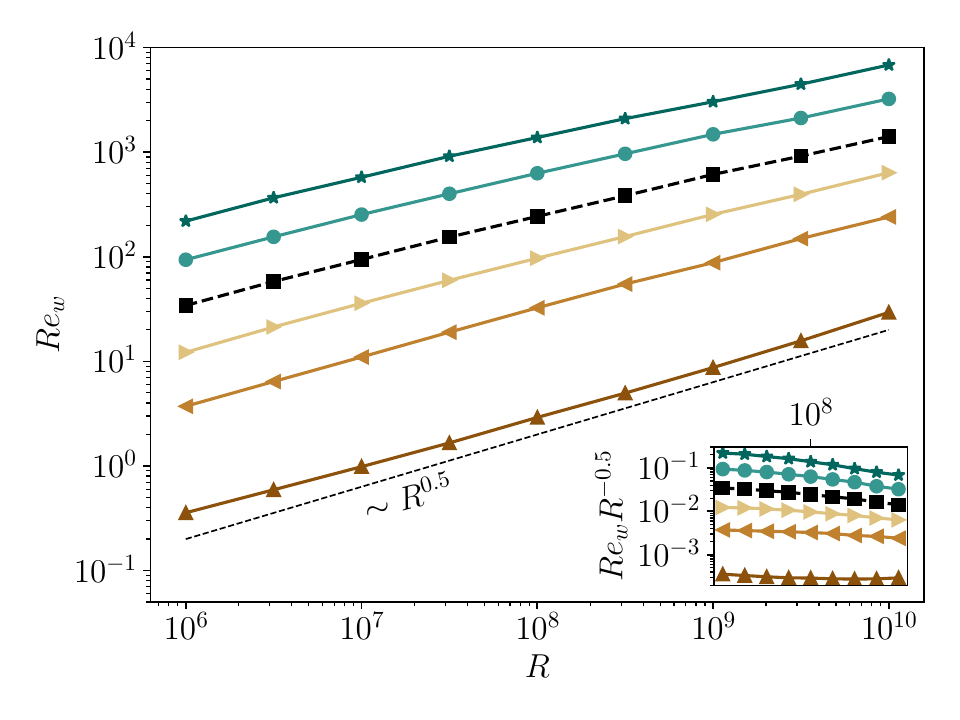}
\includegraphics[width=.49\linewidth]{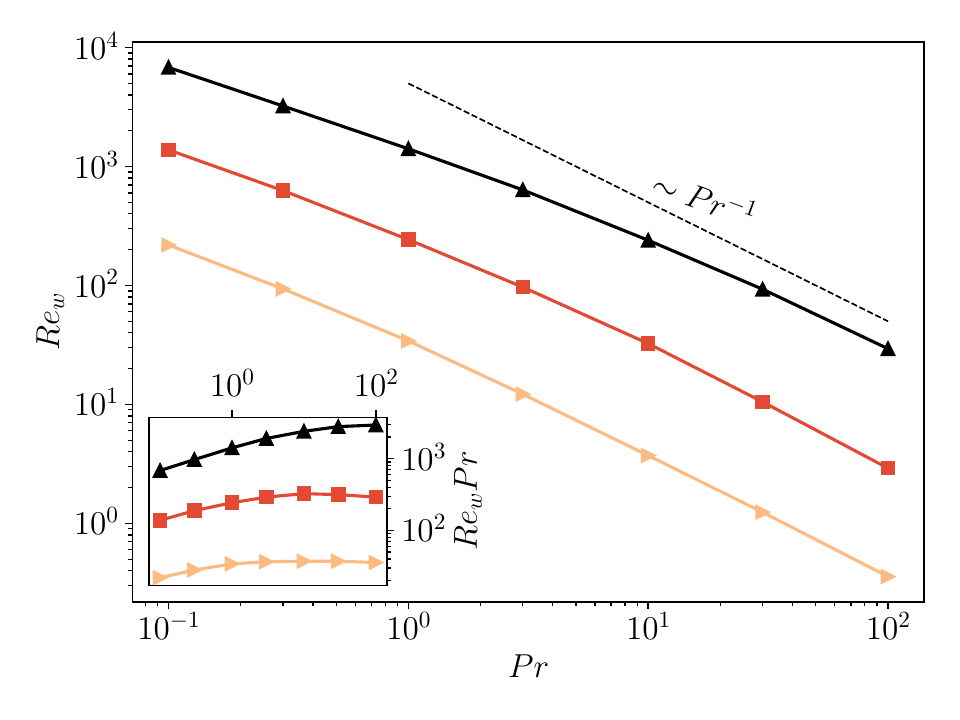}
\caption{Global responses. $\mathcal{F}_B$ (top), $\overline{\langle T \rangle}$,  (middle) and $Re_w$ (bottom) against $R$ (left) and $Pr$ (right). For clarity, only selected values of $Pr$ are shown on the left column plots.}
\label{fig:globalqs-norot}
\end{figure}

The top panels of Figure \ref{fig:globalqs-norot} depict the behaviour of $\mathcal{F}_B$ across the ($R$,$Pr$) parameter space. We note that $\mathcal{F}_B$ is the quantity conventionally used when representing mathematical upper bounds, but one could also represent $\mathcal{F}_T$, the upper heat fraction, or the vertical convection $\wT$. As in the statistically stationary state of internally heated convection, the total heat generated must be balanced by the flux through the boundaries, the three quantities are not independent but instead related through the relationships:

\begin{equation*}
 \mathcal{F}_B + \mathcal{F}_T = 1 , \qquad \mathcal{F}_B = \frac12 - \wT, \qquad \mathcal{F}_T=\frac{1}{2}+\wT\, .
\end{equation*}

\noindent Therefore, representing $\wT$ is equivalent to representing $\Fb$ (or $\mathcal{F}_T$). In general, $\wT\ge0$, due to the asymmetry introduced by convection. Consequently, $\Fb$ can be seen as a proxy for the convective heat transport efficiency, but its behaviour is independent from that of $\T$.

Across the investigated parameter space, $\Fb$ exhibits a consistent trend regardless of $Pr$: it decreases as $R$ increases, indicating that a larger proportion of heat escapes through the top boundary. This trend is consistent with enhanced vertical heat transport at higher thermal driving. While 2-D simulations have previously reported non-monotonic behaviour ($Pr\geq 1$) or weak $R$-dependence ($Pr < 1$) due to flow organisation affected by geometric constraints \citep{goluskin2016penetrative}, our 3-D results show that $\Fb$ decreases monotonically with $R$ when such constraints are removed. Furthermore, the 3-D configuration yields significantly higher vertical convective heat transfer, and thus lower $\Fb$ values, across all reported Prandtl numbers.

In general the $Pr$ dependence is such that smaller $Pr$ generally results in less efficient heat transport vertically. While the visualisations in Figure \ref{fig:flowvis-norot} suggest a more turbulent bottom boundary layer at low $Pr$, the high thermal diffusivity at these values renders vertical convection less efficient. In contrast, at high $Pr$, the flow produces intense plumes that transport heat more effectively despite weaker overall stirring. Notably, the $Pr$ dependence becomes weaker but remains unsaturated even at $Pr=100$.

The middle panels of Figure \ref{fig:globalqs-norot} show the volume-averaged temperature $\T$, which follows a decreasing trend with $R$. We find an approximate scaling of $\T\sim R^{-0.2}$ across all simulated $Pr$. This scaling, previously noted for $Pr=1$ by \cite{goluskin2016penetrative}, relates to the dominance of thermal dissipation within the top boundary layer \citep{ostilla2025ihc}. The influence of $Pr$ on the mean temperature is minimal, showing significant variation only at $Pr=0.1$, a behaviour reminiscent of the behaviour of the Nusselt number observed in standard Rayleigh-B\'enard convection of which $\T$ is a direct analog in this system.

Finally, the bottom panels of Figure \ref{fig:globalqs-norot} show the wind Reynolds number $Re_w$. As expected, $Re_w$ increases with $R$ and decreases with $Pr$ due to higher viscous dissipation. Power-law fits of the form $Re_w\sim R^\alpha$ yield exponents in the range $0.38<\alpha<0.5$, with $\alpha$ increasing alongside $Pr$. For the highest $Pr$ values, we observe a rough scaling of $Re_w\sim Pr^{-1}$. This suggests that at low $Re_w$, where the flow is nonlinear but not yet fully turbulent, the system effectively mimics infinite-$Pr$ behavior. However, as $R$ increases and hydrodynamic turbulence becomes significant, this scaling law eventually breaks down.

\subsection{Temperature and velocity statistics}
\label{sec:localstats}

\begin{figure}
\centering
\includegraphics[width=.49\linewidth]{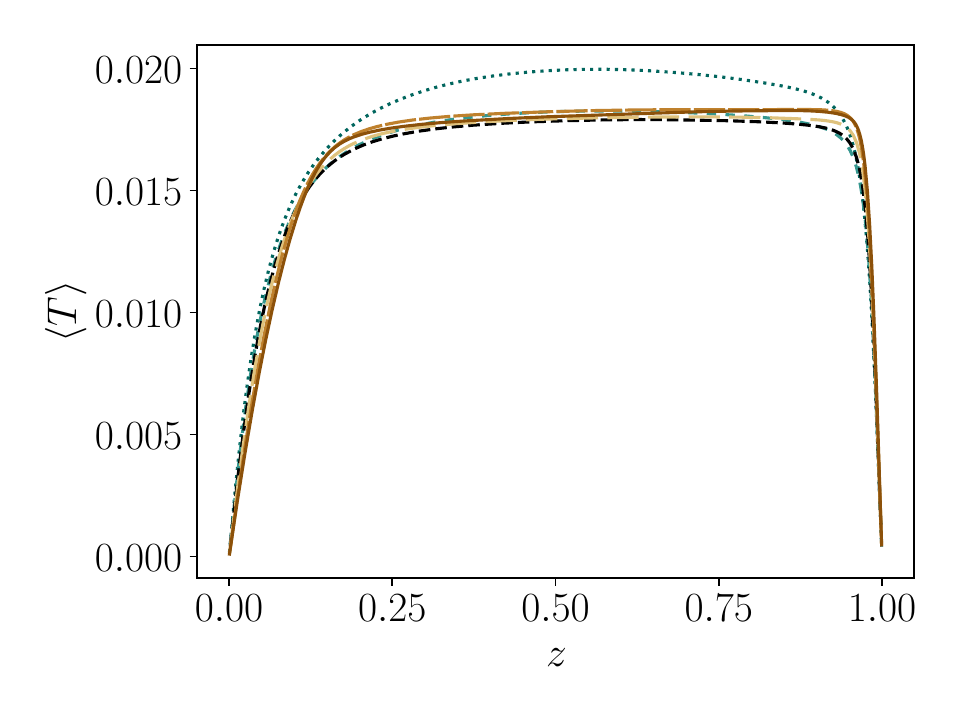}
\includegraphics[width=.49\linewidth]{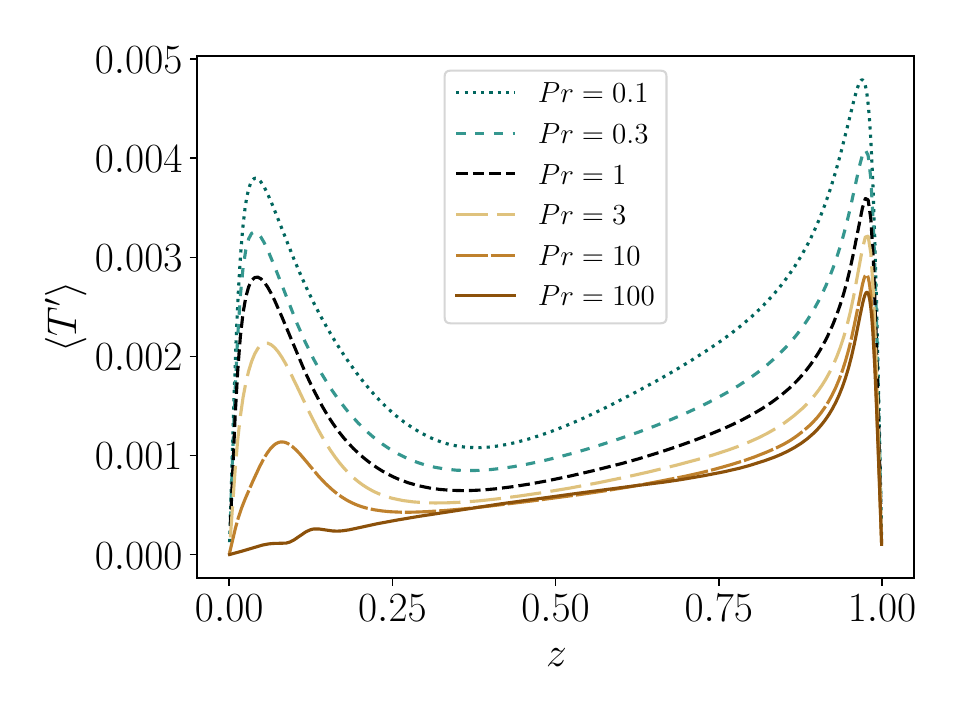}
\includegraphics[width=.49\linewidth]{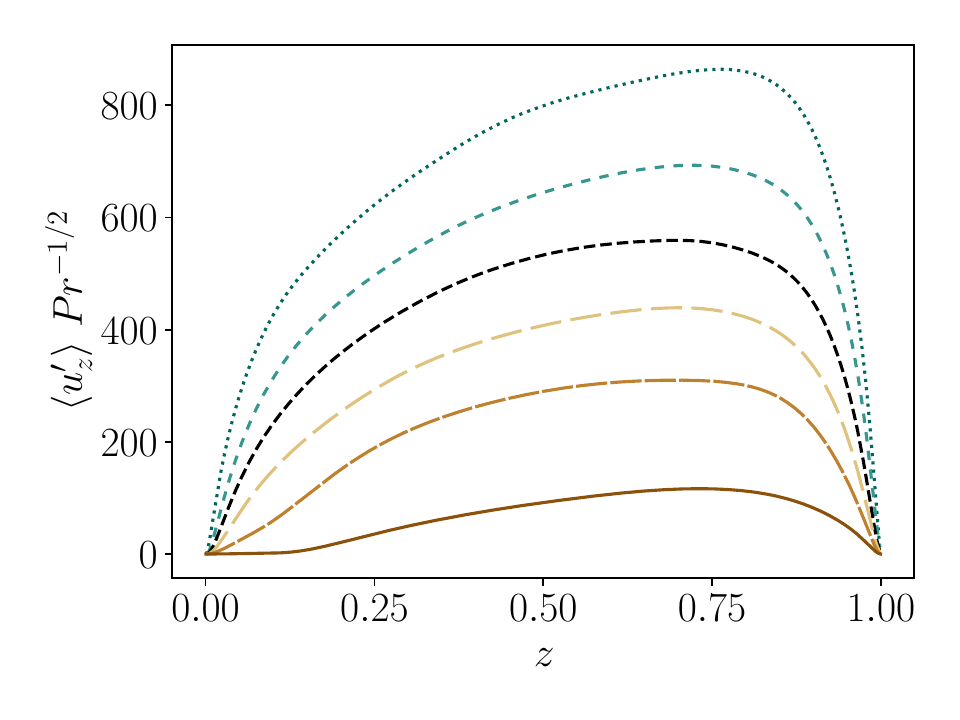}
\includegraphics[width=.49\linewidth]{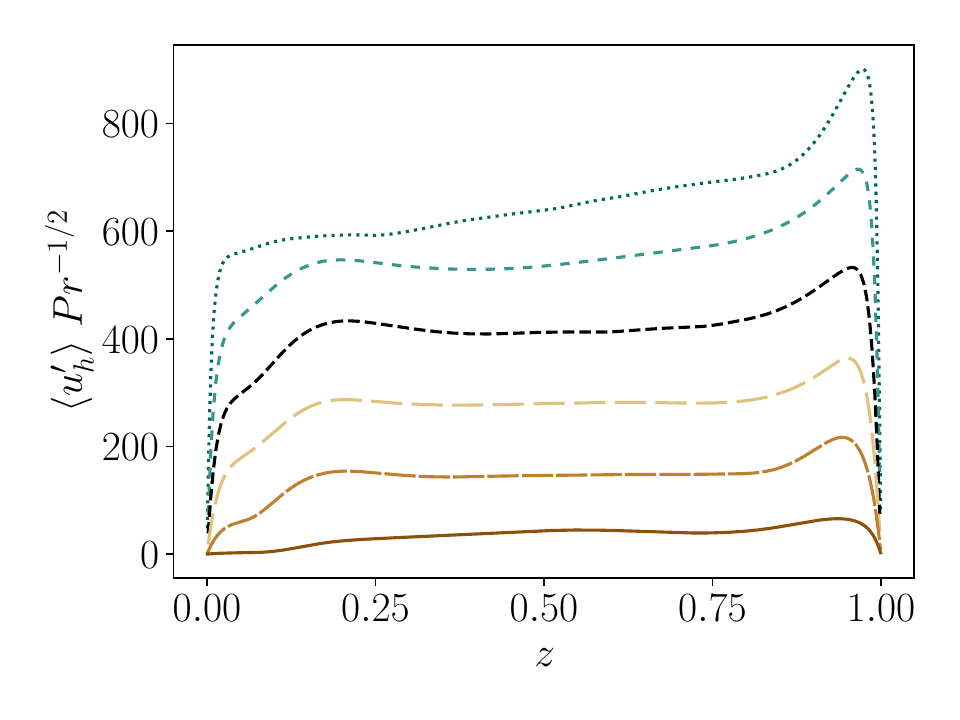}
\caption{ Plots of the horizontally averaged temperature and temperature fluctuation against $z$ (top), and scaled horizontally averaged vertical velocity and horizontal velocity fluctuations against $z$ (bottom).
All plots are for $R=10^9$ and varying Prandtl numbers.}
\label{fig:tempvel-norot}
\end{figure}

To gain further insight into the global behaviors observed, we examine the local statistics of the fluid fields. The top two panels of Figure \ref{fig:tempvel-norot} display the mean temperature $\langle T \rangle$ and RMS fluctuations $\langle T^\prime \rangle = \sqrt{\langle T^2\rangle - \langle T\rangle^2 }$ profiles for various $Pr$ at $R=10^9$. The mean temperature profiles are largely insensitive to the Prandtl number, with the exception of the low-$Pr$ regime, where slightly higher values are observed. Despite these minor magnitude differences, the profile morphology remains consistent with two distinct, steep boundary layers and a relatively flat bulk region.

Substantial differences emerge, however, when comparing the temperature fluctuations $\langle T^\prime \rangle$. Fluctuations are notably higher at small $Pr$, and the asymmetry between the top and bottom boundary layers intensifies as $Pr$ increases. This aligns with our qualitative visualisations. At low $Pr$, vigorous stirring occurs within the stably stratified bottom boundary layer, even though this ultimately corresponds to lower net heat transport. For the highest $Pr$ cases, fluctuations near the bottom plate vanish entirely as the stable stratification and high viscosity effectively damp turbulence, restricting flow activity to the bulk and the upper boundary.

Turning to the velocity statistics, the bottom panels of Figure \ref{fig:tempvel-norot} show the profiles for horizontal velocity fluctuations $\langle u_h^\prime\rangle=\langle u_x^{\prime2}+u_y^{\prime2}\rangle^{\frac{1}{2}}$ and vertical fluctuations $\langle u_z^\prime\rangle$. Note that we have scaled these profiles by $Pr^{-1/2}$ to compensate for the diffusive time-scale used in the non-dimensionalisation. This compensation allows for a more direct comparison of the profile shapes. Without this compensation, the $Pr=100$ magnitudes would dwarf the others, obscuring qualitative shifts. 

The vertical fluctuations generally exhibit a profile that is relatively independent of $Pr$, with a gradual increase from the bottom wall (which is steeper at low $Pr$), followed by a sharp drop to zero at the top boundary. Consistent with the temperature statistics, for $Pr=100$, a stagnant zone is evident at the bottom plate where fluid motion is entirely suppressed.

The horizontal velocity fluctuations reveal a more intricate structure. While all cases exhibit a flat bulk region and a prominent peak representing the velocity boundary layer near the top, the bottom stable layer demonstrates complex $Pr$-dependence. At $Pr=0.1$, there is a simple monotonic increase to the bulk level. However, at $Pr=0.3$, two distinct regions emerge within the bottom layer, an initial sharp rise followed by a more gradual increase. This bifurcation becomes more pronounced with increasing Prandtl. As suggested by \citet{ostilla2025ihc}, these transitions mark the distinction between the viscous boundary layer and the deeper-extending stably stratified region. At higher Prandtl number, the velocity boundary layer is completely damped as the stably stratified zone is stagnant.

\begin{figure}
\centering
\includegraphics[width=.49\linewidth]{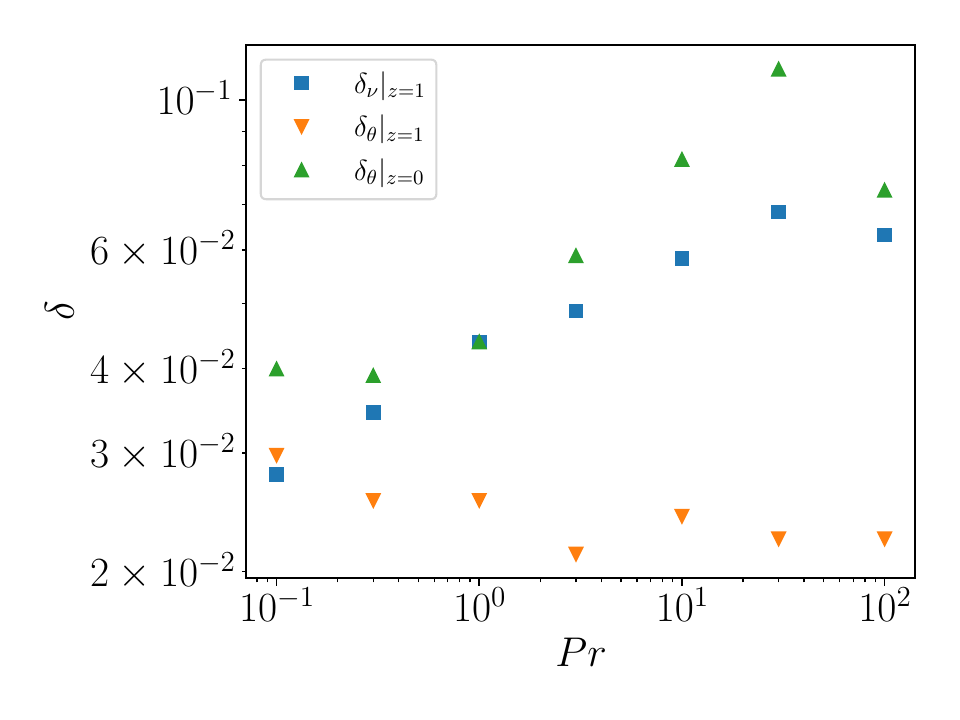}
\caption{Thermal and viscous boundary layer sizes for $R=10^9$ and all values of $Pr$ simulated.}
\label{fig:blsize-norot}
\end{figure}

Figure \ref{fig:blsize-norot} compares the thermal and (top) velocity boundary layer thicknesses, defined through the local maxima in the temperature and horizontal velocity fluctuations respectively. The thickness of the top thermal boundary layer remains nearly constant across the Prandtl number range explored, which is consistent with the nearly constant mean temperature $\T$ (analogous to the behavior of a Nusselt number). Conversely, the bottom thermal boundary layer thickens with increasing $Pr$, except at $Pr=100$ where the lack of a distinct temperature fluctuation peak makes identification difficult. Finally, the top velocity boundary layer thickens as $Pr$ increases, a direct consequence of the increased kinematic viscosity.

\subsection{Dissipation rates}
\label{sec:diss-norot}

To conclude the analysis of non-rotating IHC, we examine the local thermal ($\lepsth$) and viscous ($\lepsnu$) dissipation rates. These quantities are coupled to the global parameters $\wT$ and $\T$ via the exact relations in Eqs.~\ref{eq:exrel1}--\ref{eq:exrel2}. Identifying the relative contributions of the bulk and boundary layers to these dissipation rates is a standard approach in thermal convection studies for determining the factors that limit global transport \citep{grossmann2000scaling,Wang2020}.

\begin{figure}
\centering
\includegraphics[width=.49\linewidth]{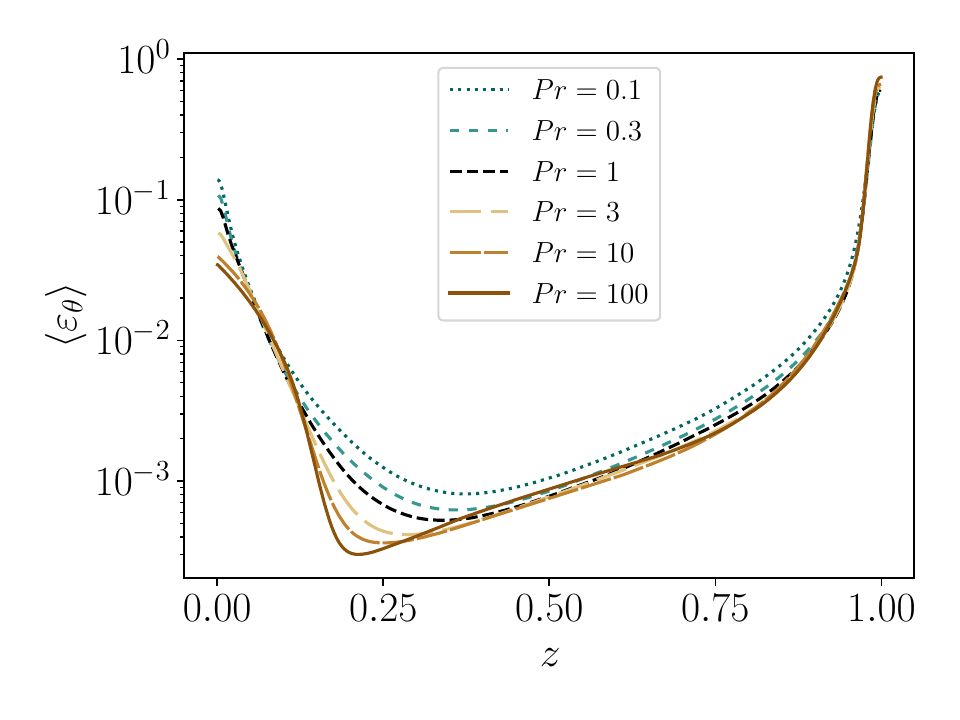}
\includegraphics[width=.49\linewidth]{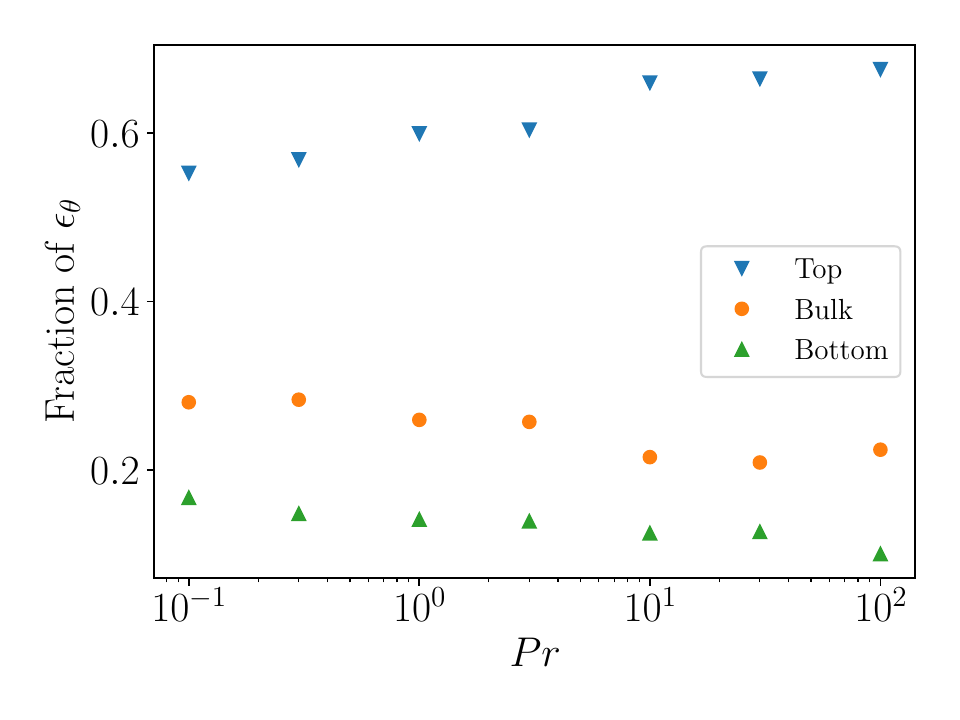}
\includegraphics[width=.49\linewidth]{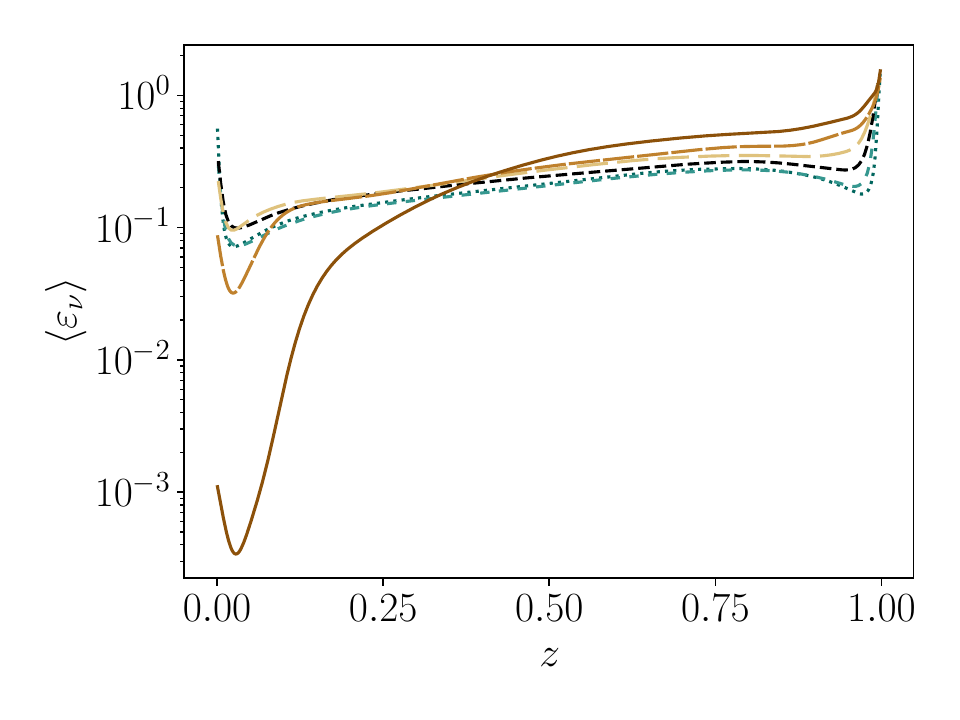}
\includegraphics[width=.49\linewidth]{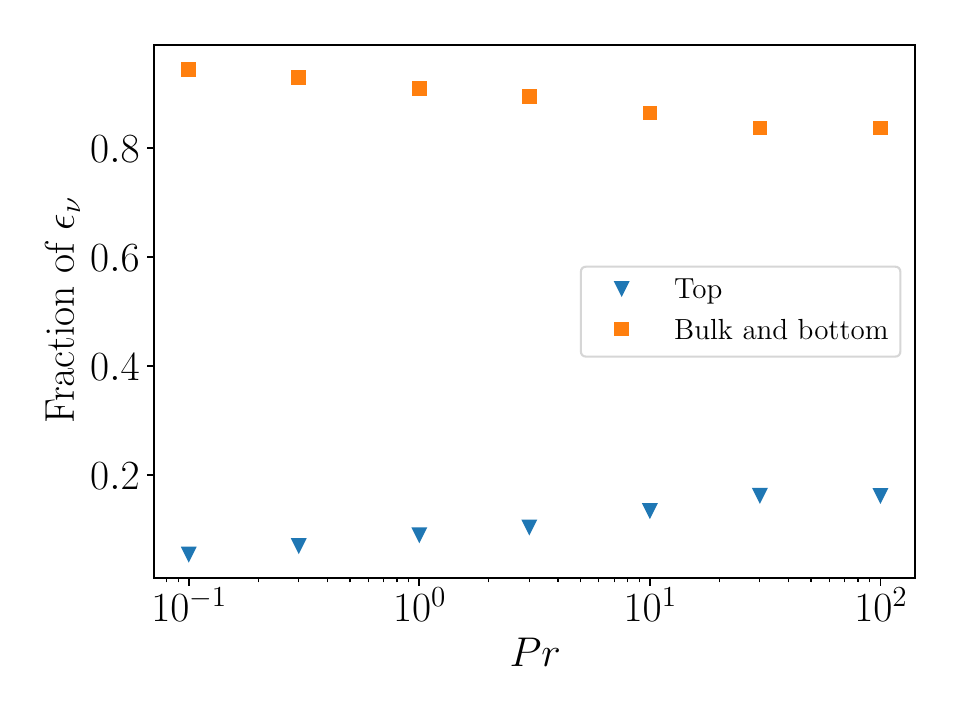}
\caption{Left column: horizontally averaged thermal and viscous energy dissipation rate profiles for $R=10^9$ and varying $Pr$. Right column: Relative contributions to the dissipation rates of the flow regions as a function of $Pr$.}
\label{fig:diss-norot}
\end{figure}

The top-left panel of Figure \ref{fig:diss-norot} shows the local thermal dissipation $\lepsth$. Remarkably, the profile of $\lepsth$ is largely independent of $Pr$. While dissipation is slightly elevated at the bottom plate for low $Pr$, the statistics are overwhelmingly dominated by the top plate. This is confirmed in the top-right panel. For all cases, the top boundary layer accounts for more than half of the total thermal dissipation. As in \citet{ostilla2025ihc}, the top-layer thermal dissipation scales with driving as $R^{-0.2}$, which directly underpins the global scaling $\T\sim R^{-0.2}$.

In contrast, the viscous dissipation $\lepsnu$ (bottom-left panel) shows large sensitivity to $Pr$. Viscous dissipation is significant at the bottom plate only for small $Pr$, and even then, it remains comparable to bulk levels. At high $Pr$, dissipation at the bottom boundary drops to near zero due to the suppression of velocity fluctuations. Meanwhile, dissipation in the top half of the domain increases, reflecting the enhanced convective efficiency $\wT$ at higher $Pr$.

Following the methodology of \citet{ostilla2025ihc}, we decompose the contributions of $\lepsth$ and $\lepsnu$ into the total dissipation rates $\epsth$ and $\epsnu$. This is done by decomposing $\epsnu$ into contributions from the top and bottom boundary layer regions, and as well as the bulk. For $\epsnu$, we divide into top boundary layer region and a combined ``bulk and bottom'' region. The results in the bottom-right panel of Figure \ref{fig:diss-norot} confirm that for all $Pr$, the total viscous dissipation is dominated by the bulk. This reinforces the finding that 3-D IHC flows are bulk-dominated, contrasting with the boundary-layer-dominated results reported for 2-D simulations \citep{Wang2020}. This can also be related to the relatively smaller values of $\mathcal{F}_B$ (larger values of $\wT$) seen for high $Pr$. The dissipation in the bulk is stronger at high $Pr$, and as $\epsnu$ is dominated by the bulk, this results in a larger $\wT$. 

\subsection{Summary}

In conclusion, our analysis of the $Pr$-dependence in non-rotating IHC reveals that Prandtl number variations primarily influence the behaviour of the stably stratified lower layer. At low $Pr$, this region is vigorously agitated by the system's turbulence, which effectively ``recovers'' some of the top-down symmetry otherwise broken by the stable stratification. Conversely, at high $Pr$, the dynamics are dominated by plumes and concentrated within the upper regions of the domain. For the highest $Pr$ value investigated, the stable stratification renders the fluid near the bottom plate completely quiescent, resulting in a ``dead zone''.

Remarkably, these profound shifts in flow morphology and local statistics are not significantly reflected in the volume-averaged temperature, $\T$. This stability is due to the mean temperature being primarily governed by the upper thermal boundary layer—a region that remains largely insensitive to $Pr$. However, the vertical convective heat flux $\wT$ (and by extension, the heat fraction $\mathcal{F}_B$) does exhibit a $Pr$-dependence; at low $Pr$, more heat is evacuated through the lower boundary as a result of the turbulence-driven symmetry recovery.

\section{Rotationally-affected IHC}
\label{sec:rihc}

Following our characterisation of the non-rotating system, we now examine how the Prandtl number influences IHC in the presence of rotation, extending the parameter space explored by \citet{ostilla2025ihc}. The introduction of the Coriolis force fundamentally alters the transport mechanisms and flow morphology, often competing with the $Pr$ number effects discussed in the previous section.

\subsection{Flow visualisation}

\begin{figure}
\centering
 \includegraphics[width=.32\linewidth]{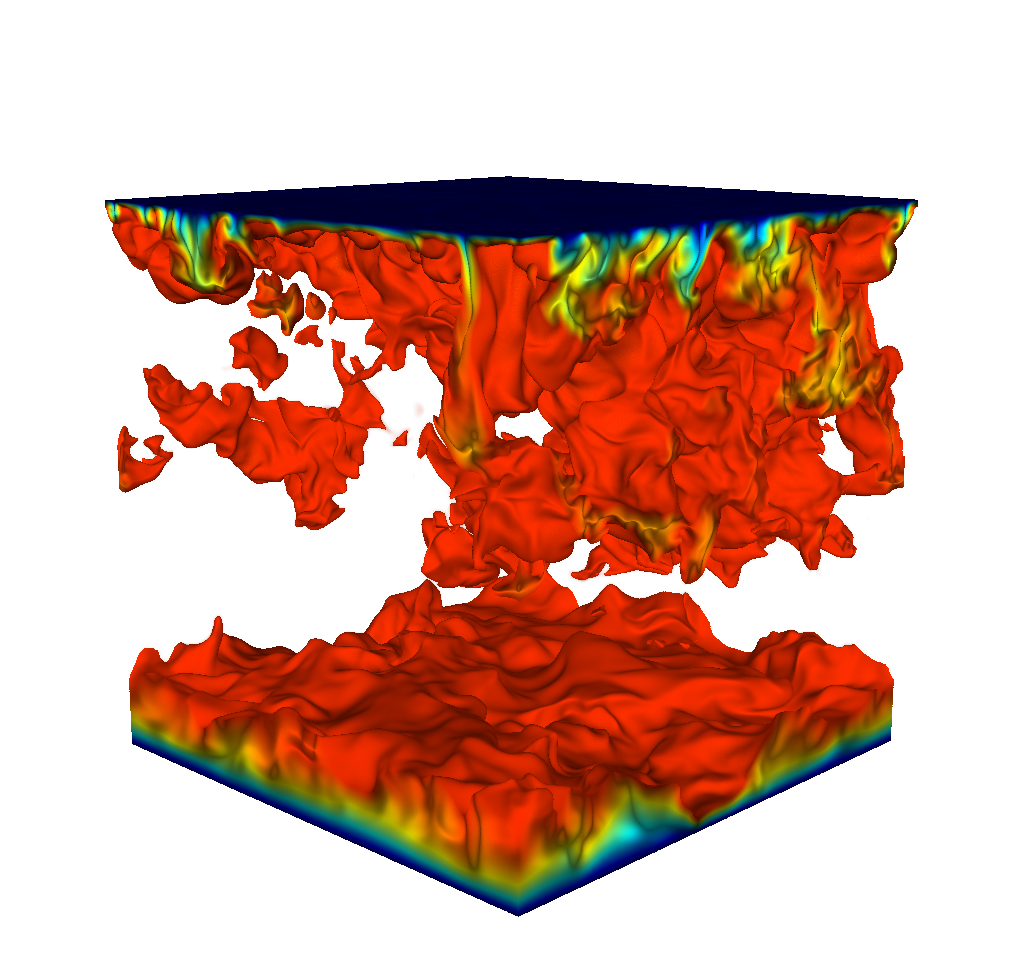}
 \includegraphics[width=.32\linewidth]{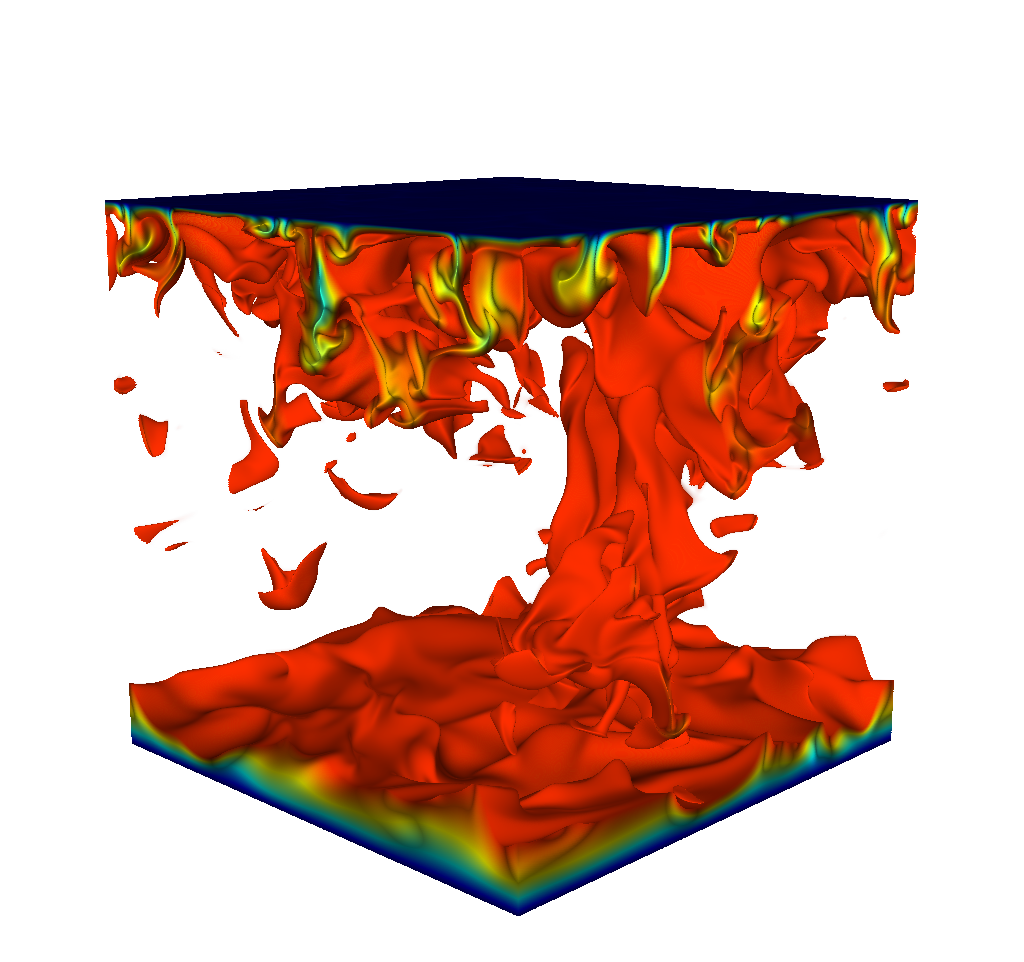}
 \includegraphics[width=.32\linewidth]{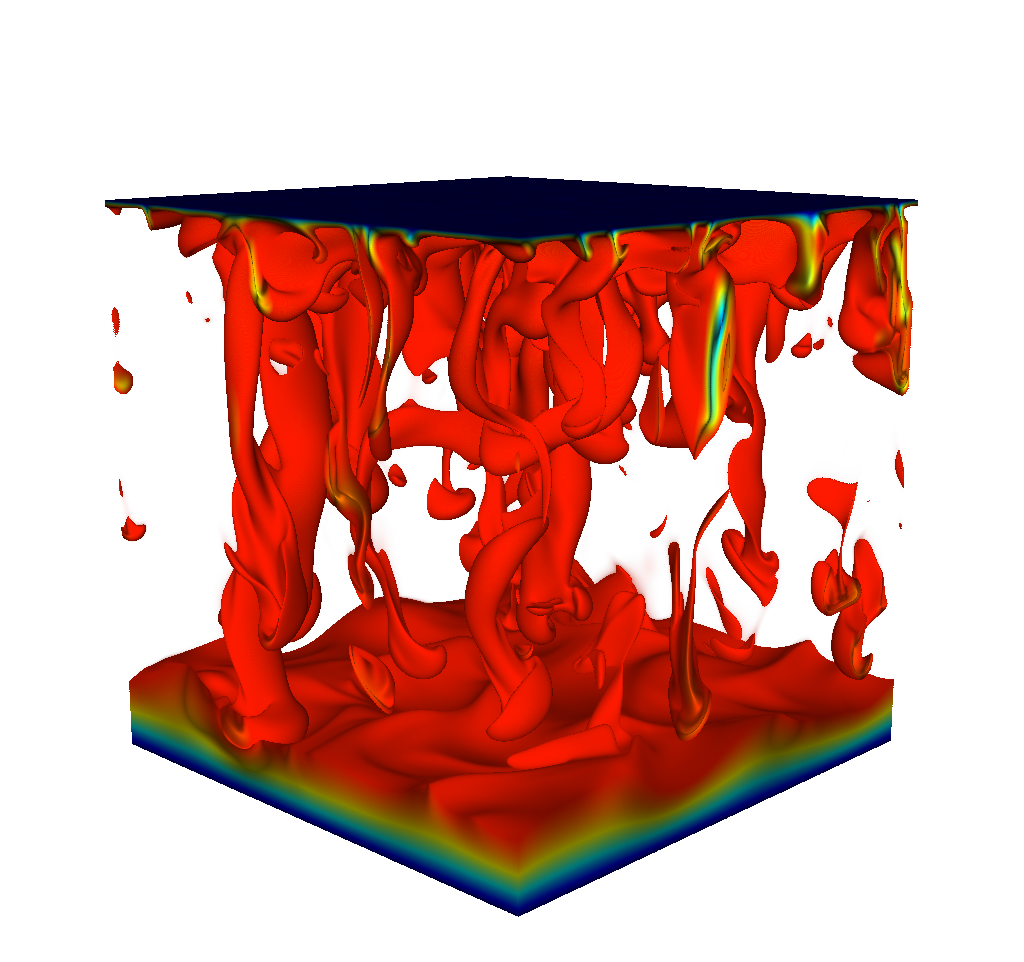}
\caption{Volumetric visualisation of the instantaneous temperature field for $R=10^{10}$, $E=10^{-4}$. From left to right: $Pr=0.1$, $Pr=1$ and $Pr=10$.}
\label{fig:flowvis-rot}
\end{figure}

We begin by assessing the impact of rotation on the global flow topology. Figure \ref{fig:flowvis-rot} displays volumetric visualisations of the instantaneous temperature field at a constant Ekman number $E=10^{-4}$ for three representative Prandtl numbers. Consistent with observations in classic rotating Rayleigh-B\'enard convection (RRBC), the degree to which rotation reorganises the flow is highly sensitive to $Pr$.

At $Pr=0.1$, the flow remains largely unaffected by rotation, it maintains a highly turbulent state characterised by significant fluctuation within the lower boundary layer, similar to the non-rotating case. In contrast, at $Pr=1$, the influence of rotation becomes apparent as the flow begins to organise into coherent structures, with plumes clustering in localised regions of the domain. The most dramatic shift occurs at $Pr=10$, where the convective structures are transformed into elongated, vertically aligned columns. This morphological transition is a hallmark of rotationally constrained flow \citep{zhong2009,Kunnen04052021}, and is driven by Ekman pumping mechanisms that enhance vertical transport within the columnar cores.

\subsection{Global Quantities}

\begin{figure}
\centering
\includegraphics[width=.32\linewidth]{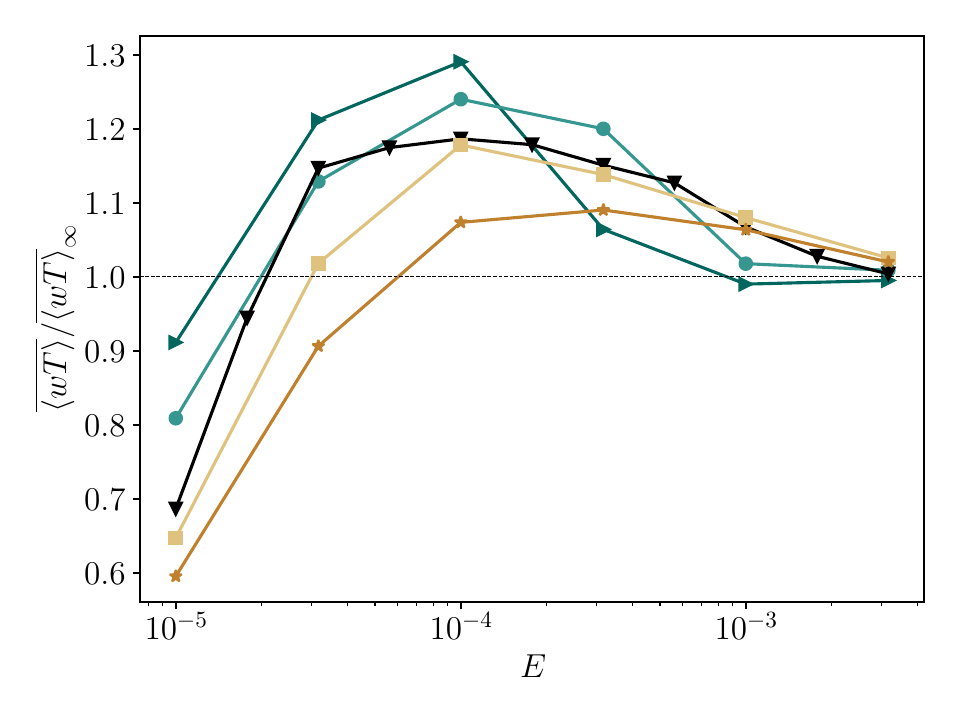}
\includegraphics[width=.32\linewidth]{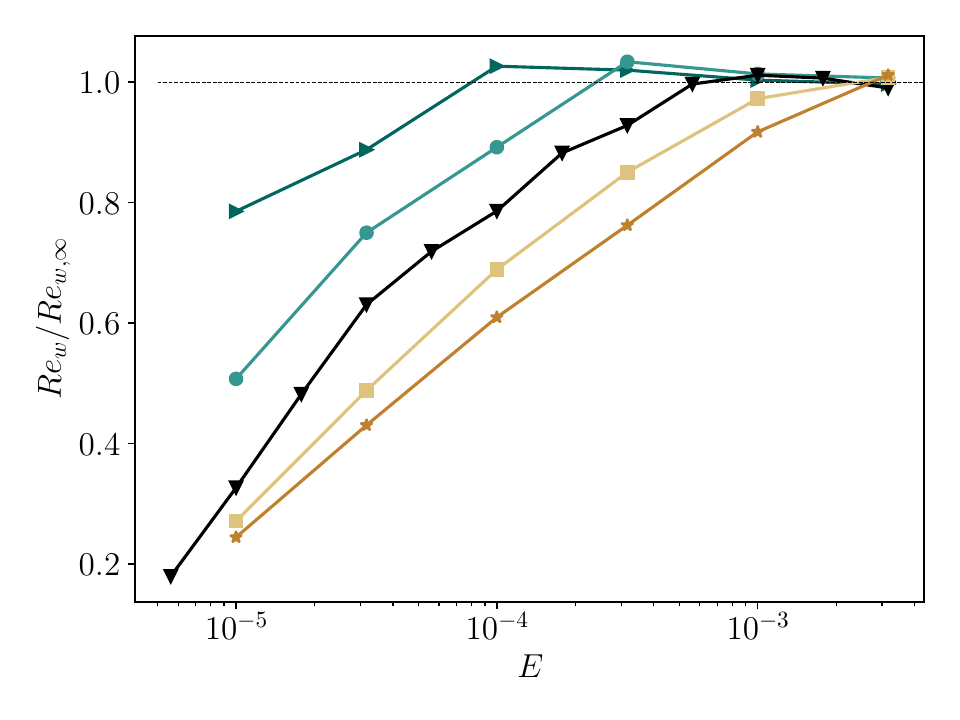}
\includegraphics[width=.32\linewidth]{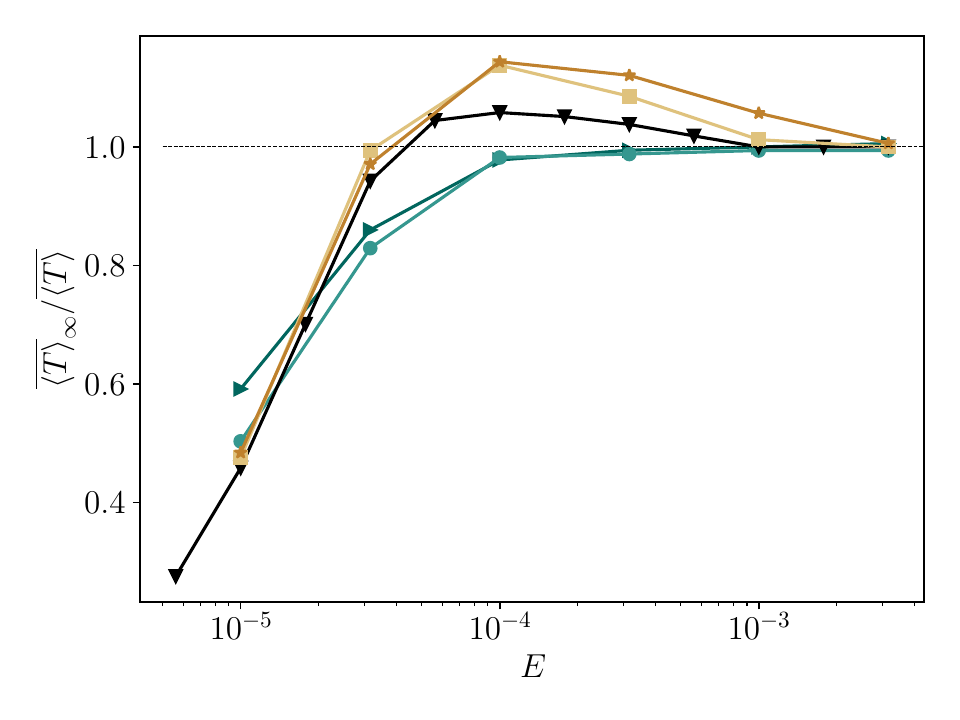}
\includegraphics[width=.32\linewidth]{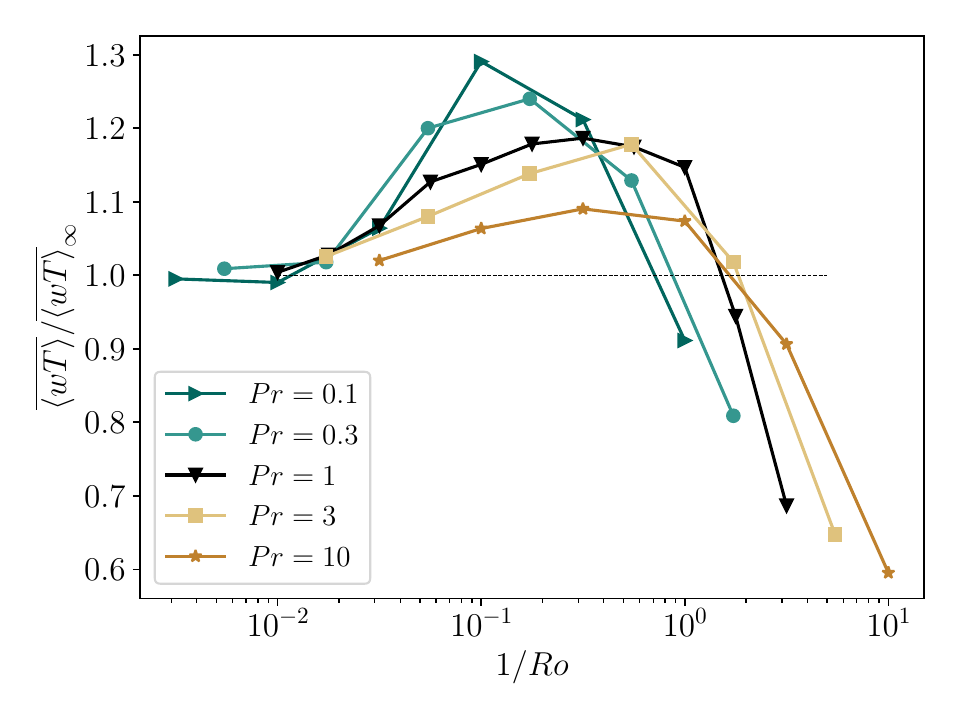}
\includegraphics[width=.32\linewidth]{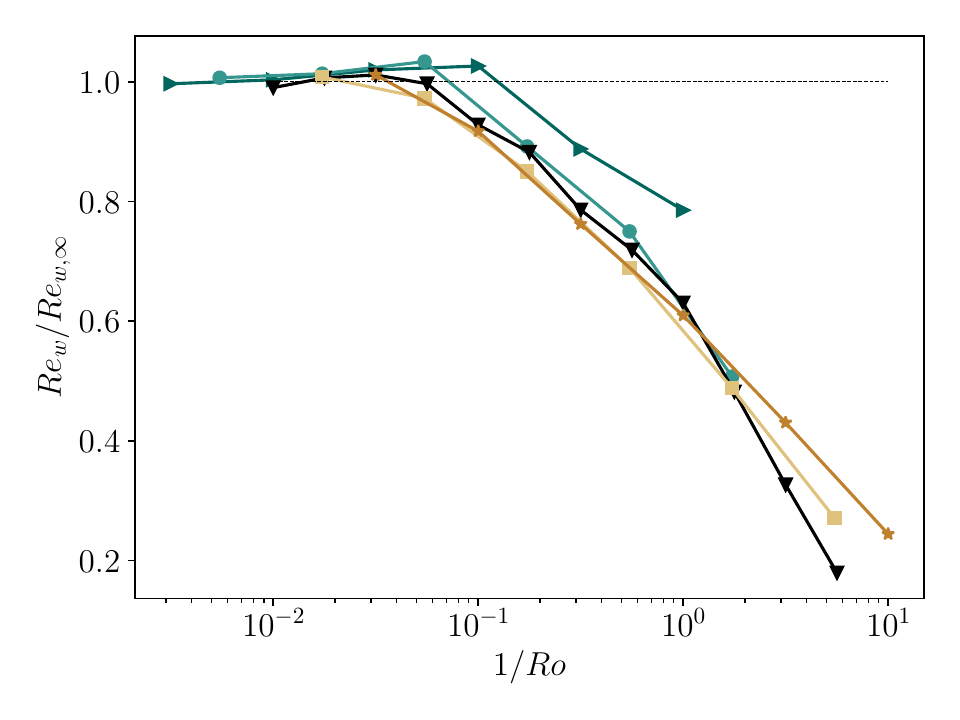}
\includegraphics[width=.32\linewidth]{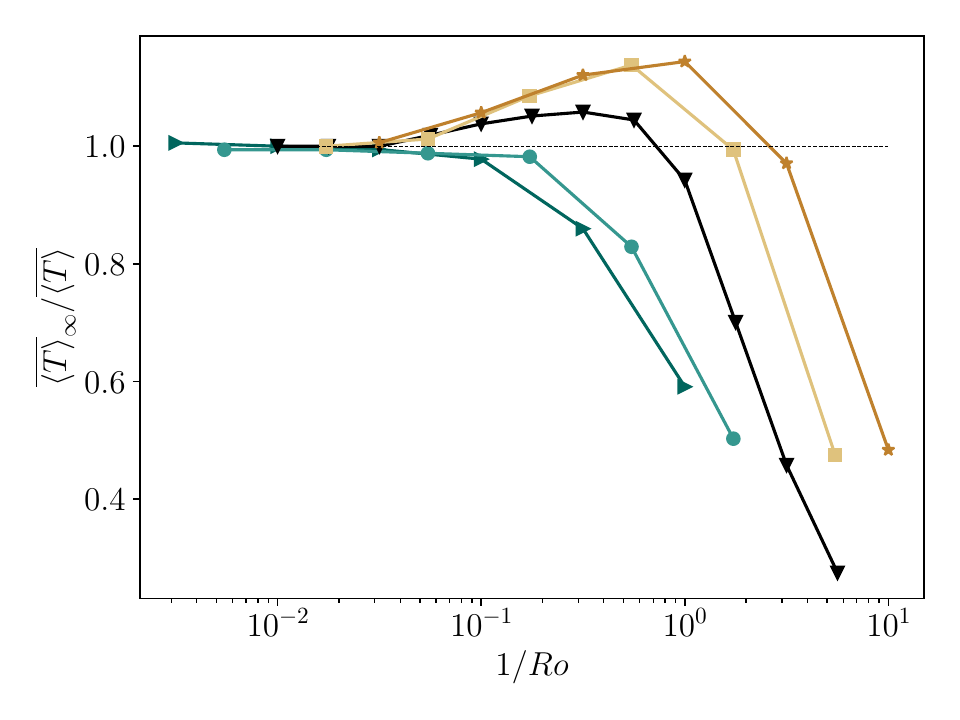}
\caption{Changes in the global responses with rotation for $R=10^9$. $\wT/\wT_\infty$ (left column), $Re_w/Re_{w,\infty}$ (middle column) and  $\T_\infty/\T$ (right column) against $E$ (top row) and $1/Ro$ (bottom row).}
\label{fig:globalqs-rot-r1e9}
\end{figure}

To evaluate the influence of rotation on the global system response, Figure \ref{fig:globalqs-rot-r1e9} presents the vertical convective flux $\wT$, the mean temperature $\T$, and the wind Reynolds number $Re_w$, all normalised by their respective non-rotating values (denoted by the subscript $\infty$). These quantities are plotted against both the Ekman number $E$ and the inverse Rossby number $1/Ro$ at a fixed Rayleigh number $R=10^9$.

When it comes to vertical convection, $\wT$, rotation enhances this magnitude across all investigated Prandtl numbers. For lower $Pr$, we observe an increase of up to 30$\%$ at intermediate rotation rates ($0.1\leq 1/Ro\leq 1$), whereas for $Pr=10$, this enhancement is more modest, peaking at approximately 10$\%$. As discussed in Appendix \ref{sec:app}, while the specific magnitude of the increase at $Pr=0.1$ depends on the domain aspect ratio $\Gamma$, the qualitative trend—characterised by an initial rise in $\wT$ followed by a subsequent decline at very high rotation rates—remains robust across different horizontal periodic lengths.

The physical origin of this convective enhancement is not immediately evident from global metrics alone. The middle column of Figure \ref{fig:globalqs-rot-r1e9} shows that the wind Reynolds number $Re_w$ decreases monotonically with rotation for all $Pr$. This suggests that $Re_w$ does not capture the fundamental shifts in flow topology, however, it does indicate that the rise in $\wT$ is not a by product of increased overall turbulence, but rather a reorganisation of the transport mechanism itself. 

Before further discussing $\wT$, we turn to the inverse mean temperature $\T_\infty/\T$ (a proxy for heat transfer efficiency). This quantity shows significant enhancement only for $Pr\ge 1$. As previously noted, the behavior of this quantity in IHC is analogous to the Nusselt number in rotating Rayleigh-B\'enard convection (RRBC). Consistent with RRBC literature, increased convective efficiency is primarily restricted to higher Prandtl numbers where Ekman pumping plays a dominant role \citep{Kunnen04052021,ecke2023}. At low $Pr$, no such optimum is observed, and the efficiency gain is absent.

The behaviour of $\wT$ and $\T$ becomes more evident at low $Pr$ when rotation is present. Total heat transport is essentially limited by the top thermal boundary layer (c.f.~$\S$\ref{sec:diss-norot}), while $\wT$ is linked to the asymmetry between top and bottom boundary layers. An increase of rotation leads to two phenomena that increase the asymmetry of heat transport. At low $Pr$, higher thermal diffusion maintains the stably stratified region at the bottom, increasing the asymmetry between top and bottom, increasing the vertical heat transport. At high $Pr$, rotation enhances heat transport at the top boundary layer through Ekman pumping, increasing both plate asymmetry. For $\T$, at low $Pr$, increased diffusion inhibits the effect of transport by columnar flows and rotation does not enhance the turbulent mixing. The opposite occurs at high $Pr$, where Ekman pumping is effective at enhancing mixing and thereby decreasing $\T$ for intermediate values of $E$.

\begin{figure}
\centering
\includegraphics[width=.32\linewidth]{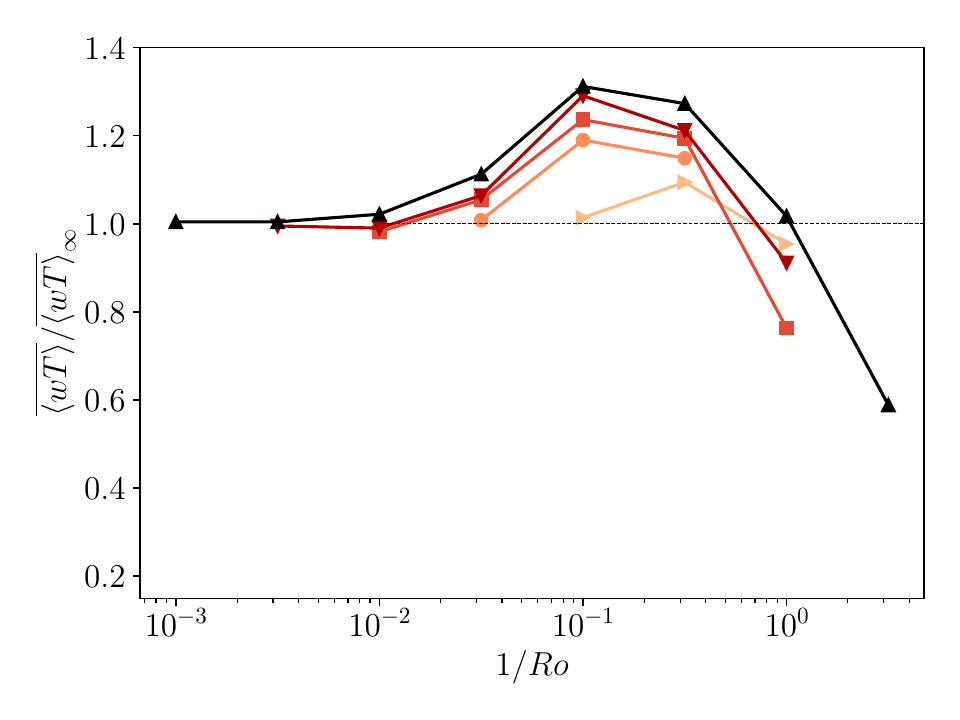}
\includegraphics[width=.32\linewidth]{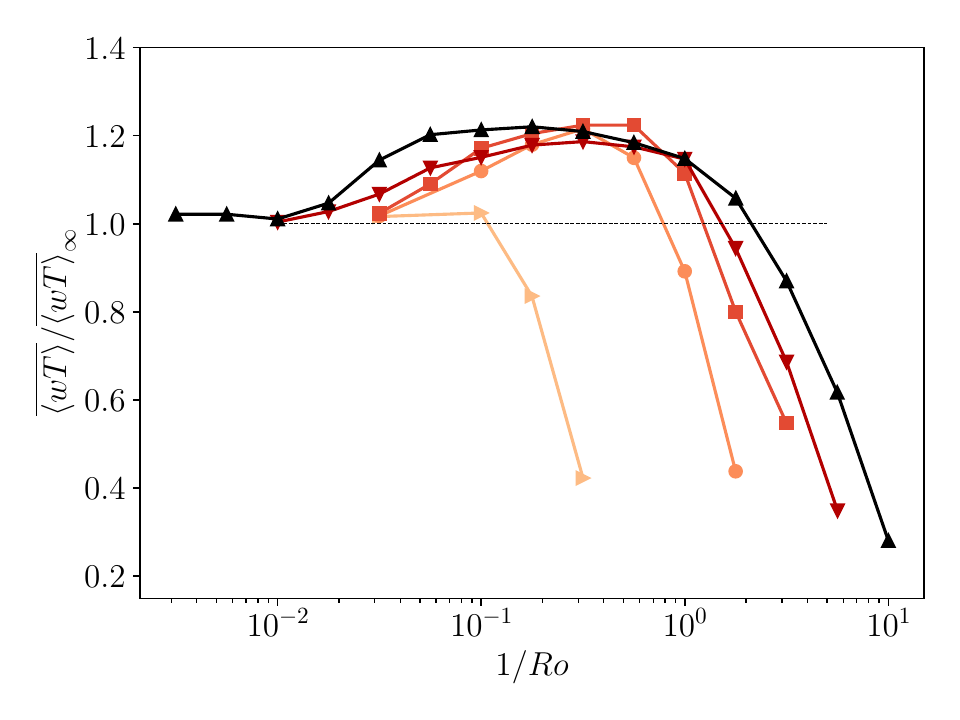}
\includegraphics[width=.32\linewidth]{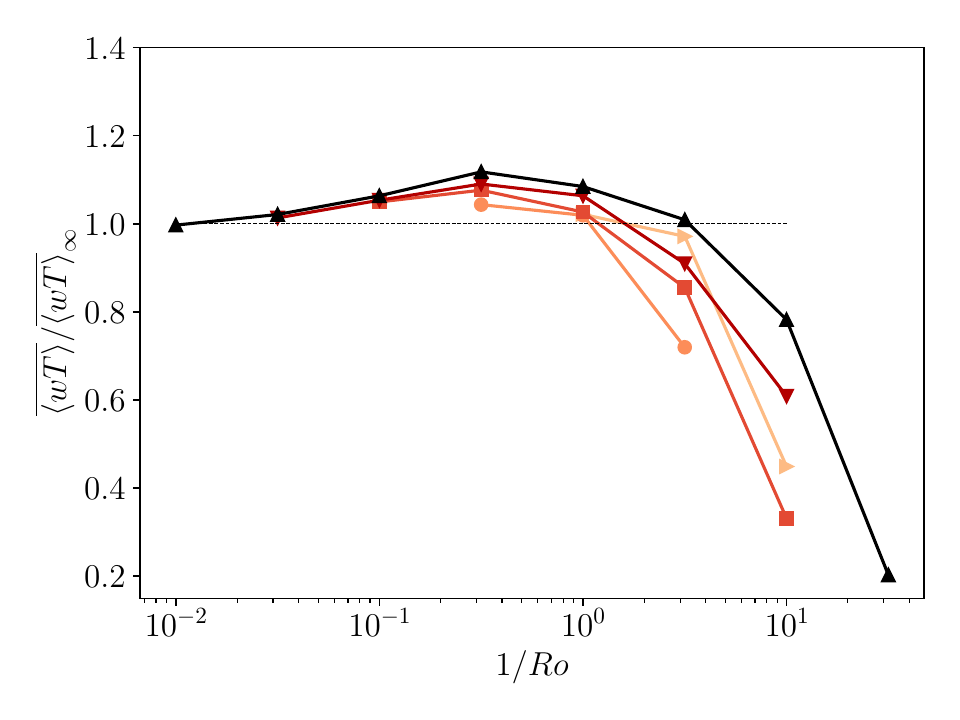}
\includegraphics[width=.32\linewidth]{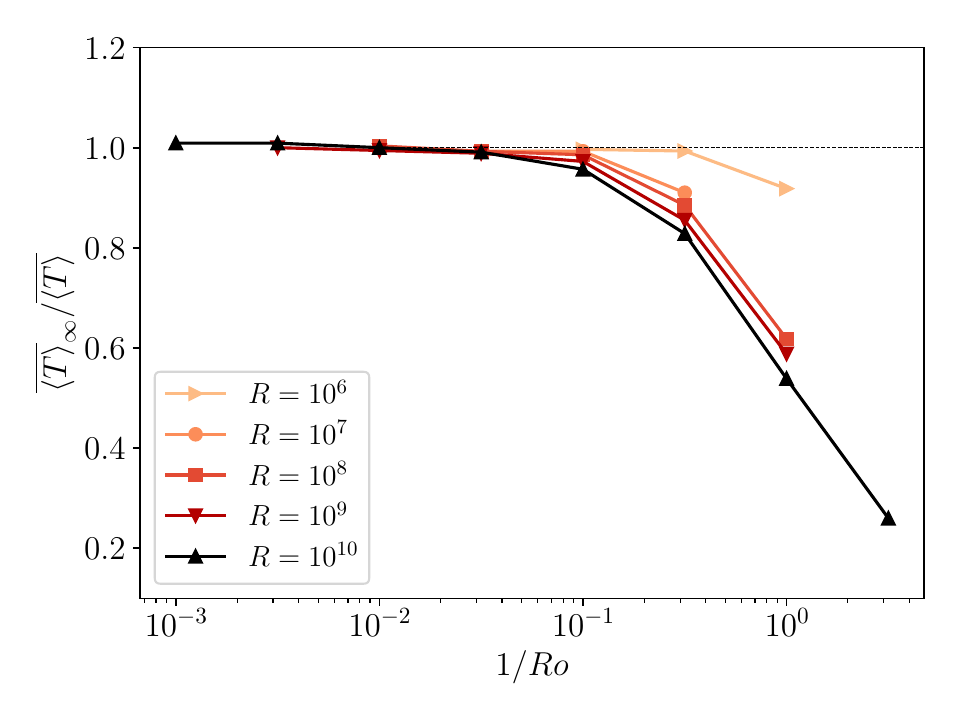}
\includegraphics[width=.32\linewidth]{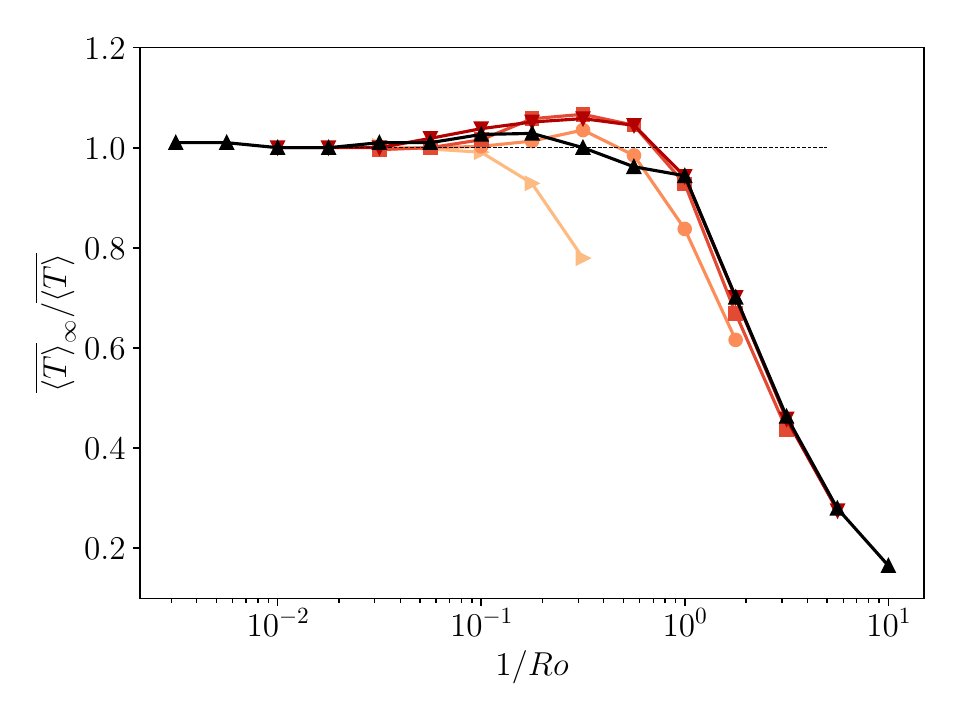}
\includegraphics[width=.32\linewidth]{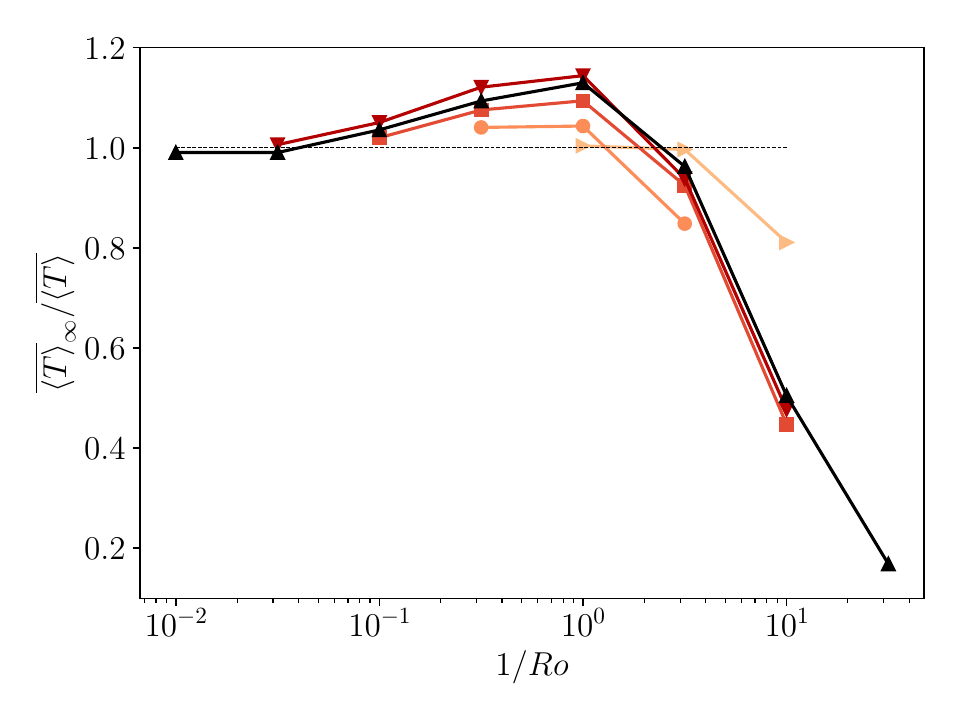}
\caption{Changes in the global responses $\wT$ and $\T$ with rotation for $Pr=0.1$ (left column), $Pr=1$ (middle column) and $Pr=10$ (right column).}
\label{fig:globalqs-rot-varyingr}
\end{figure}

Figure \ref{fig:globalqs-rot-varyingr} further illustrates the rotational dependence of $\wT$ and $\T$ across a range of Rayleigh numbers for $Pr=0.1$, 1, and 10. Here, the analogy between $\T$ and Nusselt number in RRBC becomes particularly apparent, not only is the effect limited to $Pr\ge 1$, but the decrease in $\T$ is confined to a specific range of $R$. For instance, at $Pr=1$, the effect begins to diminish at the highest Rayleigh numbers. In contrast, the enhancement of $\wT$ persists across the entire parameter space and is notably more pronounced at lower $Pr$. These divergent trends underscore that the mechanisms driving vertical convective transport are physically distinct from those governing global heat transport efficiency.

\subsection{Temperature and velocity statistics}

\begin{figure}
\centering
\includegraphics[width=.32\linewidth]{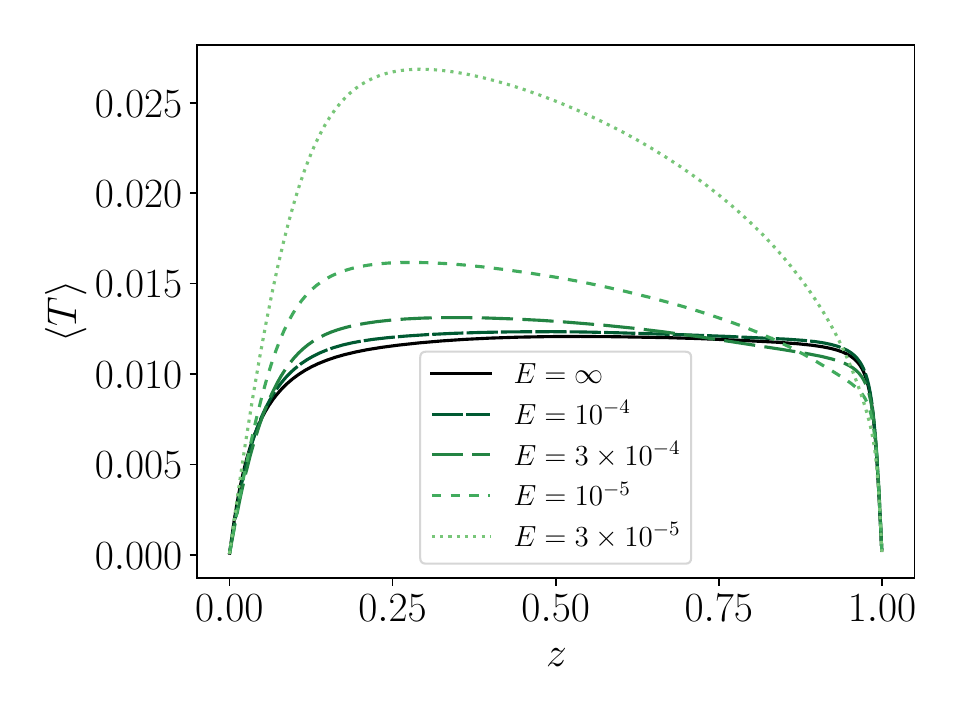}
\includegraphics[width=.32\linewidth]{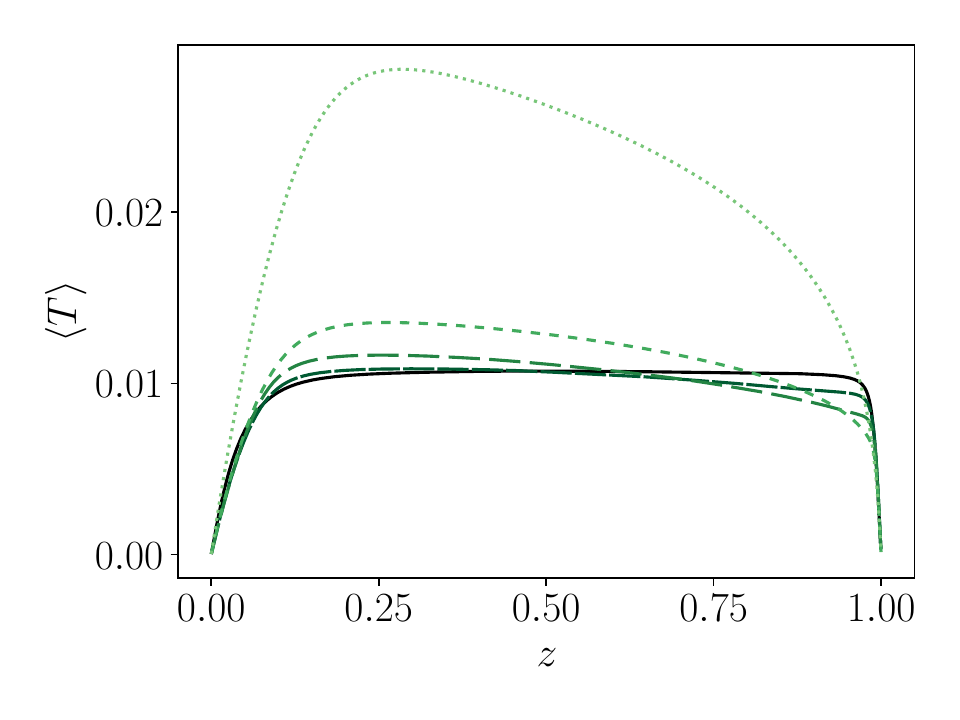}
\includegraphics[width=.32\linewidth]{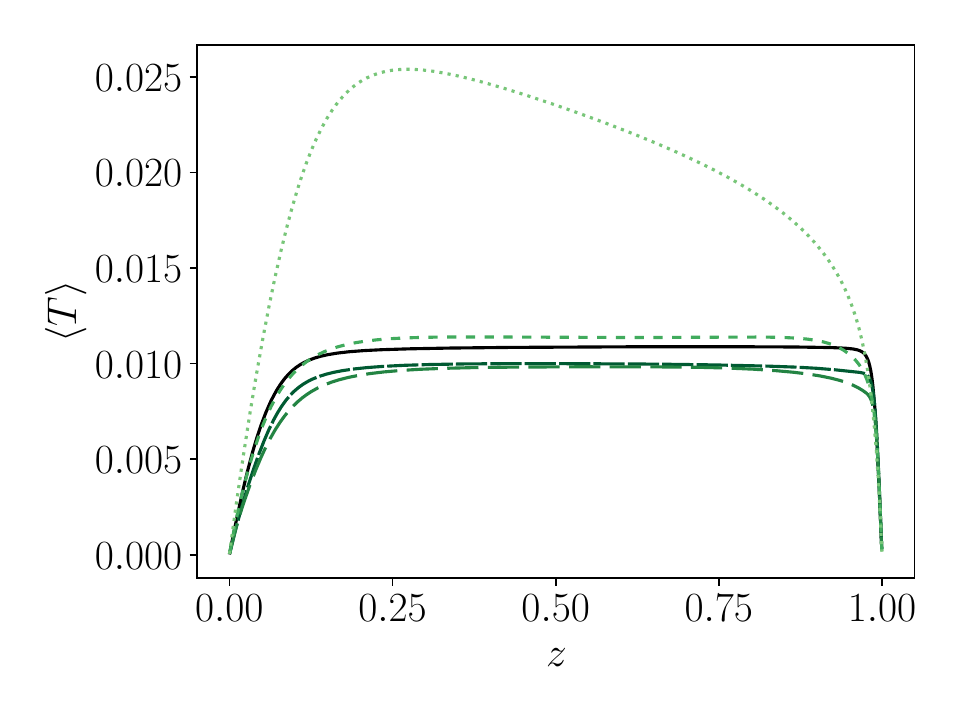}
\includegraphics[width=.32\linewidth]{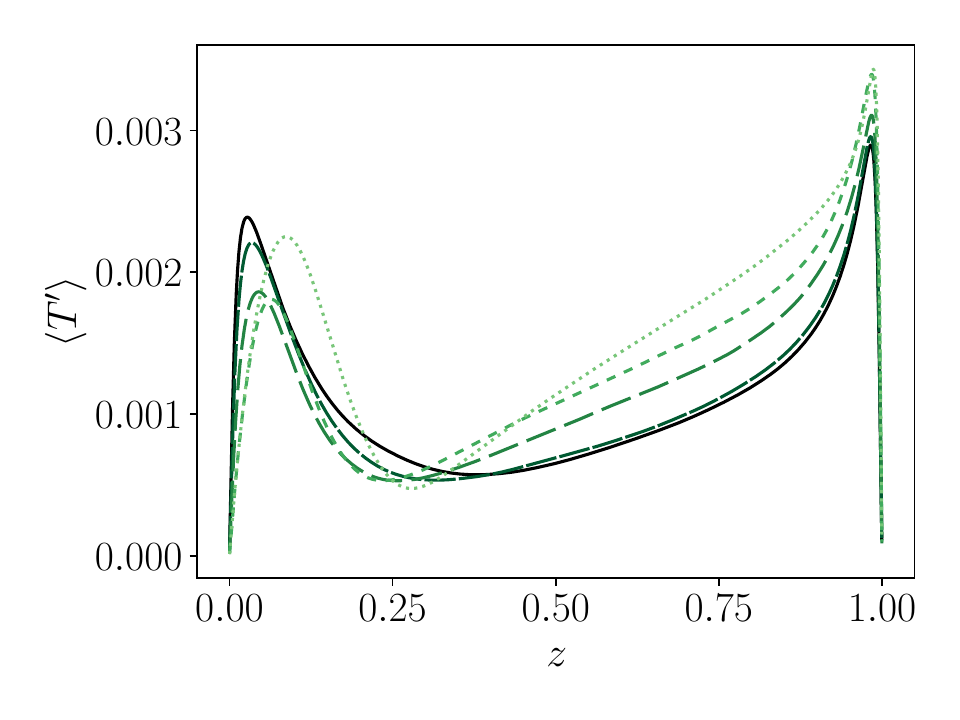}
\includegraphics[width=.32\linewidth]{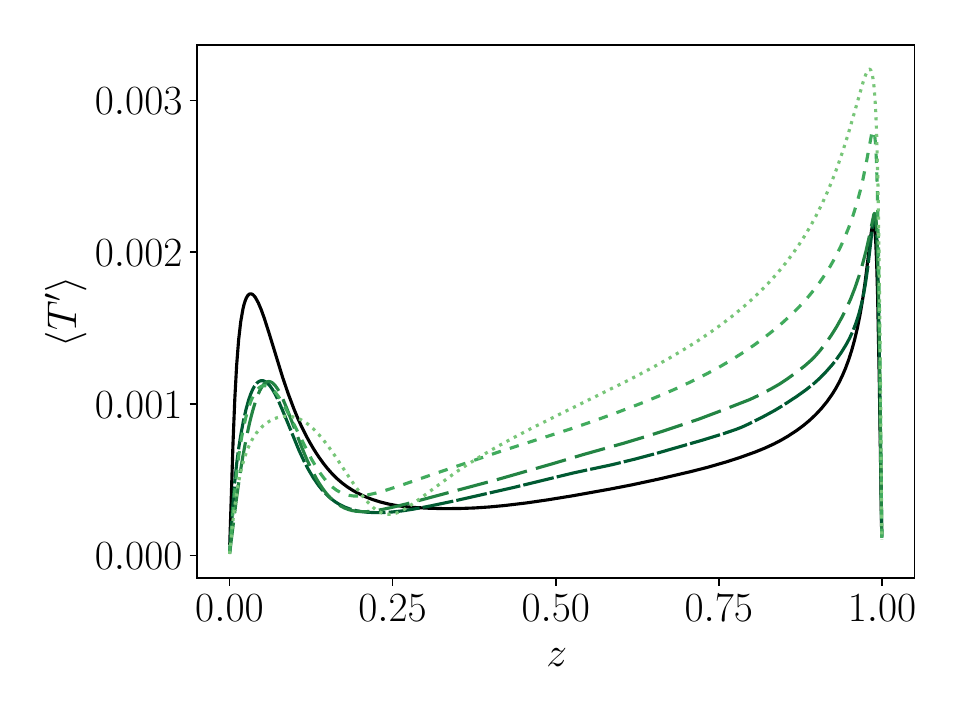}
\includegraphics[width=.32\linewidth]{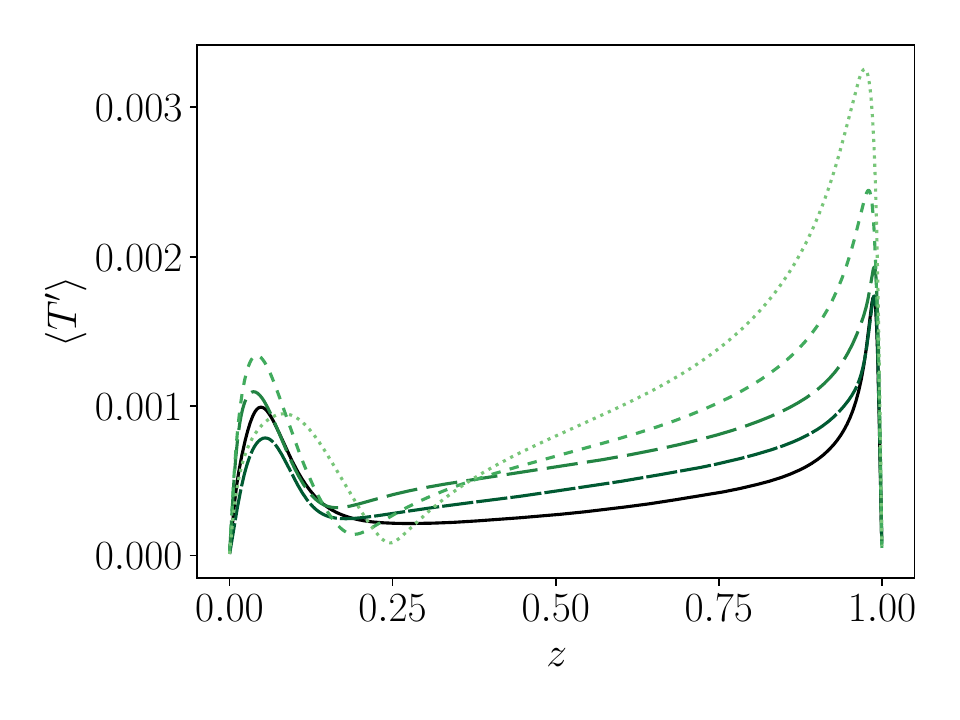}
\caption{Mean $\langle T \rangle$ (top row) and fluctuation $\langle T^\prime\rangle$ (bottom row) profiles for $R=10^{10}$ for several values of $E$ and $Pr=0.1$ (left column), $Pr=1$ (middle) and $Pr=10$ (right column).}
\label{fig:tempstats-rot-r1e10}
\end{figure}

\begin{figure}  
\centering
\includegraphics[width=.49\linewidth]{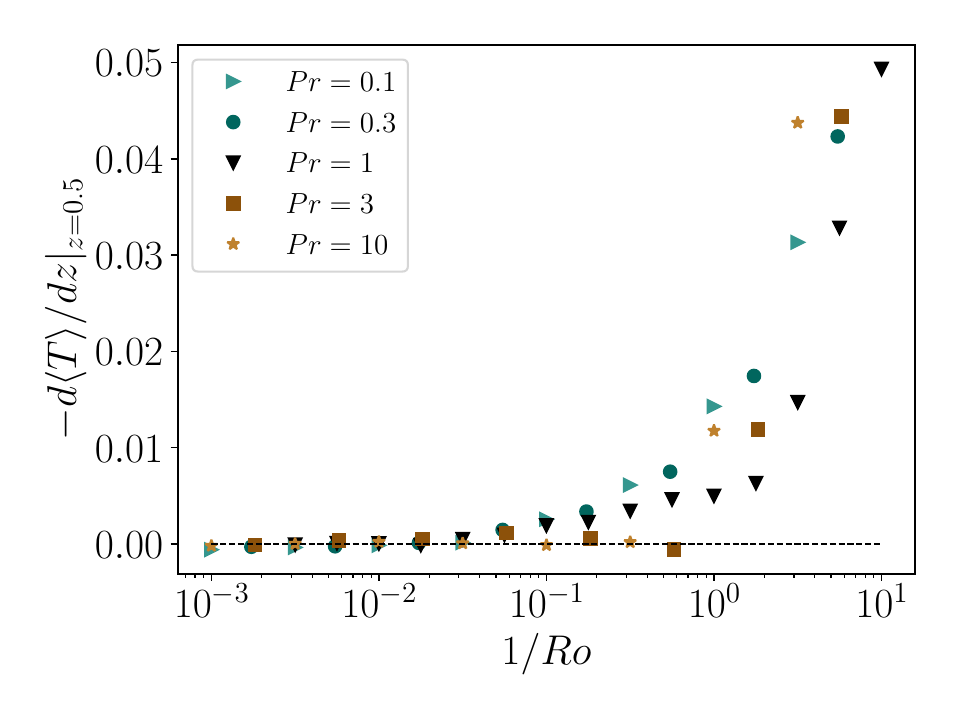}
\includegraphics[width=.49\linewidth]{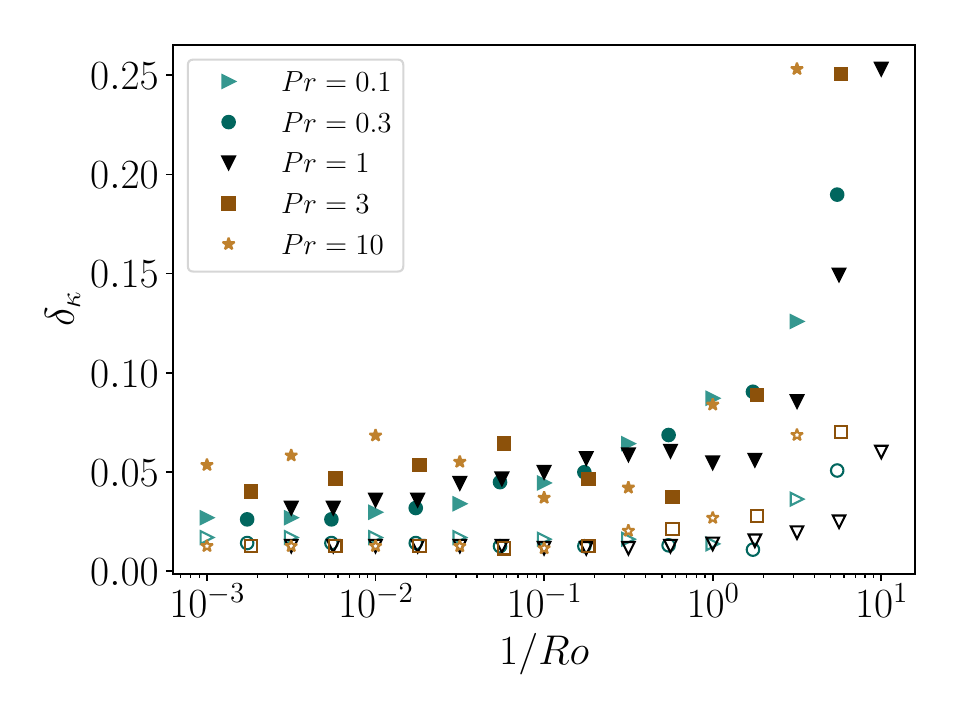}
\caption{Left: temperature gradient in the bulk as a function of rotation for $R=10^{10}$. Right: top (hollow symbols) and bottom (solid symbols) thermal boundary layer size as a function of rotation for $R=10^{10}$ and several values of $Pr$.}
\label{fig:temp-derived-rot-r1e10}
\end{figure}

We now examine the influence of rotation on local statistics. Figure \ref{fig:tempstats-rot-r1e10} illustrates the mean temperature and fluctuation profiles for $R=10^{10}$ across various $Pr$ and $E$ values. Consistent with non-rotating IHC, the Prandtl number does not fundamentally alter the morphology of the mean temperature profiles. However, the introduction of rotation induces a notable bulk temperature gradient. The profiles transition from a relatively flat central region to a distinct slope as $1/Ro$ increases. This phenomenon is quantified in the left panel of Figure \ref{fig:temp-derived-rot-r1e10}, which demonstrates that the gradient emerges once Coriolis forces become significant relative to convective forces ($1/Ro\ge1$). While the onset of this gradient shows a slight $Pr$ dependence, the inverse Rossby number remains the most robust parameter for collapsing this behavior, as it accounts for the $R$-dependence more effectively than the Ekman number \citep{ostilla2025ihc}.

The temperature fluctuations $\langle T^\prime\rangle$ exhibit a more nuanced $Pr$ dependence. In general, rotation enhances fluctuations throughout the domain, except near the lower boundary. At $E=3\times10^{-5}$, fluctuations across different $Pr$ values become comparable, contrasting with the non-rotating case where low $Pr$ significantly dominated. In the bottom boundary layer, rotation suppresses fluctuations for $Pr=0.1$ and $Pr=1$. Interestingly, for $Pr=10$, fluctuations initially decrease before rising again at high rotation rates, likely reflecting a reorganisation of the flow topology that more vigorously agitates the lower stable region.

Regarding boundary layer thickness (Figure \ref{fig:temp-derived-rot-r1e10}, right), the top thermal boundary layer remains largely independent of $Pr$ and constant in size until $1/Ro\approx1$, where it begins to thicken. The bottom thermal boundary layer shows greater variation without a singular trend, though it generally increases in size once rotation becomes dominant.

\begin{figure}
\centering
\includegraphics[width=.32\linewidth]{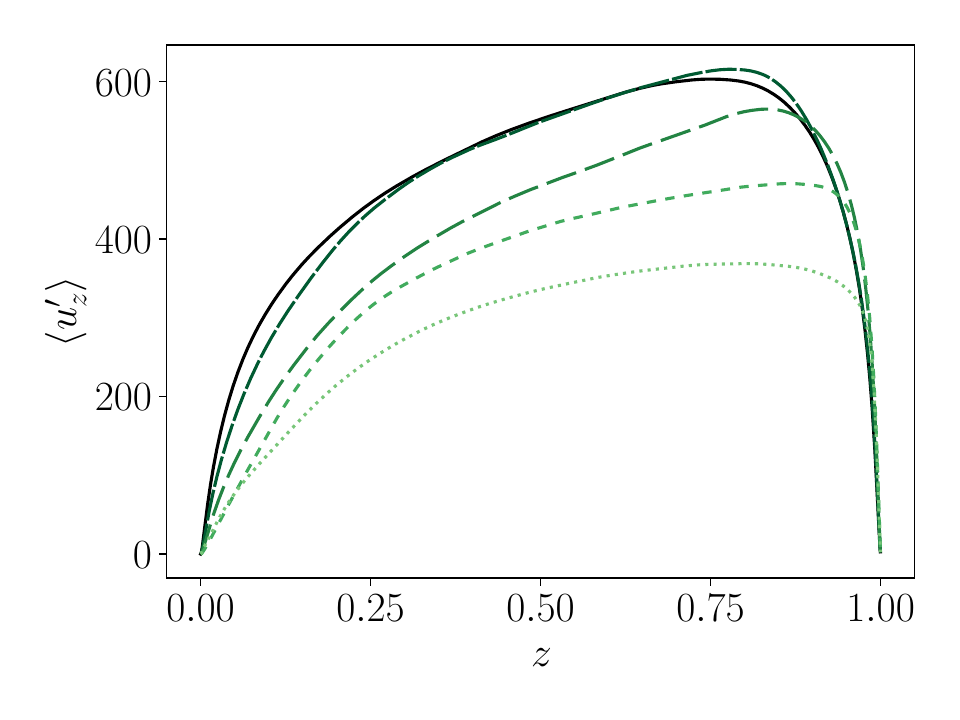}
\includegraphics[width=.32\linewidth]{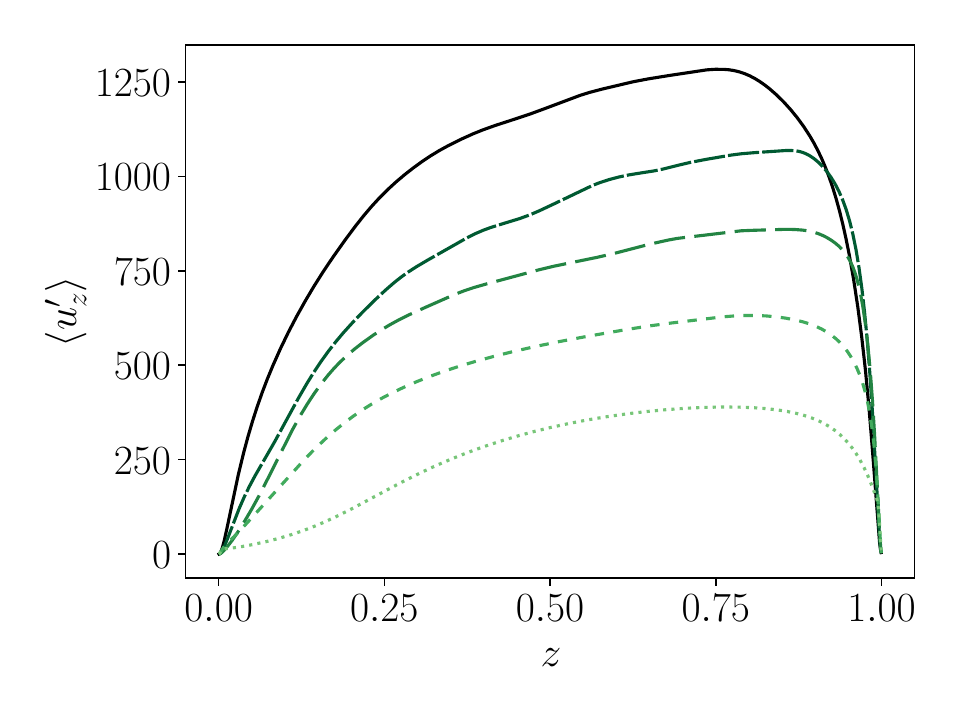}
\includegraphics[width=.32\linewidth]{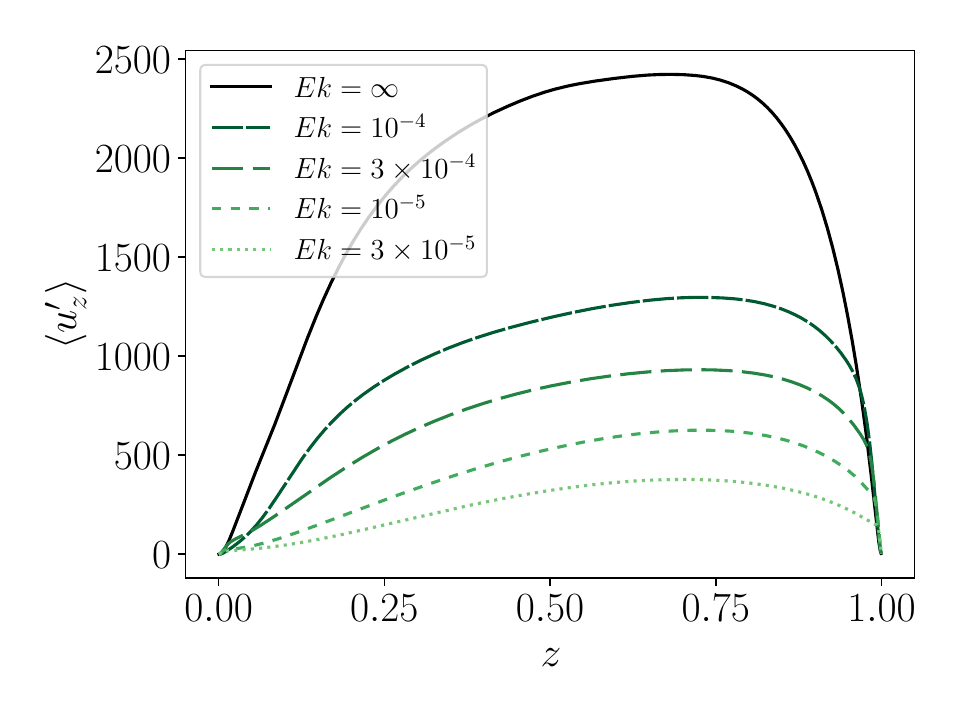}
\includegraphics[width=.32\linewidth]{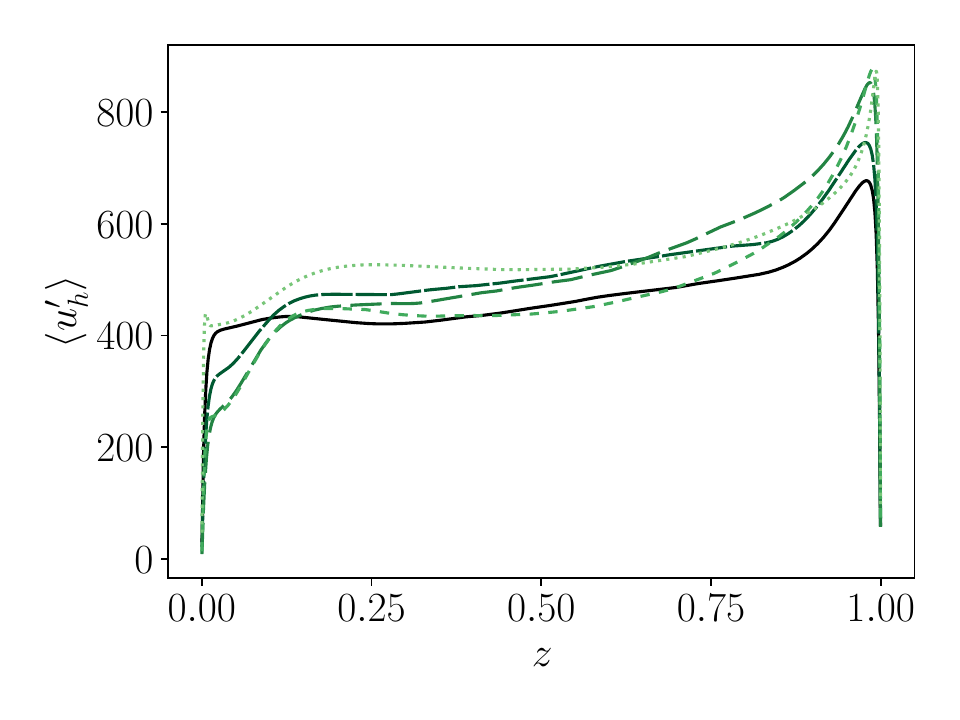}
\includegraphics[width=.32\linewidth]{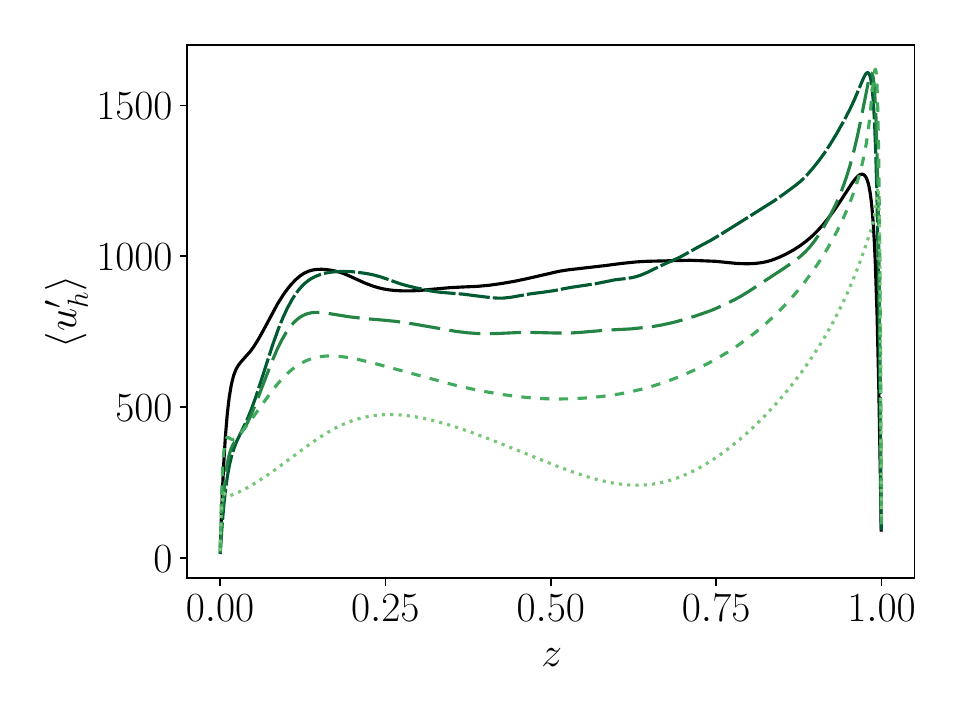}
\includegraphics[width=.32\linewidth]{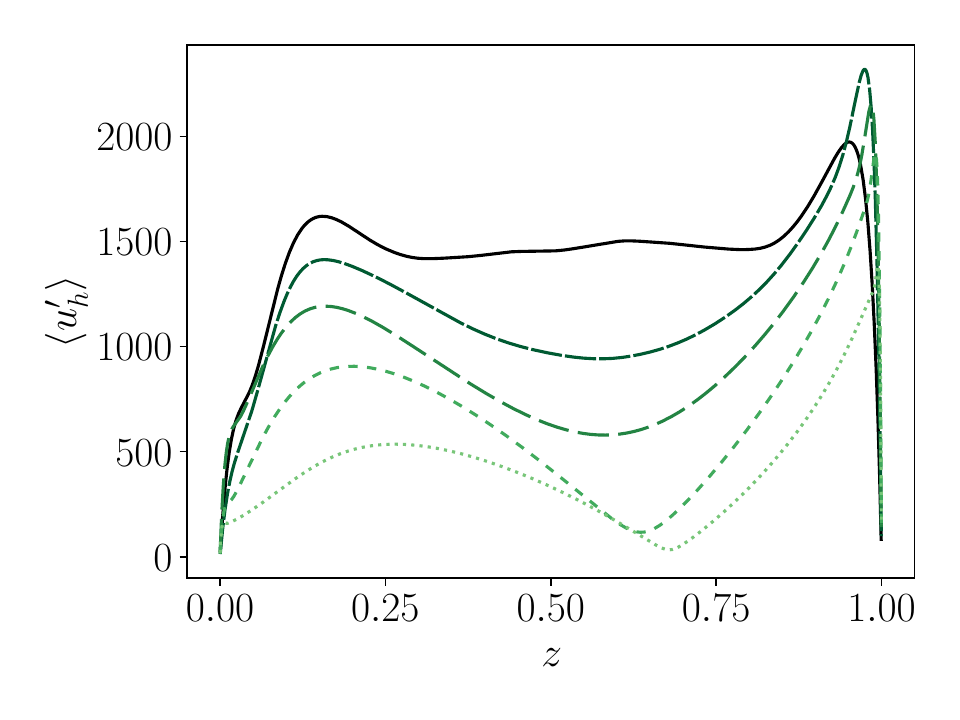}
\caption{Vertical $\langle u_z^\prime\rangle$ (top row) and horizontal $\langle u_h^\prime \rangle$ (bottom row) velocity fluctuation profiles for $R=10^{10}$ for several values of $E$ and $Pr=0.1$ (left column), $Pr=1$ (middle) and $Pr=10$ (right column).}
\label{fig:velstats-rot-r1e10}
\end{figure}

Velocity statistics (Figure \ref{fig:velstats-rot-r1e10}) reveal that vertical fluctuations $\langle u_z^\prime\rangle$ are largely independent of $Pr$, though higher $Pr$ leads to stronger suppression at comparable $E$. The presence of regions near the bottom plate where the overall value of fluctuations is small is more apparent at high $Pr$, as the stably stratified layer is even less active in these cases. 

Horizontal fluctuations $\langle u_h^\prime\rangle$ display a richer phenomenology. At small $Pr$, rotation initially stabilises the bottom region before the formation of an Ekman boundary layer triggers a secondary increase in fluctuations at very small $E$. For $Pr=1$ and $Pr=10$, this stabilisation is even more pronounced, with the bulk and lower regions becoming significantly damped before the eventual appearance of the Ekman layer, marked as a new peak close to the bottom plate, once $E$ is small enough.

\begin{figure}
\centering
\includegraphics[width=.49\linewidth]{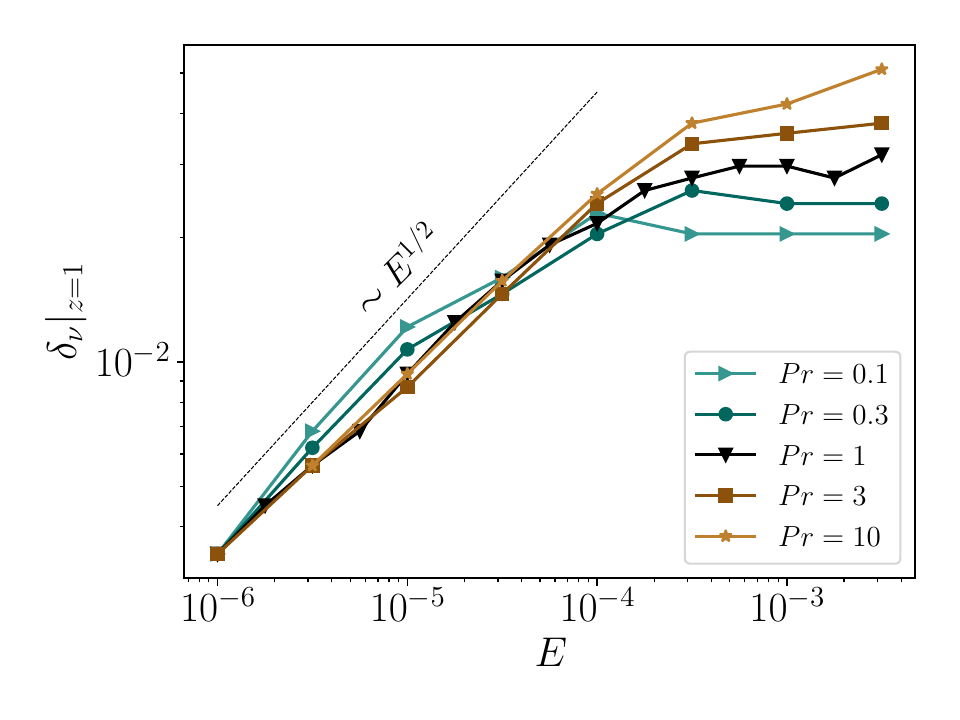}
\includegraphics[width=.49\linewidth]{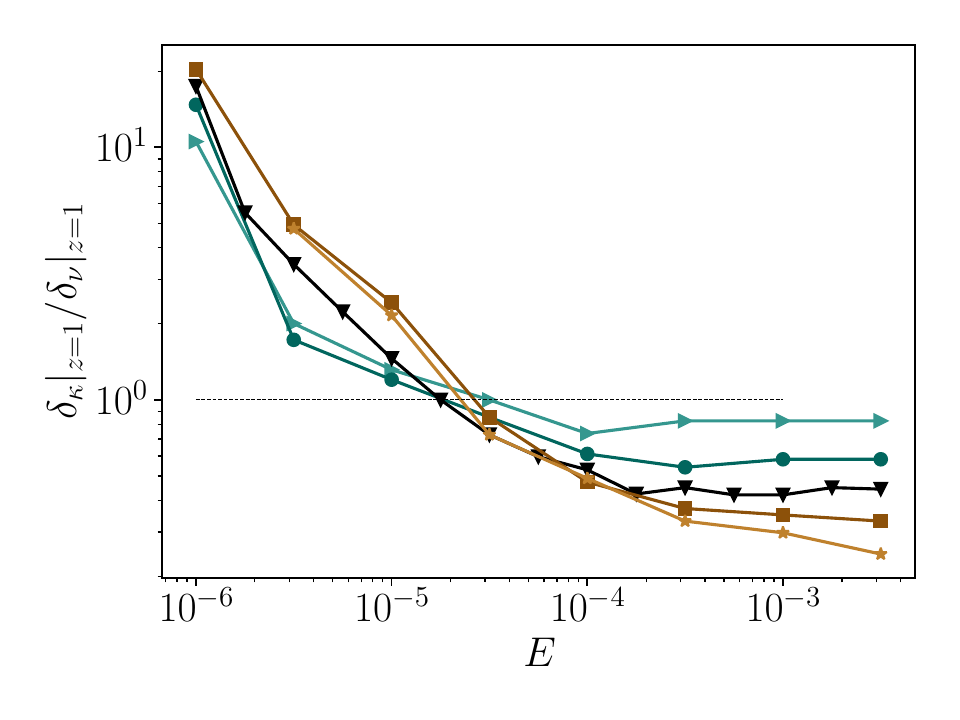}
\caption{Left: top velocity boundary layer size as a function of $E$ for $R=10^{10}$. Right: ratio of thermal and velocity boundary layer size at the top plate as a function of rotation for $R=10^{10}$ and several values of $Pr$.}
\label{fig:velderived-rot-r1e10}
\end{figure}

At the top plate, the velocity boundary layer thickness scales as $\delta_\nu|_{z=1}\sim E^{1/2}$ (Figure \ref{fig:velderived-rot-r1e10}, left), confirming the appearance of Ekman layers across all $Pr$. In addition, the ratio of thermal to velocity boundary layer thickness (right panel) shows a crossover at $E\sim2\times10^{-5}$ across all $Pr$. As discussed in \cite{ostilla2025ihc} for $Pr=1$, this diagnostic, proposed for the geostrophic transition in RRBC, does not identify this crossover in IHC regardless of the Prandtl number.

\subsection{Dissipation rates}

\begin{figure}
\centering
\includegraphics[width=.32\linewidth]{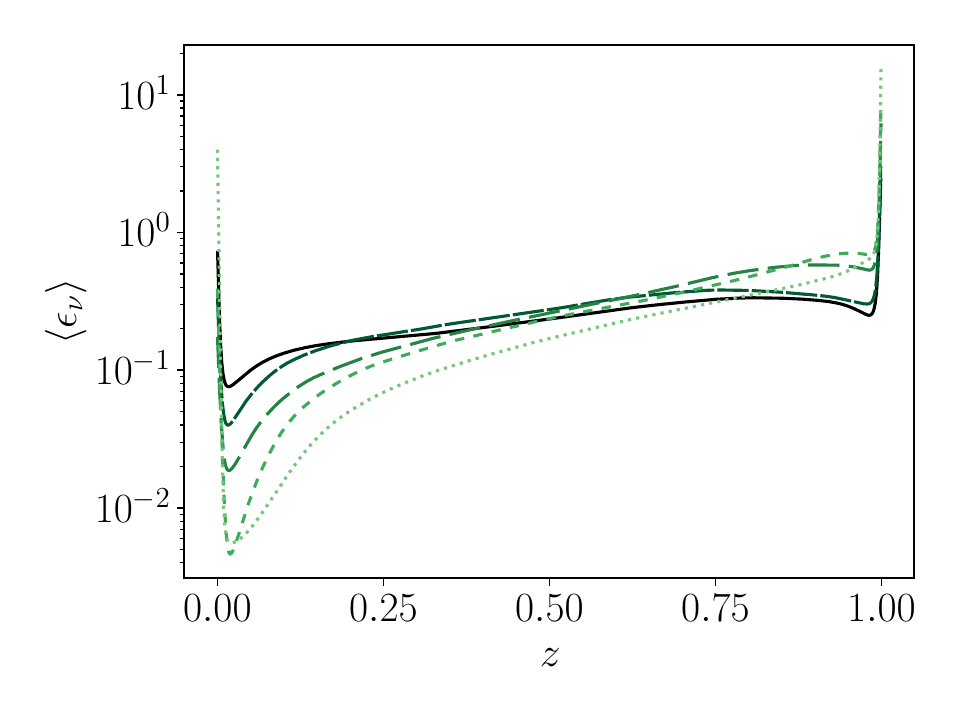}
\includegraphics[width=.32\linewidth]{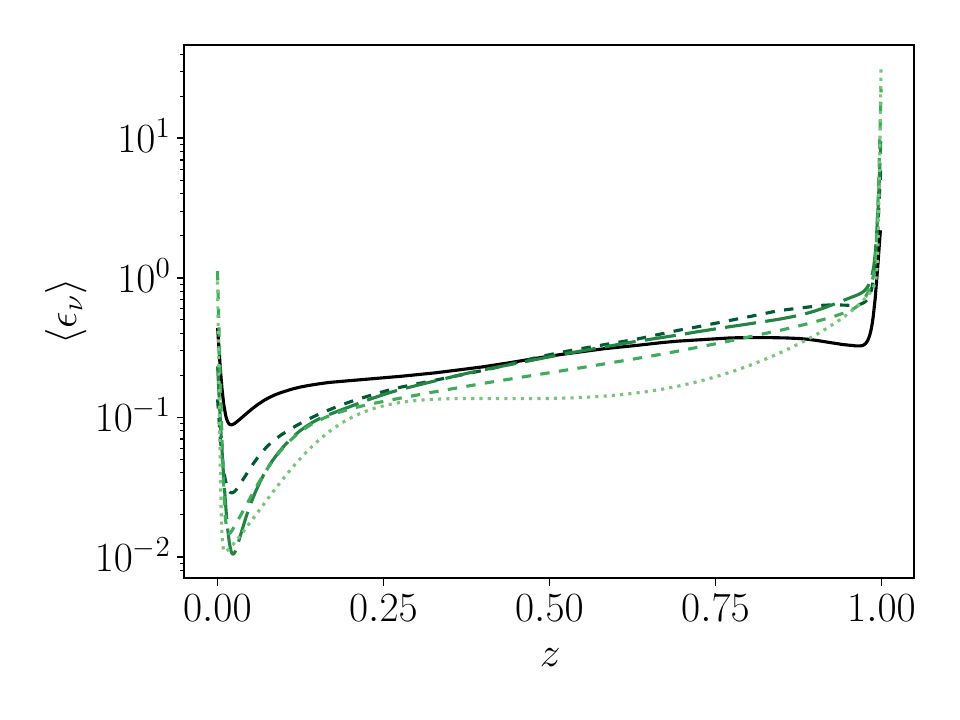}
\includegraphics[width=.32\linewidth]{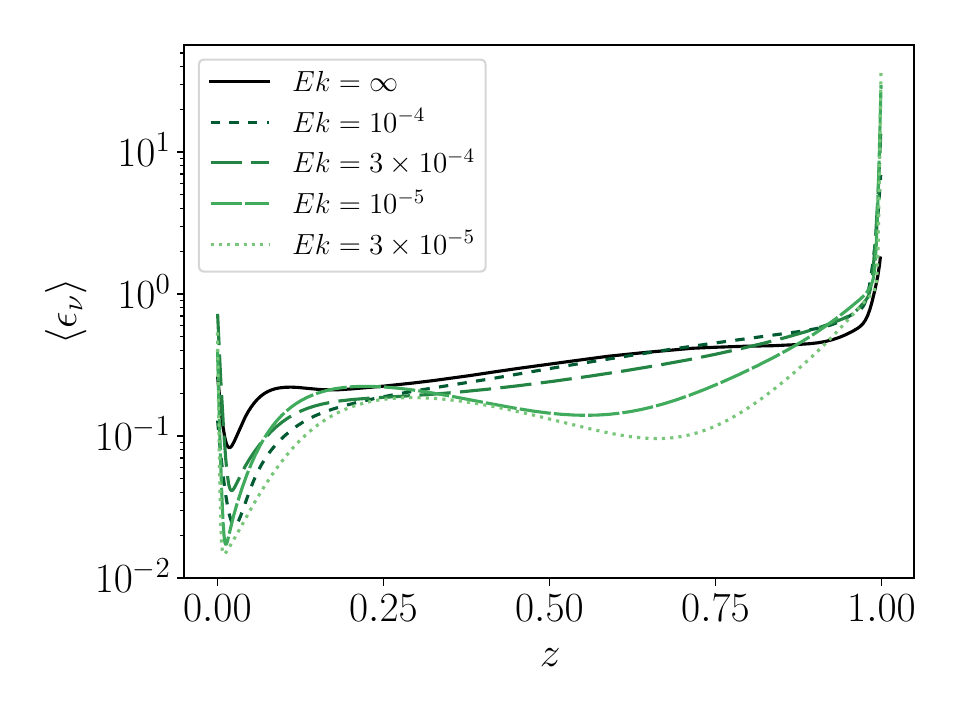}
\includegraphics[width=.32\linewidth]{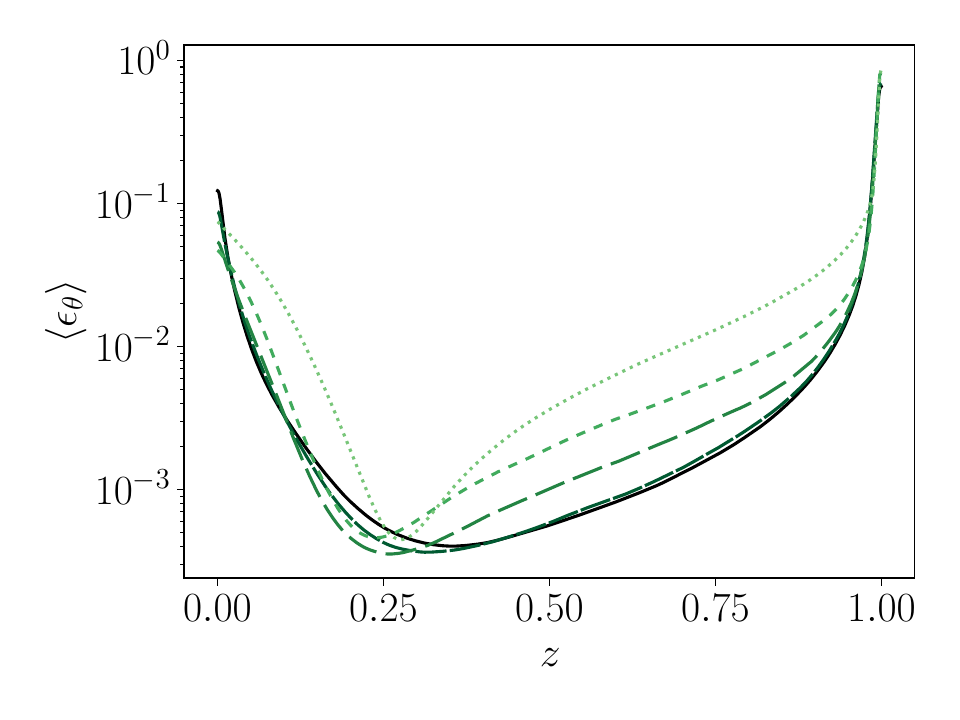}
\includegraphics[width=.32\linewidth]{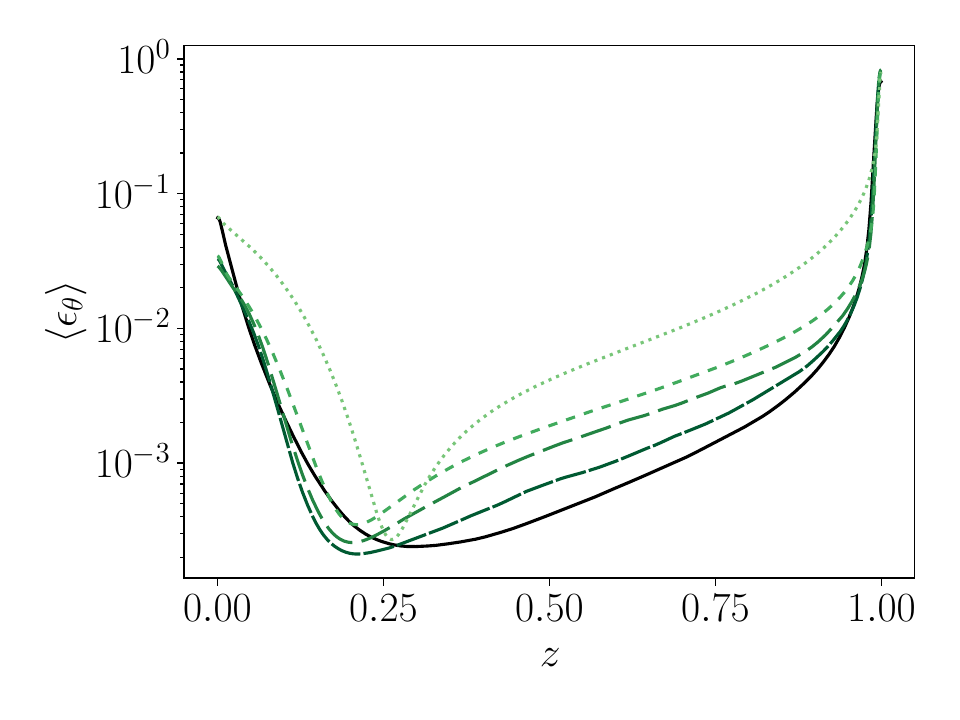}
\includegraphics[width=.32\linewidth]{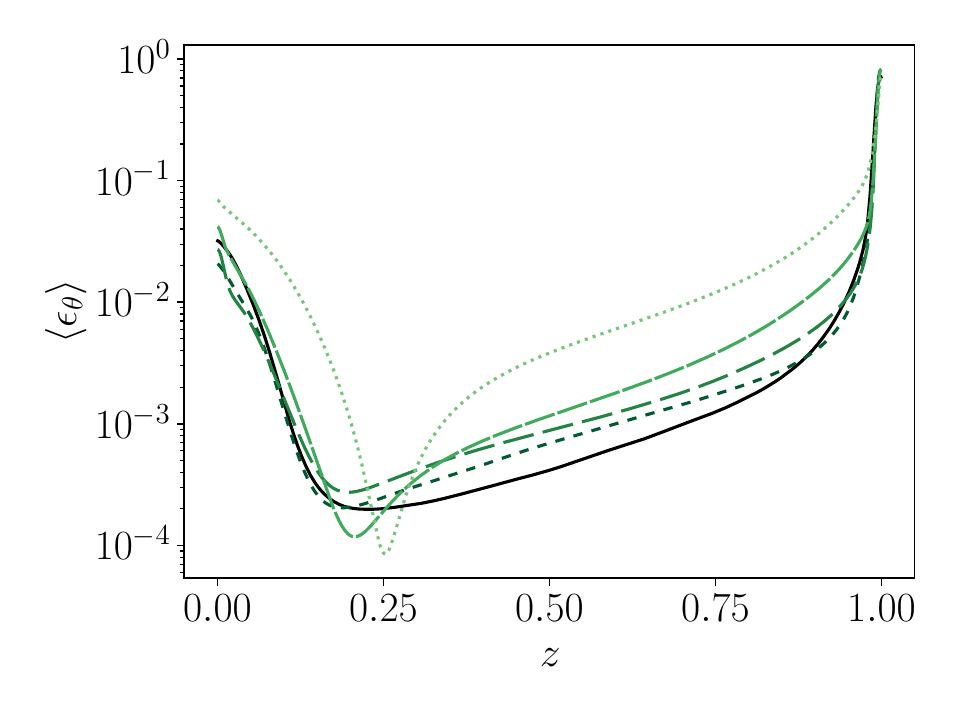}
\caption{Kinetic $\langle \epsilon_\nu \rangle$ (top row) and thermal $\langle \epsilon_\theta\rangle$ (bottom row) profiles for $R=10^{10}$ for several values of $E$ and $Pr=0.1$ (left column), $Pr=1$ (middle) and $Pr=10$ (right column).}
\label{fig:dissstats-rot-r1e10}
\end{figure}

\begin{figure}
\centering
\includegraphics[width=.48\linewidth]{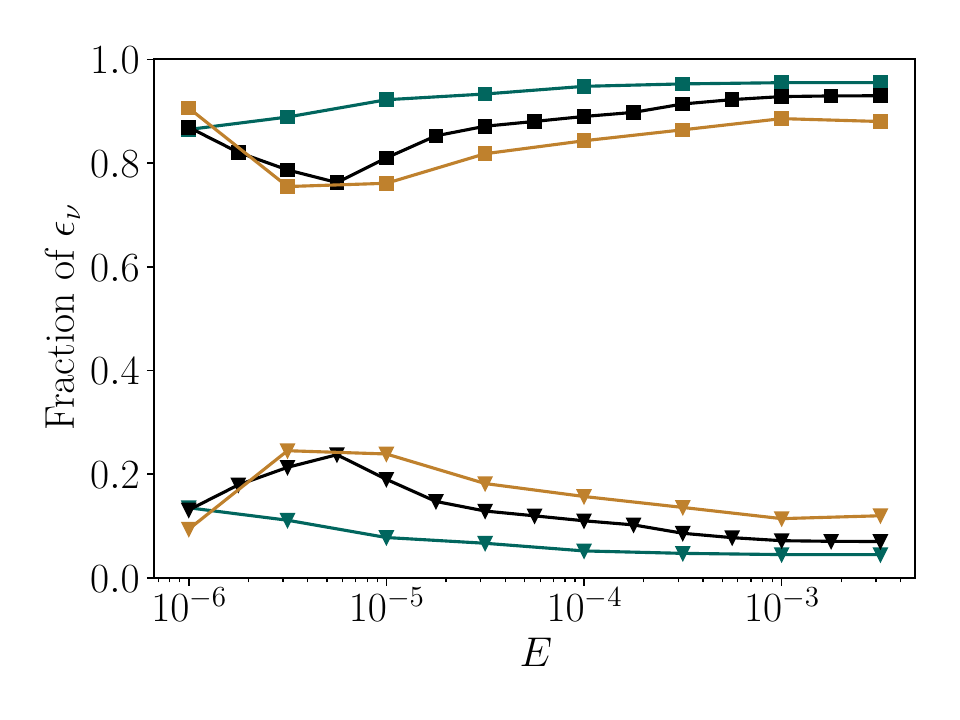}
\includegraphics[width=.48\linewidth]{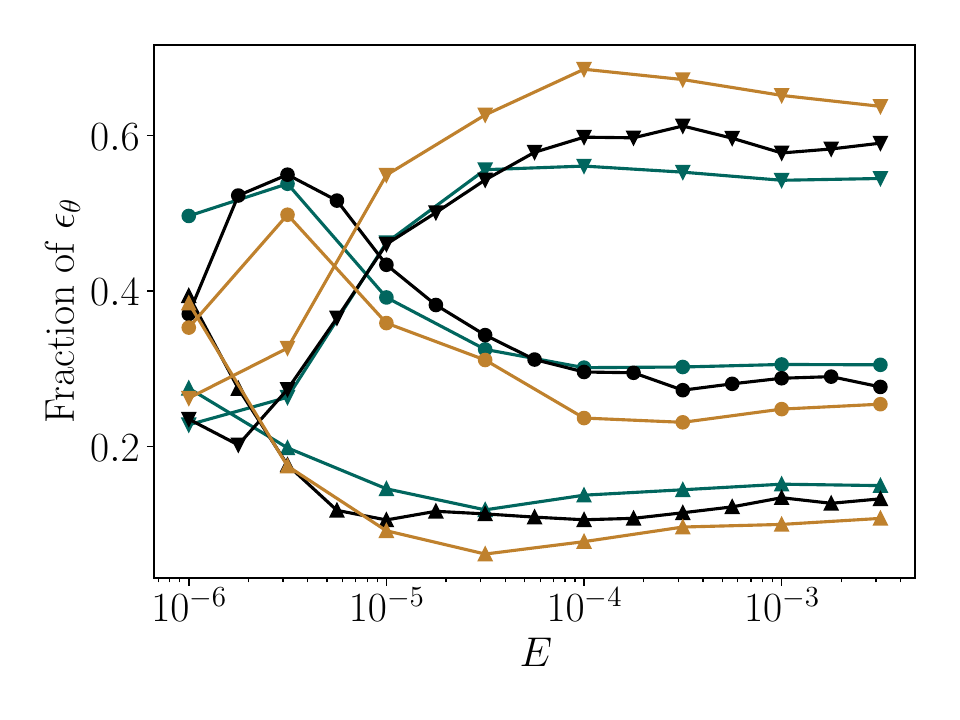}
\caption{Relative contributions to the dissipation rates of the flow regions as a function of rotation for $R=10^{10}$. Symbols: $\triangledown$, top boundary layer; $\circ$, bulk; $\triangle$, bottom boundary layer: $\square$, bulk and bottom boundary layer combined. Colors: Dark green, $Pr=0.1$; black, $Pr=1$; ochre, $Pr=10$.}
\label{fig:disssplit-rot-r1e10}
\end{figure}

To conclude the examination of rotation, we analyse the thermal ($\lepsth$) and viscous ($\lepsnu$) dissipation rates (Figure \ref{fig:dissstats-rot-r1e10}). For $\lepsnu$, the effect of rotation is qualitatively consistent across all $Pr$. Rotation suppresses dissipation in the bulk and bottom boundary layer while enhancing it near the top plate. This suppression is most visible at low $Pr$ where the non-rotating flow was initially more turbulent close to the bottom plate. On the other hand, the thermal dissipation $\lepsth$ responds to rotation by increasing in the bulk and decreasing at the boundaries, regardless of $Pr$. 

However, the impact on the total values of $\lepsth$ is more profound than on $\lepsnu$. As shown in Figure \ref{fig:disssplit-rot-r1e10}, the bulk remains the dominant source of viscous dissipation across all $E$ and $Pr$. In contrast, for thermal dissipation, the dominant contribution shifts from the top boundary layer to the bulk at high rotation rates ($E<10^{-5}$). This transition in the dissipation budget suggests a fundamental change in the heat transport bottleneck as the system enters a rotationally constrained state. This is indicative of where we expect the geostrophic regime to arise, once $R$ becomes large enough.

\section{Conclusion and Outlook}
\label{sec:conc}

In this study, we have numerically investigated the influence of the Prandtl number on internally heated convection (IHC) in both non-rotating and rotating regimes. By spanning two orders of magnitude in $Pr$ ($0.1\leq Pr\leq 10$) for rotating cases and three orders of magnitude for non-rotating cases, we have identified that while the global mean temperature $\T$ remains remarkably robust to changes in $Pr$, the underlying flow morphology and the distribution of local statistics are profoundly altered.

The most significant impact of $Pr$ is observed in the lower, stably stratified portion of the fluid layer. In the non-rotating case, low $Pr$ fluids experience a ``symmetry recovery'' where intense turbulent stirring from the bulk penetrates the stable layer, forcing it into a more active state. As $Pr$ increases, the increased viscosity and decreased thermal diffusivity allow the stable stratification to dominate, eventually leading to a ``dead zone'' of nearly quiescent fluid at $Pr=100$. This finding suggests that in geophysical contexts—where $Pr$ can vary significantly—the ``effective'' depth of a convective layer might be dynamically determined by the Prandtl number, as high-$Pr$ fluids effectively ``mask'' the lower portion of the domain from convective mixing.

The introduction of rotation introduces a competing mechanism of organisation. We find that rotation enhances vertical convective flux $\wT$ across all $Pr$, but global heat transport efficiency (indicated by $\T$) only improves for $Pr\ge1$. This confirms that the Ekman pumping mechanism, which facilitates transport in rotating fluids, requires a sufficiently high $Pr$ to overcome thermal diffusion, and is present in IHC. 

Our results demonstrate that the analogy between IHC and Rayleigh-B\'enard convection (RBC) is nuanced. While IHC shares many of the scaling behaviors of RBC at the top boundary, the stable stratification at the bottom create a unique sensitivity to the Prandtl number. Future work should explore the ``infinite Pr'' limit more rigorously to see if the trends observed at $Pr=100$ saturate, and whether the observed behaviour holds as the system approaches the fully geostrophic regime at even lower Ekman numbers. This would be particularly relevant for modeling the Earth's mantle or the deep interiors of gas giants, where both internal heating and rotation are primary drivers of fluid motion. Another direction of exploration is probing the stability of rotating IHC, particularly at low Prandtl numbers. For these control parameters, subcritical convection may be more important, significantly modifying the values of $R$ and $E$ for which convection appears. 

\noindent {\it Acknowledgments:} ROM acknowledges support from the Emergia Program of the Junta de Andalucía (Spain).
AA acknowledges funding from the ERC (agreement no. 833848-UEMHP) under the Horizon 2020 program and the SNSF (grant number 219247) under the MINT 2023 call. We also thank the Systems Unit of the Information Systems Area of the University of Cadiz for computer resources and technical support. 

\noindent {\it Declaration of Interests:} The authors report no conflict of interest.

\noindent {\it Generative AI:} Gemini has been used to assist in the writing of the manuscript.

\appendix

\section{Dependence on domain size}
\label{sec:app}

In this section, we provide a brief assessment of the independence of our results from the horizontal domain size. We first consider the non-rotating case at a fixed Rayleigh number of $R=10^9$. Simulations were conducted for $Pr=0.1$, $1$, and $10$ using aspect ratios of $\Gamma=1$ and $\Gamma=2$. To maintain consistent grid resolution, the number of grid points in the horizontal directions was doubled for the larger domain (e.g., $288^2\times144$ for $\Gamma=1$ and $576^2\times144$ for $\Gamma=2$).

The global response quantities are summarized in Table \ref{tbl:app}. The wind Reynolds number, $Re_w$, is the most sensitive parameter to the domain size, exhibiting a measurable dependence on $\Gamma$ even at $Pr=10$. In contrast, the vertical convective flux $\wT$ only shows sensitivity to $\Gamma$ at the lowest Prandtl number ($Pr=0.1$), while the mean temperature $\T$ remains largely independent of the aspect ratio across the investigated $Pr$ range.

\begin{figure}
\centering
\includegraphics[width=.49\linewidth]{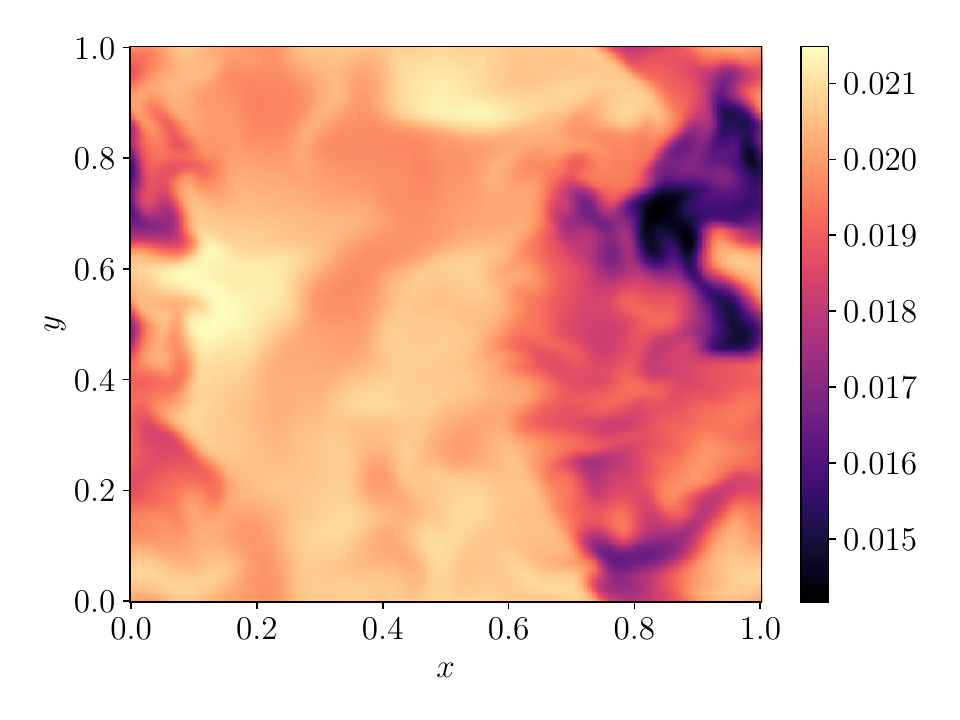}
\includegraphics[width=.49\linewidth]{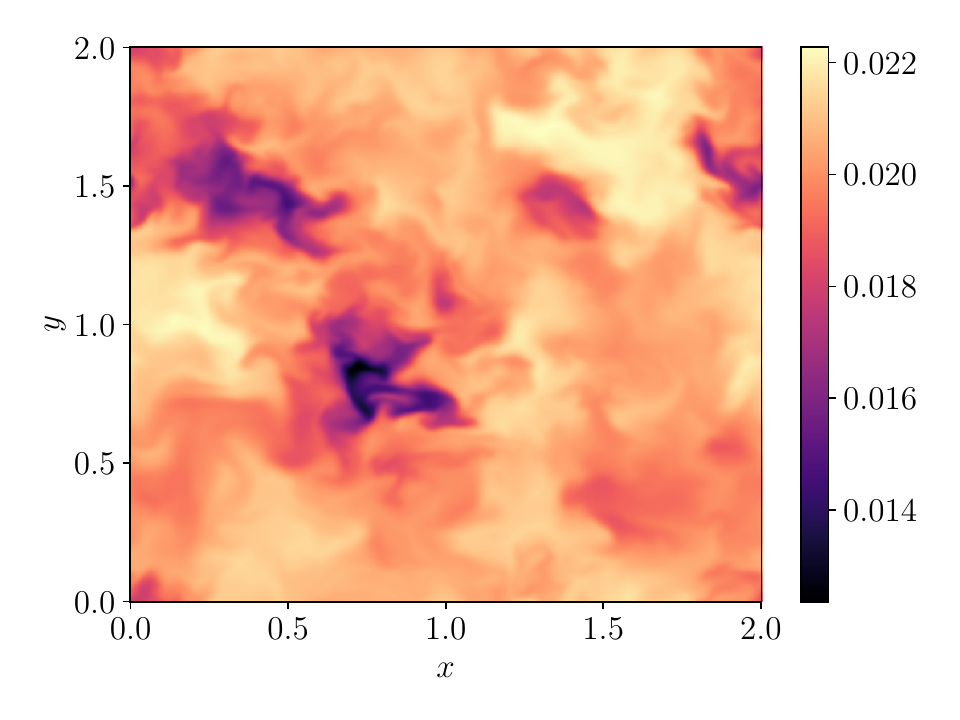}
\includegraphics[width=.49\linewidth]{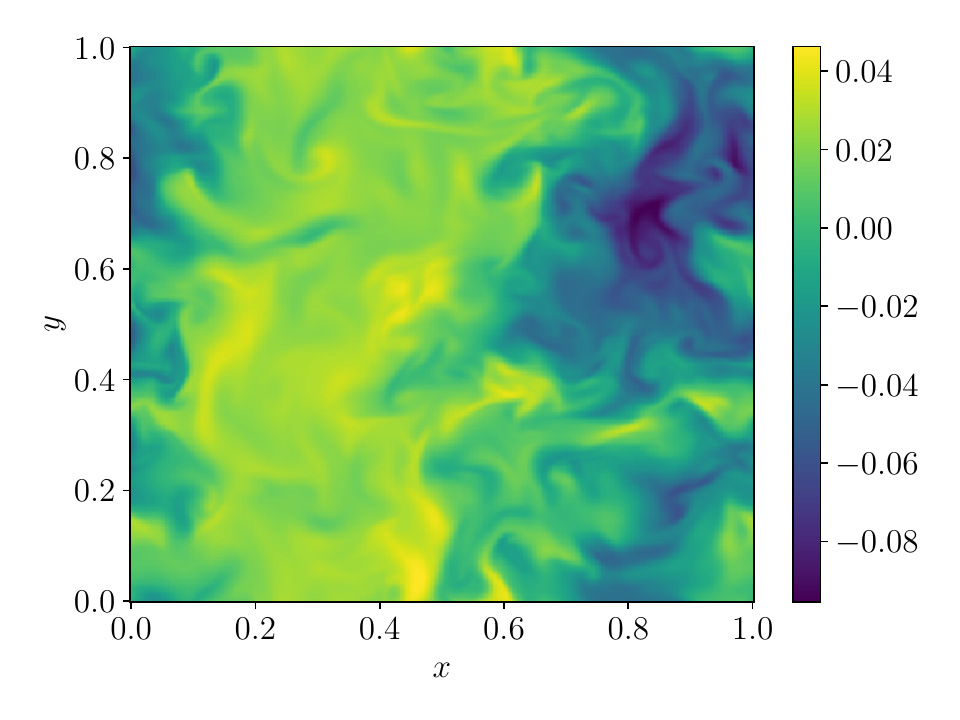}
\includegraphics[width=.49\linewidth]{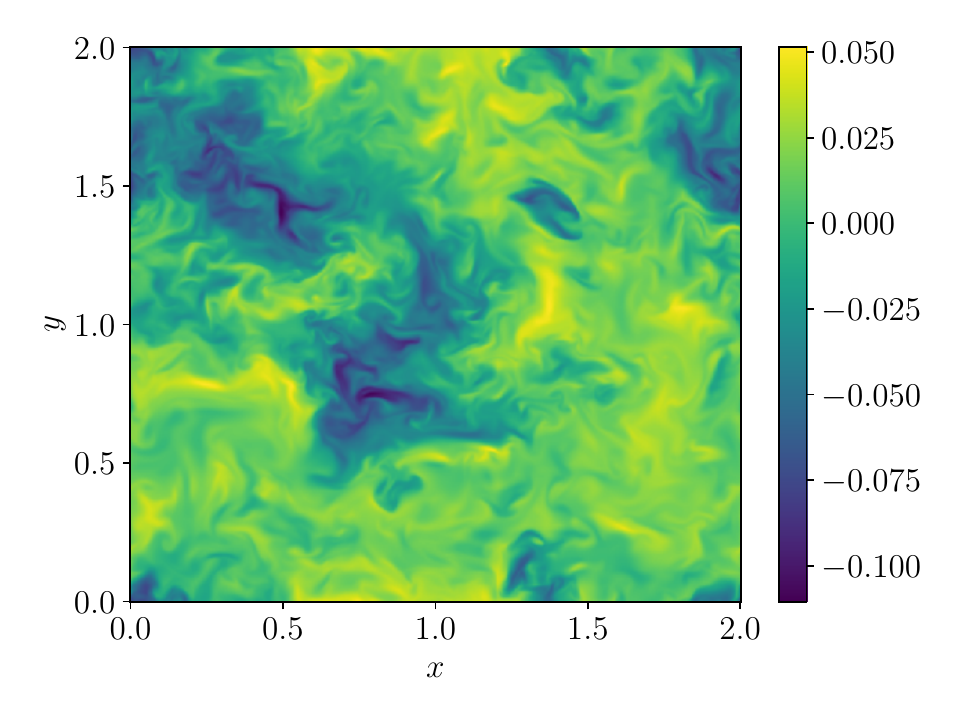}
\caption{Instantaneous temperature (top row) and vertical velocity (bottom row) for a sample case with $R=10^9$, $Pr=0.1$ and no rotation at $z=0.5$.}
\label{fig:gamma-dep-pr0p1-norot}
\end{figure}

\begin{figure}
\centering
\includegraphics[width=.49\linewidth]{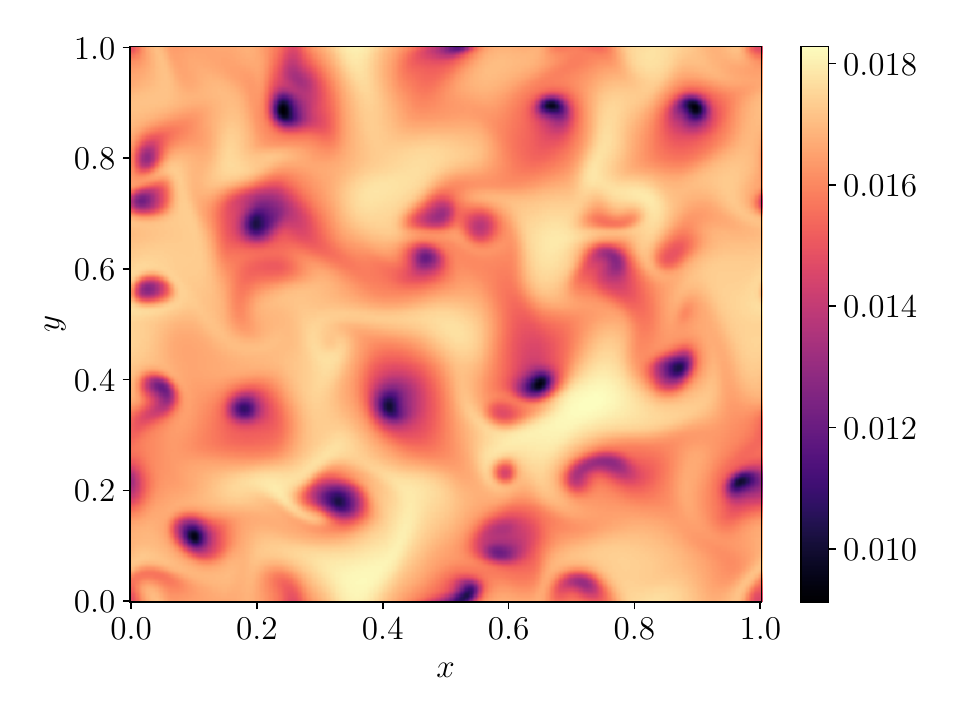}
\includegraphics[width=.49\linewidth]{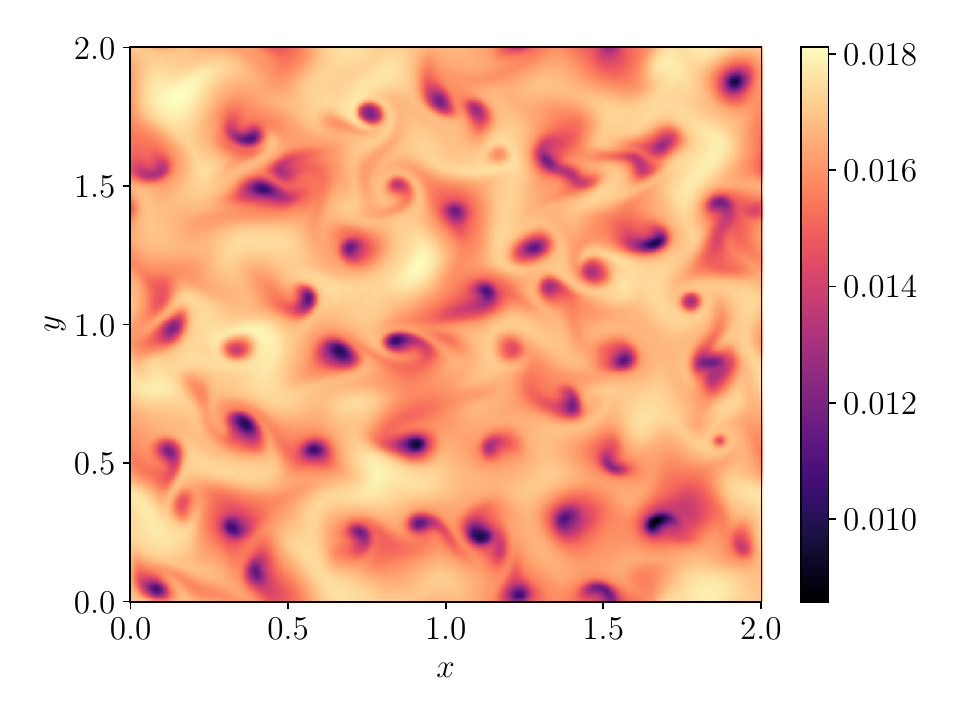}
\includegraphics[width=.49\linewidth]{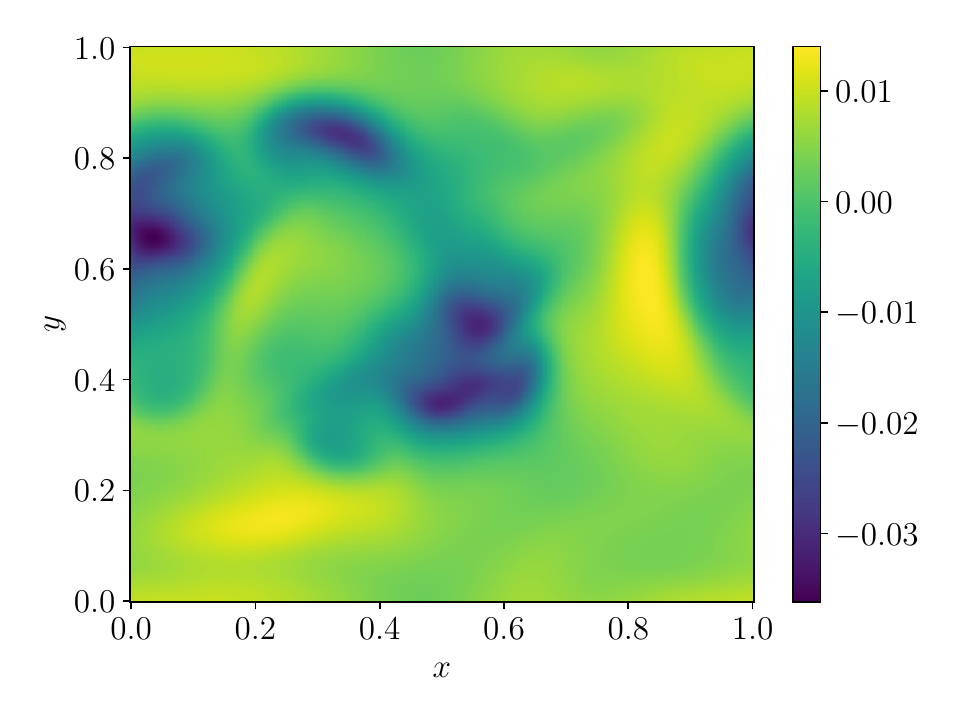}
\includegraphics[width=.49\linewidth]{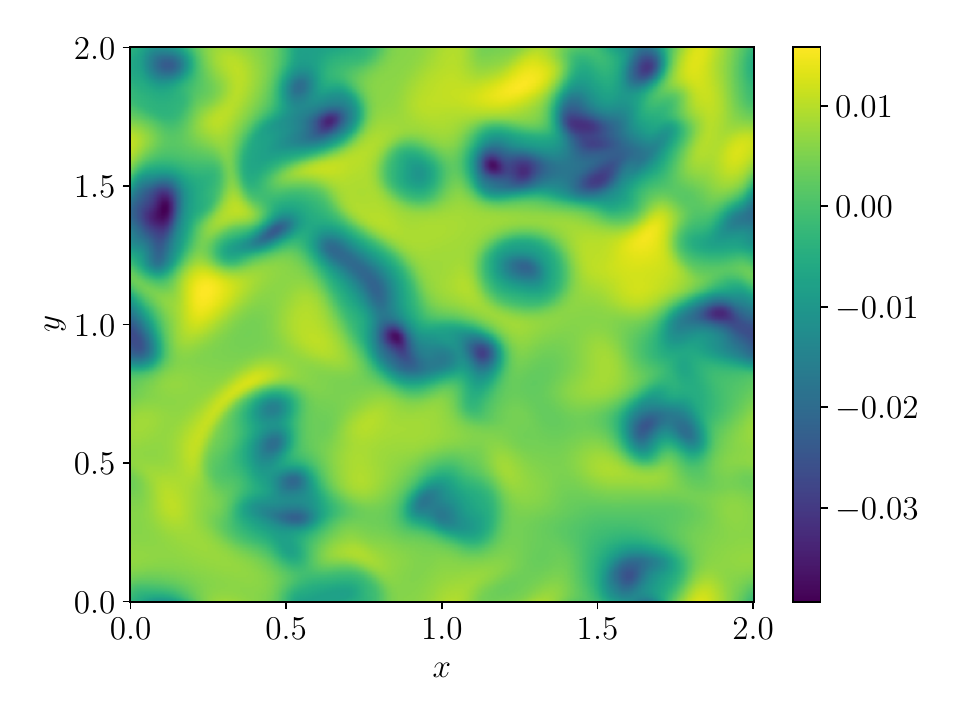}
\caption{Instantaneous temperature (top row) and vertical velocity (bottom row) for a sample case with $R=10^9$, $Pr=10$ and no rotation at $z=0.5$.}
\label{fig:gamma-dep-pr10-norot}
\end{figure}

To explain these observations, Figures \ref{fig:gamma-dep-pr0p1-norot} and \ref{fig:gamma-dep-pr10-norot} present mid-plane visualizations of the instantaneous temperature and velocity fields for $Pr=0.1$ and $Pr=10$. At $Pr=0.1$, the flow is characterized by large-scale structures that tend to fill the computational volume, whereas at $Pr=10$, the convective structures are significantly smaller and do not exhibit box-filling behavior.

\begin{figure}
\centering
\includegraphics[width=.49\linewidth]{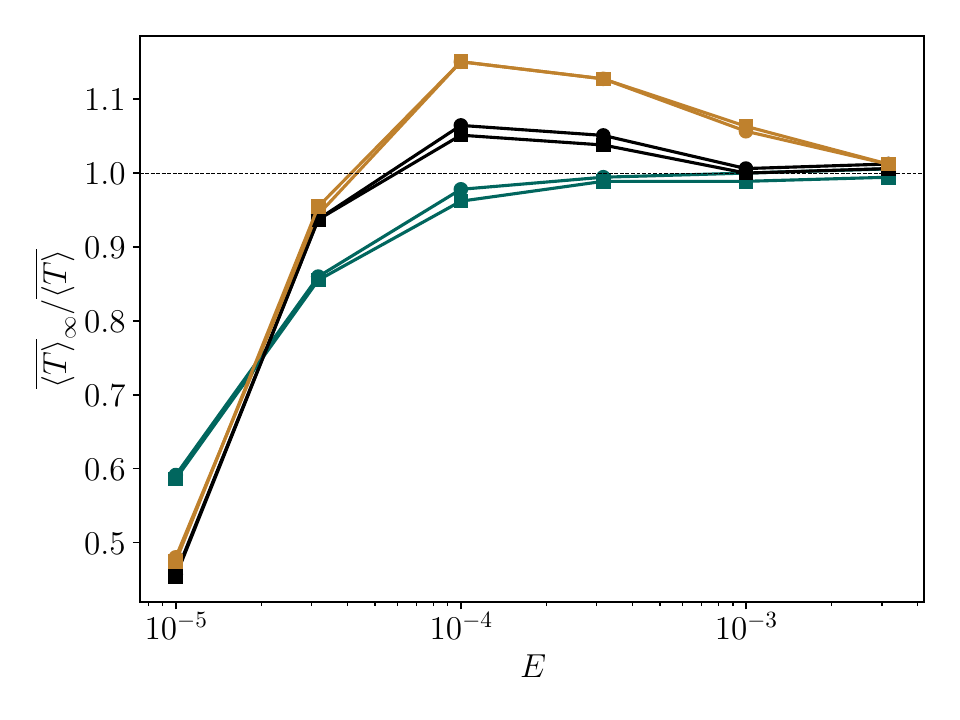}
\includegraphics[width=.49\linewidth]{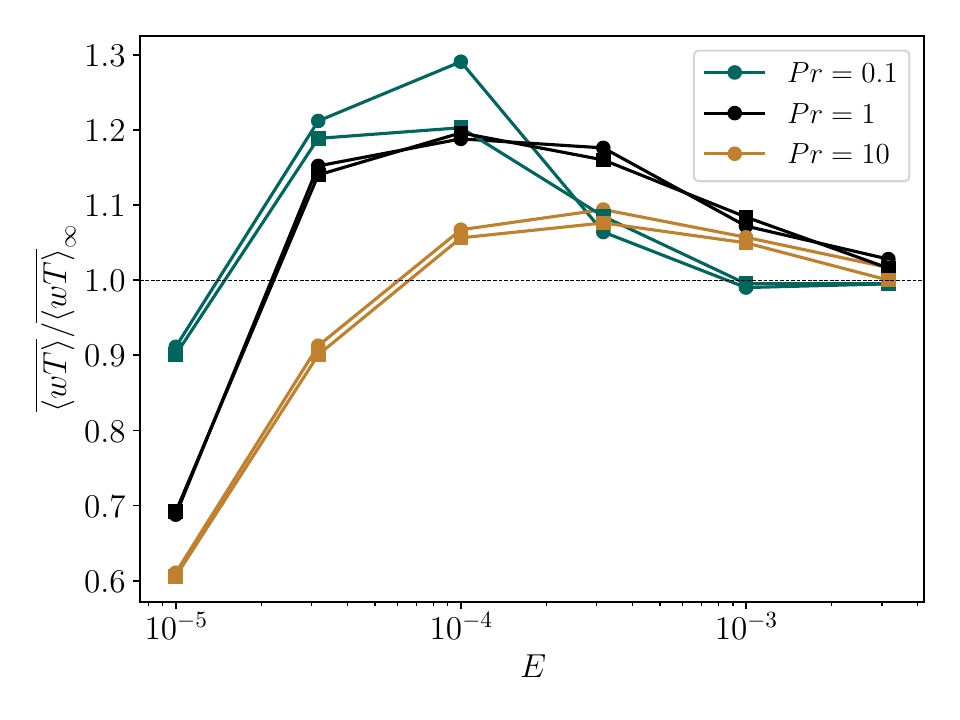}
\caption{Changes in the global responses with rotation for $R=10^9$ for $\Gamma=1$ (round symbols) and $\Gamma=2$ (square symbols).}
\label{fig:app}
\end{figure}

The influence of domain size under rotation is illustrated in Figure \ref{fig:app}, which shows $\wT$ and $\T$ (normalized by their non-rotating counterparts) for $\Gamma=1$ and $\Gamma=2$ at $R=10^{10}$. For $Pr=0.1$, the relative magnitude of the enhancement in $\wT$ is sensitive to the box size, particularly near the peak at $E=10^{-4}$. However, the qualitative physical trend—an initial rotational enhancement of convective transport—persists regardless of $\Gamma$. For $Pr=1$ and $Pr=10$, the results are remarkably robust to changes in domain size. Furthermore, the right panel of Figure \ref{fig:app} confirms that the enhancement of global heat transfer efficiency ($\T$) is nearly independent of $\Gamma$ across all Prandtl numbers.

\begin{figure}
\centering
\includegraphics[width=.49\linewidth]{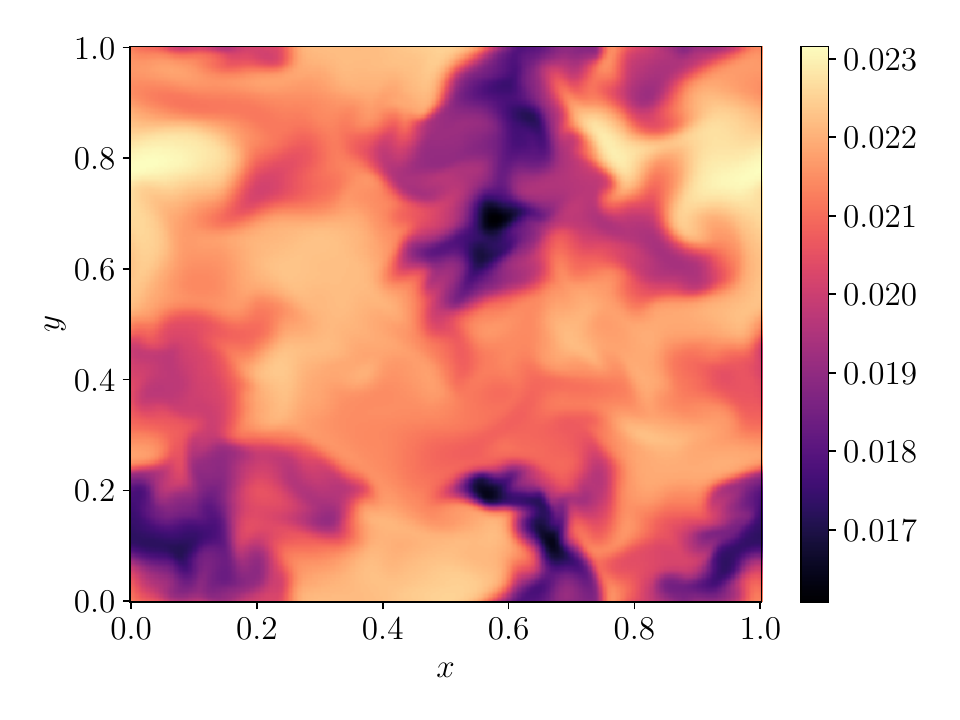}
\includegraphics[width=.49\linewidth]{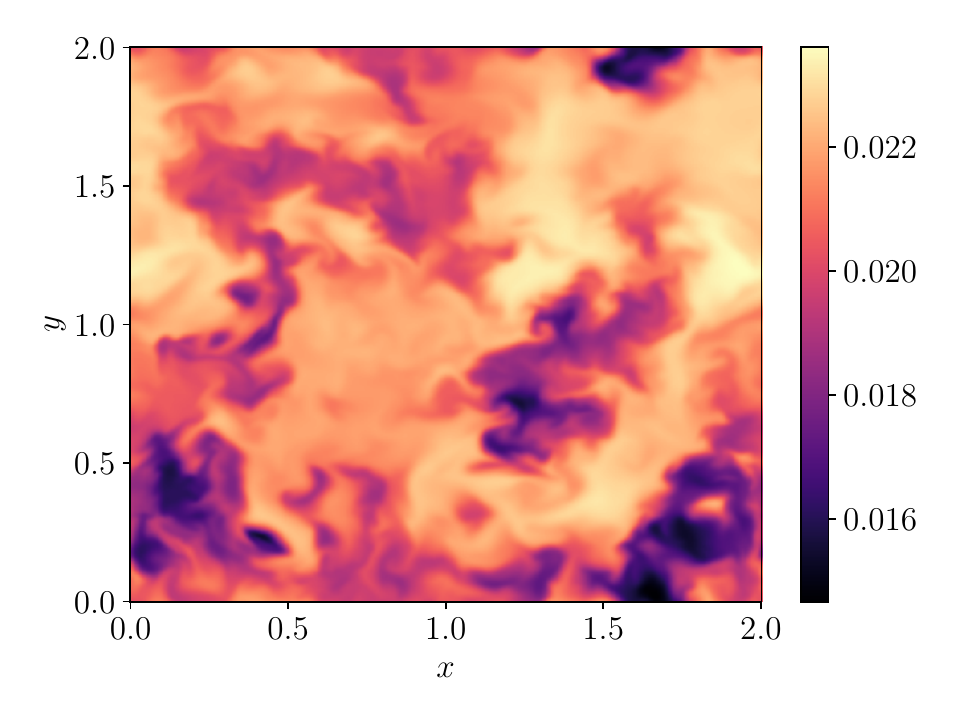}
\includegraphics[width=.49\linewidth]{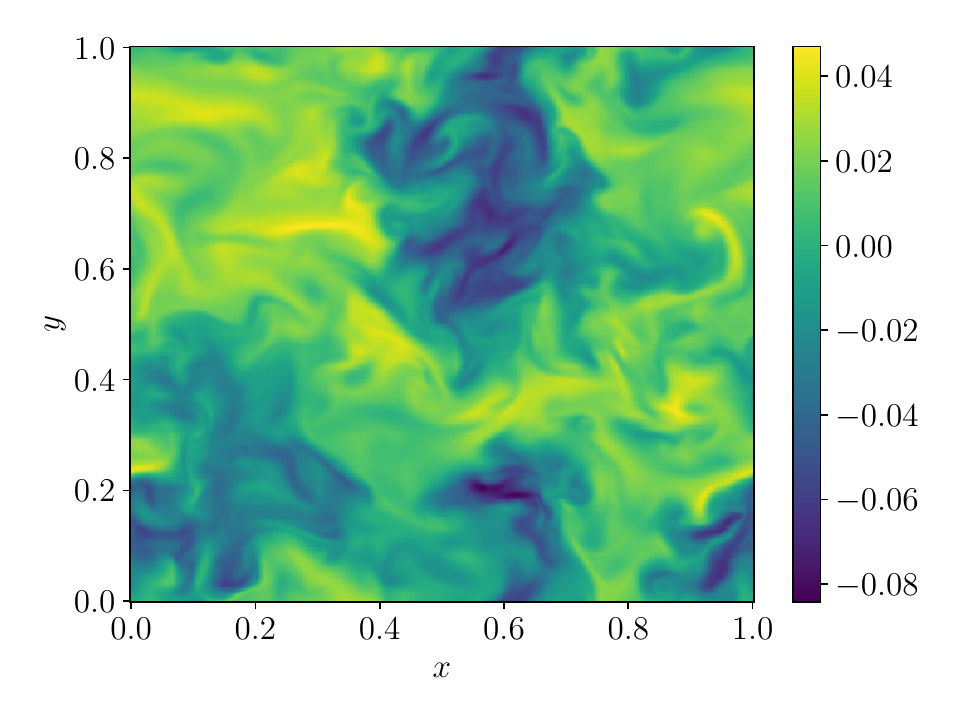}
\includegraphics[width=.49\linewidth]{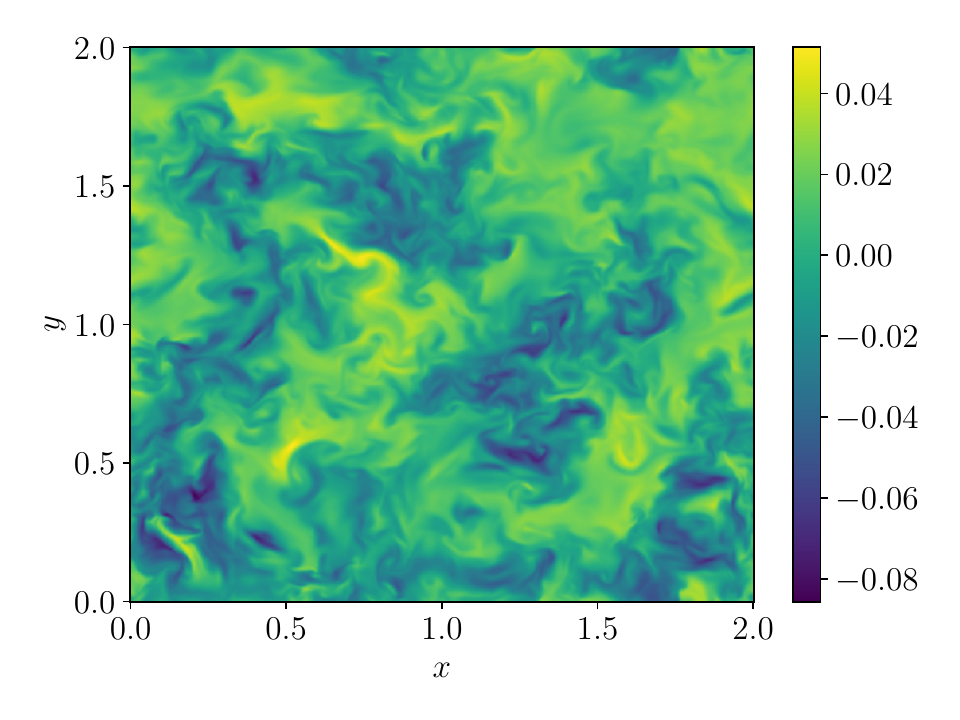}
\caption{Instantaneous temperature (top row) and vertical velocity (bottom row) for a sample case with $R=10^9$, $Pr=0.1$ and $E=10^{-4}$ at $z=0.5$.}
\label{fig:gamma-dep-pr0p1-ek1e4}
\end{figure}

\begin{figure}
\centering
\includegraphics[width=.49\linewidth]{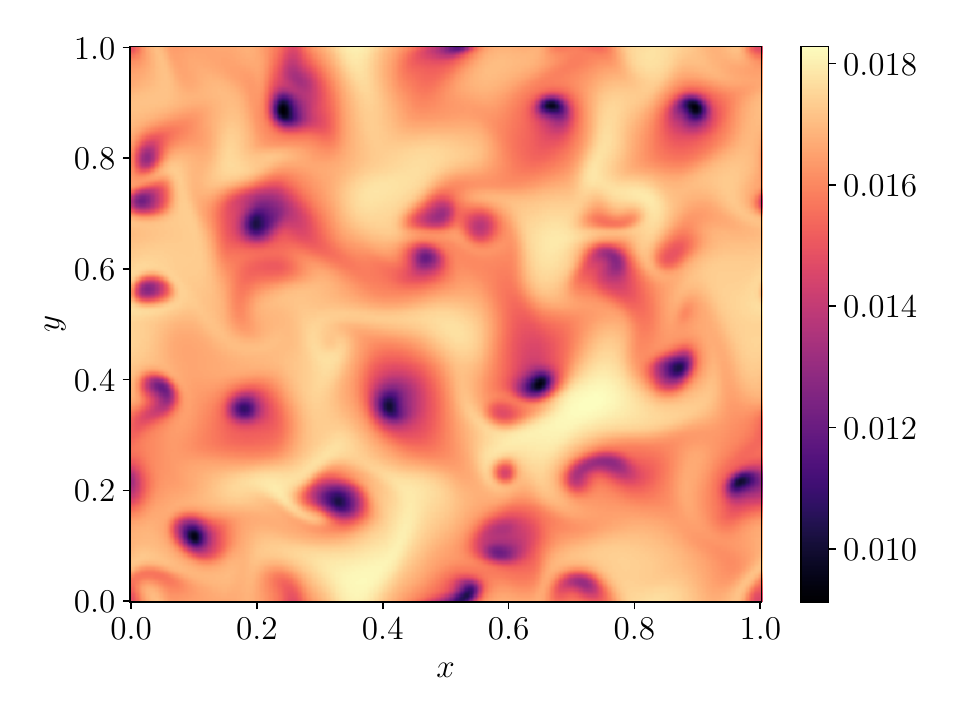}
\includegraphics[width=.49\linewidth]{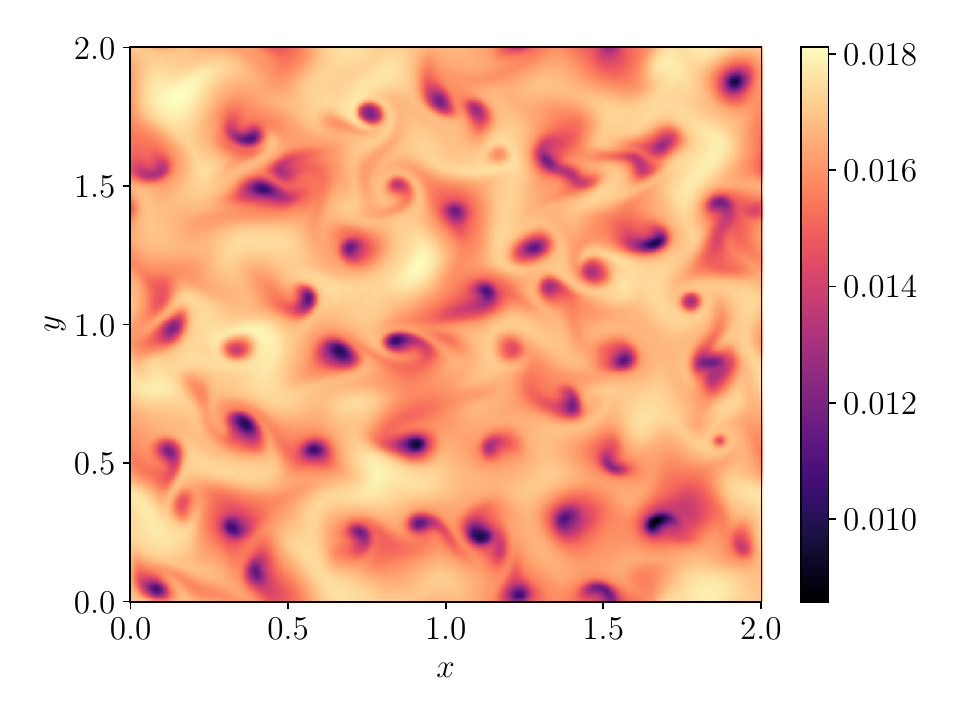}
\includegraphics[width=.49\linewidth]{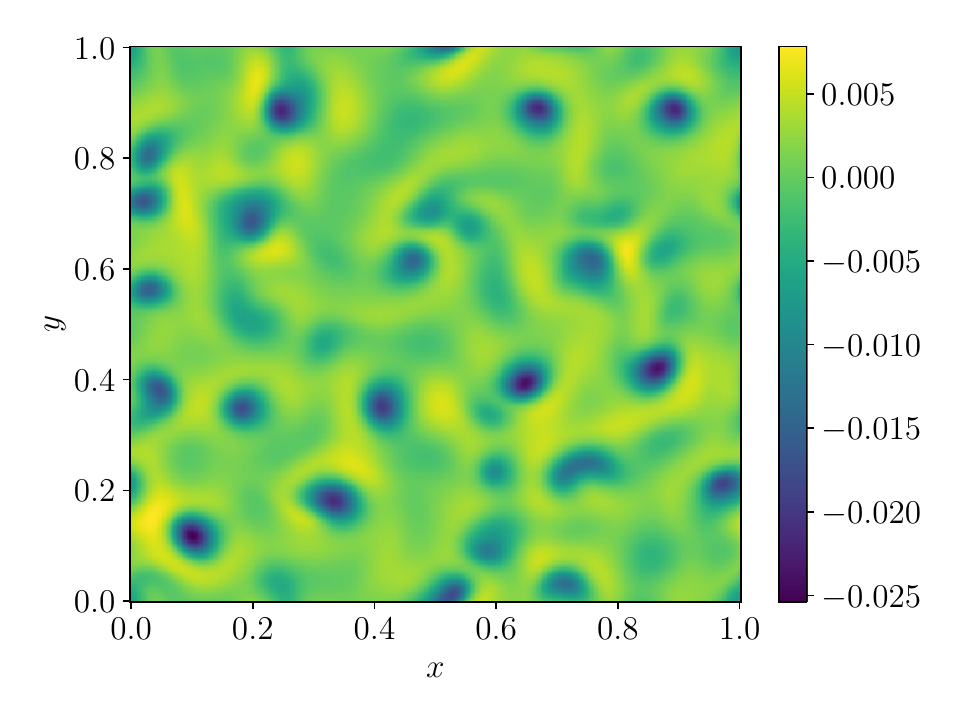}
\includegraphics[width=.49\linewidth]{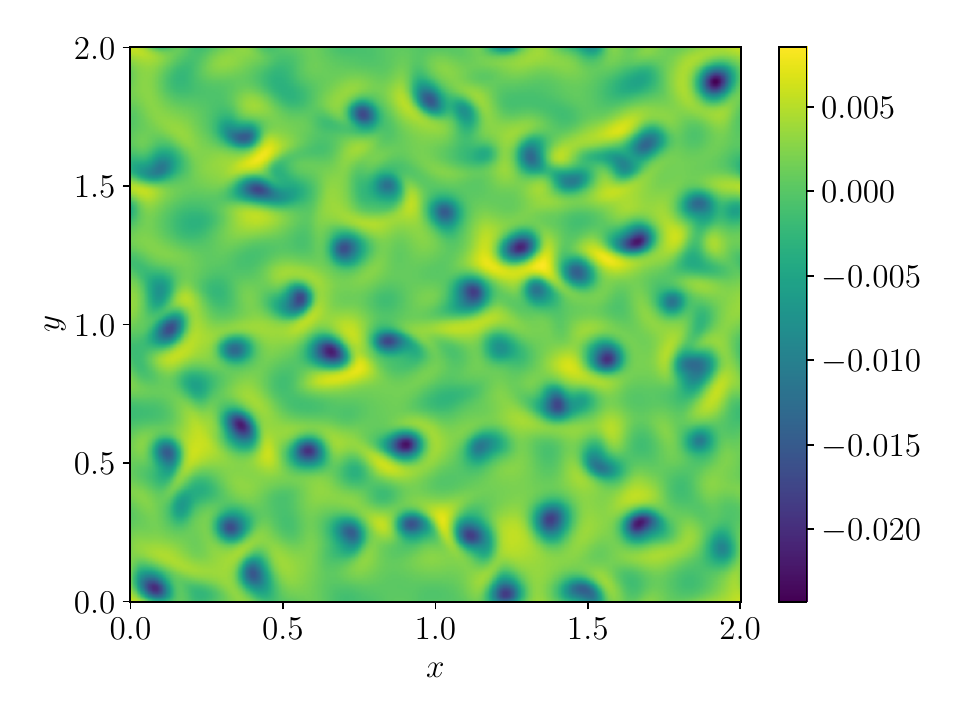}
\caption{Instantaneous temperature (top row) and vertical velocity (bottom row) for a sample case with $R=10^9$, $Pr=10$ and $E=10^{-4}$ at $z=0.5$.}
\label{fig:gamma-dep-pr10-ek1e4}
\end{figure}

Finally, visualizations in Figures \ref{fig:gamma-dep-pr0p1-ek1e4} and \ref{fig:gamma-dep-pr10-ek1e4} further illustrate the role of box size in the rotating regime. Consistent with the non-rotating cases, $Pr=0.1$ features larger convective structures; however, these do not fill the domain to the same extent as in the non-rotating simulations. The structures at $Pr=10$ remain small, thereby minimizing the influence of the periodic boundary conditions.

\begin{table}
    \centering
    \begin{tabular}{|c|c|c|c|c|c|c|c|c|}
        \hline
        $Pr$ & $\Gamma$ & $\wT$ & $\T$ & $Re_w$ & $\Gamma$ & $\wT$ & $\T$ & $Re_w$ \\
        \hline
        0.1 & 1 & $2.03\times10^{-1}$ & $1.78\times10^{-2}$ & $3.03\times10^{3}$ & 2 & $2.12\times10^{-1}$	& $1.77\times10^{-2}$ & $3.22\times10^{3}$ \\
        1 & 1 & $2.50\times10^{-1}$ & $1.65\times10^{-2}$ & $5.88\times10^{2}$ & 2 & $2.50\times10^{-1}$	& $1.64\times10^{-2}$ & $6.06\times10^{2}$ \\
        10 & 1 & $2.98\times10^{-1}$ & $1.68\times10^{-2}$ & $8.71\times10^{1}$ & 2 & $3.02\times10^{-1}$	& $1.68\times10^{-2}$ & $8.82\times10^{1}$ \\
        \hline
    \end{tabular}
    \caption{Dependence on the horizontal periodicity length of the global responses.}
    \label{tbl:app}
\end{table}

\bibliographystyle{jfm}
\bibliography{jfm}

@article{van2015pencil,
  title={A pencil distributed finite difference code for strongly turbulent wall-bounded flows},
  author={Van Der Poel, E. P. and Ostilla-M{\'o}nico, R. and Donners, J. and Verzicco, R.},
  journal={Computers \& Fluids},
  volume={116},
  pages={10--16},
  year={2015},
  publisher={Elsevier}
}

@article{tritton1967,
  title={Convection in horizontal layers with internal heat generation. Experiments},
  author={Tritton, D. J. and Zarraga, M. N.},
  journal={Journal of Fluid Mechanics},
  volume={30},
  number={1},
  pages={21--31},
  year={1967},
  publisher={Cambridge University Press}
}

@article{kulacki1972,
  title={Thermal convection in a horizontal fluid layer with uniform volumetric energy sources},
  author={Kulacki, Francis Alfred and Goldstein, RJ},
  journal={Journal of Fluid Mechanics},
  volume={55},
  number={2},
  pages={271--287},
  year={1972},
  publisher={Cambridge University Press}
}

@article{lohse2024ultimate,
  title={Ultimate rayleigh-b{\'e}nard turbulence},
  author={Lohse, Detlef and Shishkina, Olga},
  journal={Reviews of modern physics},
  volume={96},
  number={3},
  pages={035001},
  year={2024},
  publisher={APS}
}

@article{horn2017prograde,
  title={Prograde, retrograde, and oscillatory modes in rotating Rayleigh--B{\'e}nard convection},
  author={Horn, Susanne and Schmid, Peter J},
  journal={Journal of Fluid Mechanics},
  volume={831},
  pages={182--211},
  year={2017},
  publisher={Cambridge University Press}
}

@article{aurnou2018rotating,
  title={Rotating thermal convection in liquid gallium: multi-modal flow, absent steady columns},
  author={Aurnou, Jonathan M and Bertin, Vincent and Grannan, Alexander M and Horn, Susanne and Vogt, Tobias},
  journal={Journal of Fluid Mechanics},
  volume={846},
  pages={846--876},
  year={2018},
  publisher={Cambridge University Press}
}

@article{vogt2018jump,
  title={Jump rope vortex in liquid metal convection},
  author={Vogt, Tobias and Horn, Susanne and Grannan, Alexander M and Aurnou, Jonathan M},
  journal={Proceedings of the National Academy of Sciences},
  volume={115},
  number={50},
  pages={12674--12679},
  year={2018},
  publisher={National Academy of Sciences}
}

@article{fan2024scaling,
  title={Scaling behaviour of rotating convection in a spherical shell with different Prandtl numbers},
  author={Fan, Wei and Wang, Qi and Lin, Yufeng},
  journal={Journal of Fluid Mechanics},
  volume={998},
  pages={A20},
  year={2024},
  publisher={Cambridge University Press}
}

@article{xu2025thermovelocimetric,
  title={Thermovelocimetric characterization of liquid metal convection in a rotating slender cylinder},
  author={Xu, Yufan and Abbate, Jewel and David, Cy and Vogt, Tobias and Aurnou, Jonathan},
  journal={International Journal of Heat and Mass Transfer},
  volume={252},
  pages={127325},
  year={2025},
  publisher={Elsevier}
}

@article{thirlby1970,
  title={Convection in an internally heated layer},
  author={Thirlby, R.},
  journal={Journal of Fluid Mechanics},
  volume={44},
  number={4},
  pages={673--693},
  year={1970},
  publisher={Cambridge University Press}
}

@article{veronis1959,
  title={Cellular convection with finite amplitude in a rotating fluid},
  author={Veronis, G.},
  journal={Journal of Fluid Mechanics},
  volume={5},
  number={3},
  pages={401--435},
  year={1959},
  publisher={Cambridge University Press}
}

@article{chandrasekhar1953,
  title={The instability of a layer of fluid heated below and subject to Coriolis forces},
  author={Chandrasekhar, S.},
  journal={Proceedings of the Royal Society of London. Series A. Mathematical and Physical Sciences},
  volume={217},
  number={1130},
  pages={306--327},
  year={1953},
  publisher={The Royal Society London}
}

@article{marshall1999,
  title={Open-ocean convection: Observations, theory, and models},
  author={Marshall, J. and Schott, F.},
  journal={Reviews of geophysics},
  volume={37},
  number={1},
  pages={1--64},
  year={1999},
  publisher={Wiley Online Library}
}

@book{emanuel1994,
  title={Atmospheric convection},
  author={Emanuel, K. A.},
  year={1994},
  publisher={Oxford university press}
}

@article{aurnou2015,
  title={Rotating convective turbulence in Earth and planetary cores},
  author={Aurnou, J. M. and Calkins, M. A. and Cheng, J. S. and Julien, K. and King, E. M. and Nieves, D. and Soderlund, K. M. and Stellmach, S.},
  journal={Physics of the Earth and Planetary Interiors},
  volume={246},
  pages={52--71},
  year={2015},
  publisher={Elsevier}
}

@article{roberts2013,
  title={On the genesis of the Earth's magnetism},
  author={Roberts, P. H. and King, E. M.},
  journal={Reports on Progress in Physics},
  volume={76},
  number={9},
  pages={096801},
  year={2013},
  publisher={IOP Publishing}
}

@article{King2013,
author = {King, E. M. and Aurnou, J. M.},
title = {Turbulent convection in liquid metal with and without rotation},
journal = {Proceedings of the National Academy of Sciences},
volume = {110},
number = {17},
pages = {6688-6693},
year = {2013},
doi = {10.1073/pnas.1217553110},
}

@article{Wang2020,
author = {Wang, Q. and Lohse, D. and Shishkina, O.},
title = {Scaling in internally heated convection: a unifying theory},
journal = {Geophysical Research Letters},
volume = {47},
pages = {e2020GL091198},
doi = {10.1029/2020GL091198},
year = {2020}
}

@article{kazemi2022transition,
  title={Transition between boundary-limited scaling and mixing-length scaling of turbulent transport in internally heated convection},
  author={Kazemi, S. and Ostilla-M{\'o}nico, R. and Goluskin, D.},
  journal={Physical Review Letters},
  volume={129},
  number={2},
  pages={024501},
  year={2022},
  publisher={APS}
}

@article{ostilla2025ihc,
  title={Rotationally affected internally heated convection},
  author={Ostilla-M\'onico, R. and Arslan, A.},
  journal={Journal of fluid mechanics},
  volume={1018},
  pages={A31},
  year={2025},
  publisher={Cambridge University Press}
}

@article{goluskin2016penetrative,
  title={Penetrative internally heated convection in two and three dimensions},
  author={Goluskin, D. and Van der Poel, E. P.},
  journal={Journal of fluid mechanics},
  volume={791},
  pages={R6},
  year={2016},
  publisher={Cambridge University Press}
}

@article{vorobieff2002,
  title={Turbulent rotating convection: an experimental study},
  author={Vorobieff, P. and Ecke, R. E.},
  journal={Journal of Fluid Mechanics},
  volume={458},
  pages={191--218},
  year={2002},
  publisher={Cambridge University Press}
}

@article{stevens2013,
  title={Heat transport and flow structure in rotating Rayleigh--B{\'e}nard convection},
  author={Stevens, R. J. A. M. and Clercx, H. J. H. and Lohse, D.},
  journal={European Journal of Mechanics-B/Fluids},
  volume={40},
  pages={41--49},
  year={2013},
  publisher={Elsevier}
}

@article{grooms2010,
  title={Model of convective Taylor columns in rotating Rayleigh-B{\'e}nard convection},
  author={Grooms, I. and Julien, K. and Weiss, J. B. and Knobloch, E.},
  journal={Physical review letters},
  volume={104},
  number={22},
  pages={224501},
  year={2010},
  publisher={APS}
}

@article{bouillaut2021,
  title={Experimental observation of the geostrophic turbulence regime of rapidly rotating convection},
  author={Bouillaut, V. and Miquel, B. and Julien, K. and Auma{\^\i}tre, S. and Gallet, B.},
  journal={Proceedings of the National Academy of Sciences},
  volume={118},
  number={44},
  pages={e2105015118},
  year={2021},
  publisher={National Academy of Sciences}
}

@article{roberts1967convection,
  title={Convection in horizontal layers with internal heat generation. {{{Theory}}}},
  author={Roberts, P. H.},
  journal={Journal of Fluid Mechanics},
  volume={30},
  number={1},
  pages={33--49},
  year={1967},
  publisher={Cambridge University Press}
}

@book{chandrasekhar1961hydrodynamic,
  title={Hydrodynamic and hydromagnetic stability},
  author={Chandrasekhar, S.},
  year={1961},
  publisher={Oxford University Pressn}
}

@article{stevens2010radial,
  title={Radial boundary layer structure and Nusselt number in Rayleigh--B{\'e}nard convection},
  author={Stevens, R. J. A. M. and Verzicco, R. and Lohse, D.},
  journal={Journal of Fluid Mechanics},
  volume={643},
  pages={495--507},
  year={2010},
  publisher={Cambridge University Press}
}

@article{kunnen2016transition,
  title={Transition to geostrophic convection: the role of the boundary conditions},
  author={Kunnen, R. P. J. and Ostilla-M{\'o}nico, R. and Van Der Poel, E. P. and Verzicco, R. and Lohse, D.},
  journal={Journal of Fluid Mechanics},
  volume={799},
  pages={413--432},
  year={2016},
  publisher={Cambridge University Press}
}

@article{goluskin2012convection,
  title={ {Convection driven by internal heating} },
  author={Goluskin, D. and Spiegel, E. A.},
  journal={Physics Letters A},
  volume={377},
  number={1-2},
  pages={83--92},
  year={2012},
  doi={10.1016/j.physleta.2012.10.037},
  publisher={Elsevier}
}

@book{Goluskin2016book,
author = {Goluskin, D.},
doi = {10.1007/978-3-319-23941-5},
pages = {VIII, 64},
publisher = {Springer, Cham},
series = {Springer Briefs in Applied Sciences and Technologies},
title = {{Internally heated convection and Rayleigh--B\'enard convection}},
url = {https://doi.org/10.1007/978-3-319-23941-5},
year = {2016},
number = {},
}

@article{hadjerci2024,
  title={Rapidly rotating radiatively driven convection: experimental and numerical validation of the ‘geostrophic turbulence’scaling predictions},
  author={Hadjerci, G. and Bouillaut, V. and Miquel, B. and Gallet, B.},
  journal={Journal of Fluid Mechanics},
  volume={998},
  pages={A9},
  year={2024},
  publisher={Cambridge University Press}
}

@article{zhong2009,
  title = {Prandtl-, Rayleigh-, and Rossby-Number Dependence of Heat Transport in Turbulent Rotating Rayleigh-B\'enard Convection},
  author = {Zhong, J. and Stevens, R. J. A. M. and Clercx, H. J. H. and Verzicco, R. and Lohse, D. and Ahlers, G.},
  journal = {Phys. Rev. Lett.},
  volume = {102},
  issue = {4},
  pages = {044502},
  numpages = {4},
  year = {2009},
  month = {Jan},
  publisher = {American Physical Society},
  doi = {10.1103/PhysRevLett.102.044502},
  url = {https://link.aps.org/doi/10.1103/PhysRevLett.102.044502}
}

@article{Stevens_2010,
doi = {10.1088/1367-2630/12/7/075005},
url = {https://doi.org/10.1088/1367-2630/12/7/075005},
year = {2010},
month = {jul},
publisher = {},
volume = {12},
number = {7},
pages = {075005},
author = {Stevens, R. J. A. M. and Clercx, H. J. H. and Lohse, D.},
title = {Optimal Prandtl number for heat transfer in rotating Rayleigh–Bénard convection},
journal = {New Journal of Physics}
}

@article{lohse2023,
    author = {Lohse, D. and Shishkina, O. },
    title = {Ultimate turbulent thermal convection},
    journal = {Physics Today 1},
    year = {2023},
    volume = {76(11)},
    pages ={26--32}
}

@article{ecke2023,
  title={Turbulent rotating rayleigh--b{\'e}nard convection},
  author={Ecke, R. E. and Shishkina, O.},
  journal={Annual review of fluid mechanics},
  volume={55},
  pages={603--638},
  year={2023},
  publisher={Annual Reviews}
}

@article{Kunnen04052021,
author = {R. P. J. Kunnen},
title = {The geostrophic regime of rapidly rotating turbulent convection},
journal = {Journal of Turbulence},
volume = {22},
number = {4-5},
pages = {267--296},
year = {2021},
publisher = {Taylor \& Francis},
doi = {10.1080/14685248.2021.1876877},
}

@article{maffei2021inverse,
  title={On the inverse cascade and flow speed scaling behaviour in rapidly rotating Rayleigh--B{\'e}nard convection},
  author={Maffei, S. and Krouss, M. J. and Julien, K. and Calkins, M. A.},
  journal={Journal of Fluid Mechanics},
  volume={913},
  pages={A18},
  year={2021},
  publisher={Cambridge University Press}
}

@article{king2009b,
  title={Boundary layer control of rotating convection systems},
  author={King, E. M. and Stellmach, S. and Noir, J. and Hansen, U. and Aurnou, J. M.},
  journal={Nature},
  volume={457},
  number={7227},
  pages={301--304},
  year={2009},
  publisher={Nature Publishing Group UK London}
}

@article{abbate2023rotating,
  title={Rotating convective turbulence in moderate to high Prandtl number fluids},
  author={Abbate, J. A. and Aurnou, J. M.},
  journal={Geophysical \& Astrophysical Fluid Dynamics},
  volume={117},
  number={6},
  pages={397--436},
  year={2023},
  publisher={Taylor \& Francis}
}

@article{Arslan2021,
author={Arslan, A. and Fantuzzi, G. and Craske, J. and Wynn, A.},
doi={10.1017/jfm.2021.360},
journal = {Journal of Fluid Mechanics},
volume={919},
pages = {A15},
title = {{Bounds on heat transport for convection driven by internal heating}},
year = {2021},
}

@article{arslan2025a,
    author = {Arslan, A. and Rojas, R. E.},
    title = {New bounds for heat transport in internally heated convection at infinite Prandtl number},
    journal = {Journal of Mathematical Physics},
    volume = {66},
    number = {2},
    pages = {023101},
    year = {2025},
    month = {02},
    issn = {0022-2488},
    doi = {10.1063/5.0209505},
    url = {https://doi.org/10.1063/5.0209505},
}

@article{grossmann2000scaling,
  title={Scaling in thermal convection: a unifying theory},
  author={Grossmann, S. and Lohse, D.},
  journal={Journal of Fluid Mechanics},
  volume={407},
  pages={27--56},
  year={2000},
  publisher={Cambridge University Press}
}

\section{Summary of numerical resolutions and results}
\label{sec:appb}

Table \ref{tbl:summary} presents a summary of all numerical results.

\afterpage{\clearpage}
\begin{landscape}
\begin{longtable}[c]{|c|c|c|c|c|c|c|c|c|c|}
\caption{Summary of results}
\label{tbl:summary}\\

\hline
$Pr$ & $R$ & $E$ & $\Gamma$ & $N_{x,y}\times N_z$ & $t_{av}$ & $\langle T\rangle$ & $\langle wT \rangle$ & $Re_w$ \\
\hline
\endhead
0.1 & $1.00\times 10^6$ & $\infty$            & $3.14$ & $288^2\times144$  & $2.5\times10^{0}$ & $5.95\times10^{-2}$ & $7.20\times 10^{-2}$ & $2.19\times 10^{2}$ \\
0.1 & $1.00\times 10^6$ & $1.00\times10^{-2}$ & $3.14$ & $288^2\times144$  & $6.3\times10^{-1}$ & $5.97\times10^{-2}$ & $7.30\times 10^{-2}$ & $2.23\times 10^{2}$ \\
0.1 & $1.00\times 10^6$ & $3.16\times10^{-3}$ & $3.14$ & $288^2\times144$  & $1.6\times10^{0}$ & $5.99\times10^{-2}$ & $7.88\times 10^{-2}$ & $2.20\times 10^{2}$ \\
0.1 & $1.00\times 10^6$ & $1.00\times10^{-3}$ & $3.14$ & $288^2\times144$  & $1.6\times10^{0}$ & $6.48\times10^{-2}$ & $6.87\times 10^{-2}$ & $1.69\times 10^{2}$ \\
\hline 
0.1 & $3.16\times 10^6$ & $\infty$ & $3.14$ & $288^2\times144$ & $3.6\times10^{-1}$ & $4.98\times10^{-2}$ & $9.86\times 10^{-2}$ & $3.66\times 10^{2}$ \\
\hline 
0.1 & $1.00\times 10^7$ & $\infty$            & $3.14$ & $288^2\times144$  & $4.0\times10^{-1}$ & $4.15\times10^{-2}$ & $1.21\times 10^{-1}$ & $5.75\times 10^{2}$ \\
0.1 & $1.00\times 10^7$ & $3.16\times10^{-3}$ & $3.14$ & $288^2\times144$  & $2.0\times10^{-1}$ & $4.18\times10^{-2}$ & $1.22\times 10^{-1}$ & $5.77\times 10^{2}$ \\
0.1 & $1.00\times 10^7$ & $1.00\times10^{-3}$ & $3.14$ & $288^2\times144$  & $2.0\times10^{-1}$ & $4.18\times10^{-2}$ & $1.44\times 10^{-1}$ & $5.95\times 10^{2}$ \\
0.1 & $1.00\times 10^7$ & $3.16\times10^{-4}$ & $3.14$ & $288^2\times288$  & $2.0\times10^{-1}$ & $4.56\times10^{-2}$ & $1.39\times 10^{-1}$ & $4.87\times 10^{2}$ \\
\hline
0.1 & $3.16\times 10^7$ & $\infty$            & $3.14$ & $288^2\times144$  & $1.7\times10^{-1}$ & $3.41\times10^{-2}$ & $1.53\times 10^{-1}$ & $9.13\times 10^{2}$ \\ 
\hline 
0.1 & $1.00\times 10^8$ & $\infty$            & $3.14$ & $288^2\times144$  & $3.2\times10^{-1}$ & $2.77\times10^{-2}$ & $1.65\times 10^{-1}$ & $1.38\times 10^{3}$ \\
0.1 & $1.00\times 10^8$ & $3.16\times10^{-3}$ & $3.14$ & $288^2\times144$  & $3.2\times10^{-1}$ & $2.76\times10^{-2}$ & $1.62\times 10^{-1}$ & $1.36\times 10^{3}$ \\
0.1 & $1.00\times 10^8$ & $1.00\times10^{-3}$ & $3.14$ & $288^2\times144$  & $3.2\times10^{-1}$ & $2.79\times10^{-2}$ & $1.74\times 10^{-1}$ & $1.40\times 10^{3}$ \\
0.1 & $1.00\times 10^8$ & $3.16\times10^{-4}$ & $3.14$ & $288^2\times288$  & $8.4\times10^{-1}$ & $2.81\times10^{-2}$ & $2.04\times 10^{-1}$ & $1.40\times 10^{3}$ \\
0.1 & $1.00\times 10^8$ & $1.00\times10^{-4}$ & $3.14$ & $288^2\times288$  & $2.0\times10^{-1}$ & $3.13\times10^{-2}$ & $1.97\times 10^{-1}$ & $1.18\times 10^{3}$ \\
0.1 & $1.00\times 10^8$ & $3.16\times10^{-5}$ & $3.14$ & $288^2\times288$  & $2.7\times10^{-1}$ & $4.48\times10^{-2}$ & $1.26\times 10^{-1}$ & $8.97\times 10^{2}$ \\
\hline
0.1 & $3.16\times 10^8$ & $\infty$            & $1.5$ & $288^2\times144$ & $1.8\times10^{-1}$ & $2.21\times10^{-2}$ & $1.87\times 10^{-1}$ & $2.09\times 10^{3}$ \\
\hline 
0.1 & $1.00\times 10^9$ & $\infty$            & $1.5$ & $288^2\times144$  & $1.5\times10^{-1}$ & $1.78\times10^{-2}$ & $2.03\times 10^{-1}$ & $3.03\times 10^{3}$ \\
0.1 & $1.00\times 10^9$ & $3.16\times10^{-3}$ & $1.5$ & $288^2\times144$  & $1.5\times10^{-1}$ & $1.77\times10^{-2}$ & $2.02\times 10^{-1}$ & $3.02\times 10^{3}$ \\
0.1 & $1.00\times 10^9$ & $1.00\times10^{-3}$ & $1.5$ & $288^2\times144$  & $1.5\times10^{-1}$ & $1.78\times10^{-2}$ & $2.01\times 10^{-1}$ & $3.04\times 10^{3}$ \\
0.1 & $1.00\times 10^9$ & $3.16\times10^{-4}$ & $1.5$ & $288^2\times288$  & $1.2\times10^{-1}$ & $1.79\times10^{-2}$ & $2.16\times 10^{-1}$ & $3.09\times 10^{3}$ \\
0.1 & $1.00\times 10^9$ & $1.00\times10^{-4}$ & $1.5$ & $288^2\times288$  & $8.0\times10^{-2}$ & $1.82\times10^{-2}$ & $2.62\times 10^{-1}$ & $3.11\times 10^{3}$ \\
0.1 & $1.00\times 10^9$ & $3.16\times10^{-5}$ & $1.5$ & $288^2\times384$  & $3.3\times10^{-2}$ & $2.07\times10^{-2}$ & $2.46\times 10^{-1}$ & $2.69\times 10^{3}$ \\
0.1 & $1.00\times 10^9$ & $1.00\times10^{-5}$ & $1.5$ & $288^2\times512$  & $1.3\times10^{-2}$ & $3.01\times10^{-2}$ & $1.85\times 10^{-1}$ & $2.38\times 10^{3}$ \\
\hline
0.1 & $3.16\times 10^9$ & $\infty$             & $1$ & $288^2\times144$ & $6.4\times10^{-3}$ & $1.40\times10^{-2}$ & $2.12\times 10^{-1}$ & $4.46\times 10^{3}$ \\
\hline 
0.1 & $1.00\times 10^{10}$ & $\infty$            & $1$ & $384^2\times288$  & $5.0\times10^{-2}$ & $1.11\times10^{-2}$ & $2.31\times 10^{-1}$ & $6.81\times 10^{3}$ \\
0.1 & $1.00\times 10^{10}$ & $3.16\times10^{-3}$ & $1$ & $384^2\times288$  & $5.0\times10^{-2}$ & $1.10\times10^{-2}$ & $2.32\times 10^{-1}$ & $6.88\times 10^{3}$ \\
0.1 & $1.00\times 10^{10}$ & $1.00\times10^{-3}$ & $1$ & $384^2\times288$  & $3.5\times10^{-2}$ & $1.10\times10^{-2}$ & $2.32\times 10^{-1}$ & $6.88\times 10^{3}$ \\
0.1 & $1.00\times 10^{10}$ & $3.16\times10^{-4}$ & $1$ & $384^2\times288$  & $2.9\times10^{-2}$ & $1.11\times10^{-2}$ & $2.36\times 10^{-1}$ & $6.92\times 10^{3}$ \\
0.1 & $1.00\times 10^{10}$ & $1.00\times10^{-4}$ & $1$ & $384^2\times384$  & $2.7\times10^{-2}$ & $1.12\times10^{-2}$ & $2.57\times 10^{-1}$ & $7.13\times 10^{3}$ \\
0.1 & $1.00\times 10^{10}$ & $3.16\times10^{-5}$ & $1$ & $384^2\times384$  & $1.6\times10^{-2}$ & $1.16\times10^{-1}$ & $3.03\times 10^{-1}$ & $6.82\times 10^{3}$ \\
0.1 & $1.00\times 10^{10}$ & $1.00\times10^{-5}$ & $1$ & $384^2\times512$  & $1.5\times10^{-2}$ & $1.34\times10^{-2}$ & $2.94\times 10^{-1}$ & $6.20\times 10^{3}$ \\
0.1 & $1.00\times 10^{10}$ & $3.16\times10^{-6}$ & $1$ & $384^2\times512$  & $2.5\times10^{-2}$ & $2.06\times10^{-2}$ & $2.35\times 10^{-1}$ & $6.21\times 10^{3}$ \\
0.1 & $1.00\times 10^{10}$ & $1.00\times10^{-6}$ & $1$ & $384^2\times512$  & $2.2\times10^{-2}$ & $4.27\times10^{-2}$ & $1.36\times 10^{-1}$ & $4.02\times 10^{3}$ \\
\hline
\hline
0.3 & $1.00\times 10^6$ & $\infty$            & $3.14$ & $288^2\times144$  & $3.7\times10^{-1}$ & $5.61\times10^{-2}$ & $9.23\times 10^{-2}$ & $9.36\times 10^{1}$ \\
0.3 & $1.00\times 10^6$ & $1.00\times10^{-2}$ & $3.14$ & $288^2\times144$  & $3.7\times10^{-1}$ & $5.59\times10^{-2}$ & $9.59\times 10^{-2}$ & $9.48\times 10^{1}$ \\
0.3 & $1.00\times 10^6$ & $3.16\times10^{-3}$ & $3.14$ & $288^2\times144$  & $3.7\times10^{-1}$ & $5.63\times10^{-1}$ & $1.13\times 10^{-1}$ & $9.40\times 10^{1}$ \\
0.3 & $1.00\times 10^6$ & $1.00\times10^{-3}$ & $3.14$ & $288^2\times144$  & $3.7\times10^{-1}$ & $6.81\times10^{-2}$ & $5.61\times 10^{-2}$ & $4.99\times 10^{1}$ \\
\hline 
0.3 & $3.16\times 10^6$ & $\infty$            & $3.14$ & $288^2\times144$  & $2.1\times10^{-1}$ & $4.72\times10^{-2}$ & $1.16\times 10^{-2}$ & $1.55\times 10^{2}$ \\
\hline 
0.3 & $1.00\times 10^7$ & $\infty$            & $3.14$ & $288^2\times144$  & $2.3\times10^{-1}$ & $3.92\times10^{-2}$ & $1.40\times 10^{-1}$ & $2.53\times 10^{2}$ \\
0.3 & $1.00\times 10^7$ & $3.16\times10^{-3}$ & $3.14$ & $288^2\times144$  & $1.2\times10^{-1}$ & $3.90\times10^{-2}$ & $1.49\times 10^{-1}$ & $2.55\times 10^{2}$ \\
0.3 & $1.00\times 10^7$ & $1.00\times10^{-3}$ & $3.14$ & $288^2\times144$  & $1.2\times10^{-1}$ & $3.91\times10^{-2}$ & $1.76\times 10^{-1}$ & $2.50\times 10^{2}$ \\
0.3 & $1.00\times 10^7$ & $3.16\times10^{-4}$ & $3.14$ & $288^2\times288$  & $1.2\times10^{-1}$ & $4.47\times10^{-2}$ & $1.44\times 10^{-1}$ & $1.80\times 10^{2}$ \\
\hline
0.3 & $3.16\times 10^7$ & $\infty$            & $3.14$ & $288^2\times144$  & $9.7\times10^{-2}$ & $3.22\times10^{-2}$ & $1.61\times 10^{-1}$ & $4.00\times 10^{2}$ \\ 
\hline 
0.3 & $1.00\times 10^8$ & $\infty$            & $3.14$ & $288^2\times144$  & $5.0\times10^{-2}$ & $2.61\times10^{-2}$ & $1.88\times 10^{-1}$ & $6.28\times 10^{2}$ \\
0.3 & $1.00\times 10^8$ & $3.16\times10^{-3}$ & $3.14$ & $288^2\times144$  & $5.0\times10^{-2}$ & $2.60\times10^{-2}$ & $1.90\times 10^{-1}$ & $6.28\times 10^{2}$ \\
0.3 & $1.00\times 10^8$ & $1.00\times10^{-3}$ & $3.14$ & $288^2\times144$  & $5.0\times10^{-2}$ & $2.61\times10^{-2}$ & $2.10\times 10^{-1}$ & $6.52\times 10^{2}$ \\
0.3 & $1.00\times 10^8$ & $3.16\times10^{-4}$ & $3.14$ & $288^2\times288$  & $5.0\times10^{-2}$ & $2.58\times10^{-2}$ & $2.31\times 10^{-1}$ & $5.84\times 10^{2}$ \\
0.3 & $1.00\times 10^8$ & $1.00\times10^{-4}$ & $3.14$ & $288^2\times288$  & $5.0\times10^{-2}$ & $2.97\times10^{-2}$ & $2.07\times 10^{-1}$ & $4.66\times 10^{2}$ \\
0.3 & $1.00\times 10^8$ & $3.16\times10^{-5}$ & $3.14$ & $288^2\times288$  & $5.1\times10^{-2}$ & $5.30\times10^{-2}$ & $1.02\times 10^{-1}$ & $2.47\times 10^{2}$ \\
\hline
0.3 & $3.16\times 10^8$ & $\infty$            & $1.5$ & $288^2\times144$  & $5.0\times10^{-2}$ & $2.07\times10^{-2}$ & $2.04\times 10^{-1}$ & $9.64\times 10^{3}$ \\
\hline 
0.3 & $1.00\times 10^9$ & $\infty$            & $1.5$ & $288^2\times144$  & $5.8\times10^{-2}$ & $1.66\times10^{-2}$ & $2.25\times 10^{-1}$ & $1.48\times 10^{3}$ \\
0.3 & $1.00\times 10^9$ & $3.16\times10^{-3}$ & $1.5$ & $288^2\times144$  & $5.8\times10^{-2}$ & $1.66\times10^{-2}$ & $2.27\times 10^{-1}$ & $1.49\times 10^{3}$ \\
0.3 & $1.00\times 10^9$ & $1.00\times10^{-3}$ & $1.5$ & $288^2\times144$  & $5.8\times10^{-2}$ & $1.66\times10^{-2}$ & $2.26\times 10^{-1}$ & $1.50\times 10^{3}$ \\
0.3 & $1.00\times 10^9$ & $3.16\times10^{-4}$ & $1.5$ & $288^2\times288$  & $5.9\times10^{-2}$ & $1.67\times10^{-2}$ & $2.70\times 10^{-1}$ & $1.53\times 10^{3}$ \\
0.3 & $1.00\times 10^9$ & $1.00\times10^{-4}$ & $1.5$ & $288^2\times288$  & $2.3\times10^{-2}$ & $1.68\times10^{-2}$ & $2.79\times 10^{-1}$ & $1.32\times 10^{3}$ \\
0.3 & $1.00\times 10^9$ & $3.16\times10^{-5}$ & $1.5$ & $288^2\times384$  & $4.6\times10^{-2}$ & $1.99\times10^{-2}$ & $2.55\times 10^{-1}$ & $1.11\times 10^{3}$ \\
0.3 & $1.00\times 10^9$ & $1.00\times10^{-5}$ & $1.5$ & $288^2\times512$  & $1.7\times10^{-2}$ & $3.28\times10^{-2}$ & $1.82\times 10^{-1}$ & $7.51\times 10^{2}$ \\
\hline
0.3 & $3.16\times 10^9$ & $\infty$             & $1$ & $288^2\times144$    & $1.7\times10^{-2}$ & $1.31\times10^{-2}$ & $2.30\times 10^{-1}$ & $2.12\times 10^{3}$ \\
\hline 
0.3 & $1.00\times 10^{10}$ & $\infty$            & $1$ & $384^2\times288$  & $1.1\times10^{-2}$ & $1.02\times10^{-2}$ & $2.41\times 10^{-1}$ & $3.23\times 10^{3}$ \\
0.3 & $1.00\times 10^{10}$ & $3.16\times10^{-3}$ & $1$ & $384^2\times288$  & $2.1\times10^{-2}$ & $1.02\times10^{-2}$ & $2.47\times 10^{-1}$ & $3.24\times 10^{3}$ \\
0.3 & $1.00\times 10^{10}$ & $1.00\times10^{-3}$ & $1$ & $384^2\times288$  & $2.9\times10^{-2}$ & $1.03\times10^{-2}$ & $2.55\times 10^{-1}$ & $3.27\times 10^{3}$ \\
0.3 & $1.00\times 10^{10}$ & $3.16\times10^{-4}$ & $1$ & $384^2\times288$  & $1.4\times10^{-2}$ & $1.04\times10^{-2}$ & $2.77\times 10^{-1}$ & $3.41\times 10^{3}$ \\
0.3 & $1.00\times 10^{10}$ & $1.00\times10^{-4}$ & $1$ & $384^2\times384$  & $1.0\times10^{-2}$ & $1.04\times10^{-2}$ & $3.15\times 10^{-1}$ & $3.36\times 10^{3}$ \\
0.3 & $1.00\times 10^{10}$ & $3.16\times10^{-5}$ & $1$ & $384^2\times384$  & $8.1\times10^{-3}$ & $1.08\times10^{-2}$ & $3.24\times 10^{-1}$ & $2.91\times 10^{3}$ \\
0.3 & $1.00\times 10^{10}$ & $1.00\times10^{-5}$ & $1$ & $384^2\times512$  & $3.2\times10^{-3}$ & $1.30\times10^{-2}$ & $3.00\times 10^{-1}$ & $2.70\times 10^{3}$ \\
0.3 & $1.00\times 10^{10}$ & $3.16\times10^{-6}$ & $1$ & $384^2\times512$  & $1.2\times10^{-2}$ & $2.16\times10^{-2}$ & $2.37\times 10^{-1}$ & $1.95\times 10^{3}$ \\
0.3 & $1.00\times 10^{10}$ & $1.00\times10^{-6}$ & $1$ & $384^2\times512$  & $9.1\times10^{-3}$ & $5.57\times10^{-2}$ & $9.40\times 10^{-2}$ & $8.71\times 10^{2}$ \\
\hline
\hline
3 & $1.00\times 10^6$ & $\infty$            & $3.14$ & $288^2\times144$  & $8.0\times10^{-1}$ & $5.57\times10^{-2}$ & $1.30\times 10^{-1}$ & $1.22\times 10^{1}$ \\
3 & $1.00\times 10^6$ & $1.00\times10^{-2}$ & $3.14$ & $288^2\times144$  & $4.0\times10^{-1}$ & $5.48\times10^{-2}$ & $1.37\times 10^{-1}$ & $1.18\times 10^{1}$ \\
3 & $1.00\times 10^6$ & $3.16\times10^{-3}$ & $3.14$ & $288^2\times144$  & $2.0\times10^{-1}$ & $5.55\times10^{-1}$ & $1.32\times 10^{-1}$ & $1.02\times 10^{1}$ \\
3 & $1.00\times 10^6$ & $1.00\times10^{-3}$ & $3.14$ & $288^2\times144$  & $2.0\times10^{-1}$ & $6.81\times10^{-2}$ & $6.16\times 10^{-2}$ & $5.32\times 10^{0}$ \\
\hline 
3 & $3.16\times 10^6$ & $\infty$            & $3.14$ & $288^2\times144$  & $3.2\times10^{-1}$ & $4.62\times10^{-2}$ & $1.65\times 10^{-1}$ & $2.13\times 10^{1}$ \\
\hline 
3 & $1.00\times 10^7$ & $\infty$            & $3.14$ & $288^2\times144$  & $1.8\times10^{-1}$ & $3.83\times10^{-2}$ & $1.95\times 10^{-1}$ & $3.59\times 10^{1}$ \\
3 & $1.00\times 10^7$ & $3.16\times10^{-3}$ & $3.14$ & $288^2\times144$  & $1.8\times10^{-1}$ & $3.69\times10^{-2}$ & $2.10\times 10^{-1}$ & $3.37\times 10^{1}$ \\
3 & $1.00\times 10^7$ & $1.00\times10^{-3}$ & $3.14$ & $288^2\times144$  & $1.8\times10^{-1}$ & $3.70\times10^{-2}$ & $2.11\times 10^{-1}$ & $2.89\times 10^{1}$ \\
3 & $1.00\times 10^7$ & $3.16\times10^{-4}$ & $3.14$ & $288^2\times288$  & $1.1\times10^{-1}$ & $4.49\times10^{-2}$ & $1.55\times 10^{-1}$ & $1.95\times 10^{1}$ \\
\hline
3 & $3.16\times 10^7$ & $\infty$            & $3.14$ & $288^2\times144$  & $1.0\times10^{-1}$ & $3.16\times10^{-2}$ & $2.17\times 10^{-1}$ & $5.95\times 10^{1}$ \\ 
\hline 
3 & $1.00\times 10^8$ & $\infty$            & $3.14$ & $288^2\times144$  & $8.0\times10^{-2}$ & $2.57\times10^{-2}$ & $2.38\times 10^{-1}$ & $9.69\times 10^{1}$ \\
3 & $1.00\times 10^8$ & $3.16\times10^{-3}$ & $3.14$ & $288^2\times144$  & $8.0\times10^{-2}$ & $2.56\times10^{-2}$ & $2.49\times 10^{-1}$ & $9.68\times 10^{1}$ \\
3 & $1.00\times 10^8$ & $1.00\times10^{-3}$ & $3.14$ & $288^2\times144$  & $5.8\times10^{-2}$ & $2.40\times10^{-2}$ & $2.68\times 10^{-1}$ & $8.75\times 10^{1}$ \\
3 & $1.00\times 10^8$ & $3.16\times10^{-4}$ & $3.14$ & $288^2\times288$  & $2.9\times10^{-2}$ & $2.34\times10^{-2}$ & $2.74\times 10^{-1}$ & $7.31\times 10^{1}$ \\
3 & $1.00\times 10^8$ & $1.00\times10^{-4}$ & $3.14$ & $288^2\times288$  & $3.5\times10^{-2}$ & $2.79\times10^{-2}$ & $2.26\times 10^{-1}$ & $5.21\times 10^{1}$ \\
3 & $1.00\times 10^8$ & $3.16\times10^{-5}$ & $3.14$ & $288^2\times288$  & $1.2\times10^{-2}$ & $5.75\times10^{-2}$ & $8.99\times 10^{-2}$ & $2.18\times 10^{1}$ \\
\hline
3 & $3.16\times 10^8$ & $\infty$            & $1.5$ & $288^2\times144$  & $9.7\times10^{-2}$ & $2.08\times10^{-2}$ & $2.56\times 10^{-1}$ & $1.56\times 10^{2}$ \\
\hline 
3 & $1.00\times 10^9$ & $\infty$            & $1.5$ & $288^2\times144$  & $1.8\times10^{-2}$ & $1.65\times10^{-2}$ & $2.75\times 10^{-1}$ & $2.54\times 10^{2}$ \\
3 & $1.00\times 10^9$ & $3.16\times10^{-3}$ & $1.5$ & $288^2\times144$  & $1.8\times10^{-2}$ & $1.65\times10^{-2}$ & $2.82\times 10^{-1}$ & $2.56\times 10^{2}$ \\
3 & $1.00\times 10^9$ & $1.00\times10^{-3}$ & $1.5$ & $288^2\times144$  & $1.8\times10^{-2}$ & $1.63\times10^{-2}$ & $2.97\times 10^{-1}$ & $2.47\times 10^{2}$ \\
3 & $1.00\times 10^9$ & $3.16\times10^{-4}$ & $1.5$ & $288^2\times288$  & $3.7\times10^{-2}$ & $1.52\times10^{-2}$ & $3.13\times 10^{-1}$ & $2.16\times 10^{2}$ \\
3 & $1.00\times 10^9$ & $1.00\times10^{-4}$ & $1.5$ & $288^2\times288$  & $1.5\times10^{-2}$ & $1.45\times10^{-2}$ & $3.24\times 10^{-1}$ & $1.75\times 10^{2}$ \\
3 & $1.00\times 10^9$ & $3.16\times10^{-5}$ & $1.5$ & $288^2\times384$  & $1.5\times10^{-2}$ & $1.66\times10^{-2}$ & $2.80\times 10^{-1}$ & $1.24\times 10^{2}$ \\
3 & $1.00\times 10^9$ & $1.00\times10^{-5}$ & $1.5$ & $288^2\times512$  & $1.5\times10^{-2}$ & $3.47\times10^{-2}$ & $1.78\times 10^{-1}$ & $6.89\times 10^{1}$ \\
\hline
3 & $3.16\times 10^9$ & $\infty$             & $1$ & $288^2\times144$    & $1.2\times10^{-2}$ & $1.31\times10^{-2}$ & $2.91\times 10^{-1}$ & $3.95\times 10^{2}$ \\
\hline 
3 & $1.00\times 10^{10}$ & $\infty$            & $1$ & $384^2\times288$  & $4.6\times10^{-3}$ & $1.02\times10^{-2}$ & $3.03\times 10^{-1}$ & $6.36\times 10^{2}$ \\
3 & $1.00\times 10^{10}$ & $3.16\times10^{-3}$ & $1$ & $384^2\times288$  & $3.5\times10^{-3}$ & $1.02\times10^{-2}$ & $3.09\times 10^{-1}$ & $6.28\times 10^{2}$ \\
3 & $1.00\times 10^{10}$ & $1.00\times10^{-3}$ & $1$ & $384^2\times288$  & $3.5\times10^{-3}$ & $1.02\times10^{-2}$ & $3.30\times 10^{-1}$ & $6.27\times 10^{2}$ \\
3 & $1.00\times 10^{10}$ & $3.16\times10^{-4}$ & $1$ & $384^2\times288$  & $4.6\times10^{-3}$ & $9.99\times10^{-3}$ & $3.49\times 10^{-1}$ & $6.05\times 10^{2}$ \\
3 & $1.00\times 10^{10}$ & $1.00\times10^{-4}$ & $1$ & $384^2\times384$  & $1.7\times10^{-3}$ & $9.58\times10^{-3}$ & $3.55\times 10^{-1}$ & $5.08\times 10^{2}$ \\
3 & $1.00\times 10^{10}$ & $3.16\times10^{-5}$ & $1$ & $384^2\times384$  & $4.6\times10^{-3}$ & $9.16\times10^{-3}$ & $3.54\times 10^{-1}$ & $4.02\times 10^{2}$ \\
3 & $1.00\times 10^{10}$ & $1.00\times10^{-5}$ & $1$ & $384^2\times512$  & $1.8\times10^{-3}$ & $1.08\times10^{-2}$ & $3.07\times 10^{-1}$ & $2.92\times 10^{2}$ \\
3 & $1.00\times 10^{10}$ & $3.16\times10^{-6}$ & $1$ & $384^2\times512$  & $3.2\times10^{-3}$ & $2.06\times10^{-2}$ & $2.51\times 10^{-1}$ & $1.78\times 10^{2}$ \\
3 & $1.00\times 10^{10}$ & $1.00\times10^{-6}$ & $1$ & $384^2\times512$  & $2.9\times10^{-4}$ & $6.05\times10^{-2}$ & $7.86\times 10^{-2}$ & $6.48\times 10^{1}$ \\
\hline
\hline
10 & $1.00\times 10^6$ & $\infty$            & $3.14$ & $288^2\times144$  & $1.6\times10^{-1}$ & $5.49\times10^{-2}$ & $1.40\times 10^{-1}$ & $3.72\times 10^{0}$ \\
10 & $1.00\times 10^6$ & $1.00\times10^{-2}$ & $3.14$ & $288^2\times144$  & $3.2\times10^{-1}$ & $5.47\times10^{-2}$ & $1.43\times 10^{-1}$ & $3.59\times 10^{0}$ \\
10 & $1.00\times 10^6$ & $3.16\times10^{-3}$ & $3.14$ & $288^2\times144$  & $3.2\times10^{-1}$ & $5.51\times10^{-1}$ & $1.36\times 10^{-1}$ & $3.13\times 10^{0}$ \\
10 & $1.00\times 10^6$ & $1.00\times10^{-3}$ & $3.14$ & $288^2\times144$  & $3.2\times10^{-1}$ & $6.77\times10^{-2}$ & $6.29\times 10^{-2}$ & $1.62\times 10^{0}$ \\
\hline 
10 & $3.16\times 10^6$ & $\infty$            & $3.14$ & $288^2\times144$  & $1.8\times10^{-1}$ & $4.67\times10^{-2}$ & $1.75\times 10^{-1}$ & $6.41\times 10^{0}$ \\
\hline 
10 & $1.00\times 10^7$ & $\infty$            & $3.14$ & $288^2\times144$  & $1.0\times10^{-1}$ & $3.86\times10^{-2}$ & $2.07\times 10^{-1}$ & $1.10\times 10^{1}$ \\
10 & $1.00\times 10^7$ & $3.16\times10^{-3}$ & $3.14$ & $288^2\times144$  & $1.0\times10^{-1}$ & $3.71\times10^{-2}$ & $2.16\times 10^{-1}$ & $1.04\times 10^{1}$ \\
10 & $1.00\times 10^7$ & $1.00\times10^{-3}$ & $3.14$ & $288^2\times144$  & $1.0\times10^{-1}$ & $3.70\times10^{-2}$ & $2.11\times 10^{-1}$ & $8.87\times 10^{0}$ \\
10 & $1.00\times 10^7$ & $3.16\times10^{-4}$ & $3.14$ & $288^2\times288$  & $6.0\times10^{-2}$ & $4.55\times10^{-2}$ & $1.49\times 10^{-1}$ & $5.81\times 10^{0}$ \\
\hline
10 & $3.16\times 10^7$ & $\infty$            & $3.14$ & $288^2\times144$  & $5.6\times10^{-2}$ & $3.16\times10^{-2}$ & $2.36\times 10^{-1}$ & $1.90\times 10^{1}$ \\ 
\hline 
10 & $1.00\times 10^8$ & $\infty$            & $3.14$ & $288^2\times144$  & $6.3\times10^{-2}$ & $2.57\times10^{-2}$ & $2.62\times 10^{-1}$ & $3.25\times 10^{1}$ \\
10 & $1.00\times 10^8$ & $3.16\times10^{-3}$ & $3.14$ & $288^2\times144$  & $3.2\times10^{-2}$ & $2.52\times10^{-2}$ & $2.75\times 10^{-1}$ & $3.11\times 10^{1}$ \\
10 & $1.00\times 10^8$ & $1.00\times10^{-3}$ & $3.14$ & $288^2\times144$  & $3.2\times10^{-2}$ & $2.39\times10^{-2}$ & $2.82\times 10^{-1}$ & $2.73\times 10^{1}$ \\
10 & $1.00\times 10^8$ & $3.16\times10^{-4}$ & $3.14$ & $288^2\times288$  & $3.2\times10^{-2}$ & $2.35\times10^{-2}$ & $2.69\times 10^{-1}$ & $2.19\times 10^{1}$ \\
10 & $1.00\times 10^8$ & $1.00\times10^{-4}$ & $3.14$ & $288^2\times288$  & $1.6\times10^{-2}$ & $2.78\times10^{-2}$ & $2.24\times 10^{-1}$ & $1.55\times 10^{1}$ \\
10 & $1.00\times 10^8$ & $3.16\times10^{-5}$ & $3.14$ & $288^2\times288$  & $1.9\times10^{-2}$ & $5.74\times10^{-2}$ & $8.69\times 10^{-2}$ & $6.52\times 10^{0}$ \\
\hline
10 & $3.16\times 10^8$ & $\infty$            & $1.5$ & $288^2\times144$  & $1.8\times10^{-2}$ & $2.10\times10^{-2}$ & $2.81\times 10^{-1}$ & $5.48\times 10^{1}$ \\
\hline 
10 & $1.00\times 10^9$ & $\infty$            & $1.5$ & $288^2\times144$  & $1.0\times10^{-2}$ & $1.68\times10^{-2}$ & $2.98\times 10^{-1}$ & $8.71\times 10^{2}$ \\
10 & $1.00\times 10^9$ & $3.16\times10^{-3}$ & $1.5$ & $288^2\times144$  & $1.0\times10^{-2}$ & $1.66\times10^{-2}$ & $3.03\times 10^{-1}$ & $8.81\times 10^{2}$ \\
10 & $1.00\times 10^9$ & $1.00\times10^{-3}$ & $1.5$ & $288^2\times144$  & $1.0\times10^{-2}$ & $1.59\times10^{-2}$ & $3.15\times 10^{-1}$ & $7.99\times 10^{2}$ \\
10 & $1.00\times 10^9$ & $3.16\times10^{-4}$ & $1.5$ & $288^2\times288$  & $1.0\times10^{-2}$ & $1.49\times10^{-2}$ & $3.26\times 10^{-1}$ & $6.64\times 10^{2}$ \\
10 & $1.00\times 10^9$ & $1.00\times10^{-4}$ & $1.5$ & $288^2\times288$  & $1.0\times10^{-2}$ & $1.46\times10^{-2}$ & $3.18\times 10^{-1}$ & $5.31\times 10^{2}$ \\
10 & $1.00\times 10^9$ & $3.16\times10^{-5}$ & $1.5$ & $288^2\times384$  & $1.0\times10^{-2}$ & $1.78\times10^{-2}$ & $2.72\times 10^{-1}$ & $3.75\times 10^{2}$ \\
10 & $1.00\times 10^9$ & $1.00\times10^{-5}$ & $1.5$ & $288^2\times512$  & $8.1\times10^{-3}$ & $3.50\times10^{-2}$ & $1.82\times 10^{-1}$ & $2.13\times 10^{1}$ \\
\hline
10 & $3.16\times 10^9$ & $\infty$             & $1$ & $288^2\times144$    & $3.4\times10^{-3}$ & $1.32\times10^{-2}$ & $3.10\times 10^{-1}$ & $1.49\times 10^{2}$ \\
\hline 
10 & $1.00\times 10^{10}$ & $\infty$            & $1$ & $384^2\times288$  & $1.0\times10^{-2}$ & $1.02\times10^{-2}$ & $3.31\times 10^{-1}$ & $2.40\times 10^{2}$ \\
10 & $1.00\times 10^{10}$ & $3.16\times10^{-3}$ & $1$ & $384^2\times288$  & $1.0\times10^{-2}$ & $1.02\times10^{-2}$ & $3.30\times 10^{-1}$ & $2.38\times 10^{2}$ \\
10 & $1.00\times 10^{10}$ & $1.00\times10^{-3}$ & $1$ & $384^2\times288$  & $1.0\times10^{-2}$ & $1.03\times10^{-2}$ & $3.38\times 10^{-1}$ & $2.25\times 10^{2}$ \\
10 & $1.00\times 10^{10}$ & $3.16\times10^{-4}$ & $1$ & $384^2\times288$  & $2.0\times10^{-2}$ & $9.85\times10^{-3}$ & $3.52\times 10^{-1}$ & $1.96\times 10^{2}$ \\
10 & $1.00\times 10^{10}$ & $1.00\times10^{-4}$ & $1$ & $384^2\times384$  & $1.0\times10^{-2}$ & $9.33\times10^{-3}$ & $3.70\times 10^{-1}$ & $1.62\times 10^{2}$ \\
10 & $1.00\times 10^{10}$ & $3.16\times10^{-5}$ & $1$ & $384^2\times384$  & $1.0\times10^{-2}$ & $9.03\times10^{-3}$ & $3.59\times 10^{-1}$ & $1.24\times 10^{2}$ \\
10 & $1.00\times 10^{10}$ & $1.00\times10^{-5}$ & $1$ & $384^2\times512$  & $8.1\times10^{-3}$ & $1.06\times10^{-2}$ & $3.34\times 10^{-1}$ & $8.88\times 10^{1}$ \\
10 & $1.00\times 10^{10}$ & $3.16\times10^{-6}$ & $1$ & $384^2\times512$  & $5.1\times10^{-3}$ & $2.02\times10^{-2}$ & $2.59\times 10^{-1}$ & $5.52\times 10^{1}$ \\
10 & $1.00\times 10^{10}$ & $1.00\times10^{-6}$ & $1$ & $384^2\times512$  & $6.3\times10^{-3}$ & $6.05\times10^{-2}$ & $6.72\times 10^{-2}$ & $1.79\times 10^{1}$ \\
\hline
\hline
30 & $1.00\times 10^6$    & $\infty$            & $3.14$ & $288^2\times144$  & $9.1\times10^{-2}$ & $5.46\times10^{-2}$ & $1.46\times 10^{-1}$ & $1.24\times 10^{0}$ \\
30 & $3.16\times 10^6$    & $\infty$            & $3.14$ & $288^2\times144$  & $1.0\times10^{-1}$ & $4.62\times10^{-2}$ & $1.81\times 10^{-1}$ & $2.01\times 10^{0}$ \\
30 & $1.00\times 10^7$    & $\infty$            & $3.14$ & $288^2\times144$  & $5.8\times10^{-2}$ & $3.84\times10^{-2}$ & $2.12\times 10^{-1}$ & $3.46\times 10^{0}$ \\
30 & $3.16\times 10^7$    & $\infty$            & $3.14$ & $288^2\times144$  & $3.2\times10^{-2}$ & $3.14\times10^{-2}$ & $2.40\times 10^{-1}$ & $6.00\times 10^{0}$ \\ 
30 & $1.00\times 10^8$    & $\infty$            & $3.14$ & $288^2\times144$  & $1.8\times10^{-2}$ & $2.54\times10^{-2}$ & $2.71\times 10^{-1}$ & $1.05\times 10^{1}$ \\
30 & $3.16\times 10^8$    & $\infty$            & $1.5$  & $288^2\times144$  & $1.0\times10^{-2}$ & $2.07\times10^{-2}$ & $2.92\times 10^{-1}$ & $1.89\times 10^{1}$ \\
30 & $1.00\times 10^9$    & $\infty$            & $1.5$  & $288^2\times144$  & $5.8\times10^{-3}$ & $1.67\times10^{-2}$ & $3.19\times 10^{-1}$ & $3.19\times 10^{1}$ \\
30 & $3.16\times 10^9$    & $\infty$            & $1$    & $288^2\times144$  & $1.9\times10^{-3}$ & $1.32\times10^{-2}$ & $3.34\times 10^{-1}$ & $5.53\times 10^{1}$ \\
30 & $1.00\times 10^{10}$ & $\infty$            & $1$    & $384^2\times288$  & $1.5\times10^{-3}$ & $1.03\times10^{-2}$ & $3.45\times 10^{-1}$ & $9.09\times 10^{1}$ \\
\hline
\hline
100 & $1.00\times 10^6$    & $\infty$            & $3.14$ & $288^2\times144$  & $1.0\times10^{-1}$ & $5.48\times10^{-2}$ & $1.48\times 10^{-1}$ & $3.57\times 10^{-1}$ \\
100 & $3.16\times 10^6$    & $\infty$            & $3.14$ & $288^2\times144$  & $5.6\times10^{-2}$ & $4.59\times10^{-2}$ & $1.83\times 10^{-1}$ & $5.92\times 10^{-1}$ \\
100 & $1.00\times 10^7$    & $\infty$            & $3.14$ & $288^2\times144$  & $3.2\times10^{-2}$ & $3.81\times10^{-2}$ & $2.15\times 10^{-1}$ & $9.84\times 10^{-1}$ \\
100 & $3.16\times 10^7$    & $\infty$            & $3.14$ & $288^2\times144$  & $1.8\times10^{-2}$ & $3.13\times10^{-2}$ & $2.43\times 10^{-1}$ & $1.66\times 10^{0}$ \\ 
100 & $1.00\times 10^8$    & $\infty$            & $3.14$ & $288^2\times144$  & $1.0\times10^{-2}$ & $2.52\times10^{-2}$ & $2.72\times 10^{-1}$ & $2.91\times 10^{0}$ \\
100 & $3.16\times 10^8$    & $\infty$            & $1.5$  & $288^2\times144$  & $1.0\times10^{-2}$ & $2.04\times10^{-2}$ & $2.98\times 10^{-1}$ & $4.97\times 10^{0}$ \\
100 & $1.00\times 10^9$    & $\infty$            & $1.5$  & $288^2\times144$  & $3.2\times10^{-3}$ & $1.64\times10^{-2}$ & $3.23\times 10^{-1}$ & $9.17\times 10^{0}$ \\
100 & $3.16\times 10^9$    & $\infty$            & $1$    & $288^2\times144$  & $1.1\times10^{-3}$ & $1.31\times10^{-2}$ & $3.46\times 10^{-1}$ & $1.57\times 10^{1}$ \\
100 & $1.00\times 10^{10}$ & $\infty$            & $1$    & $384^2\times288$  & $8.0\times10^{-4}$ & $1.03\times10^{-2}$ & $3.57\times 10^{-1}$ & $2.94\times 10^{1}$ \\
\hline
\caption{Summary of results}
\label{tbl:summary}
\end{longtable}
\end{landscape}

\end{document}